\documentclass{aastex631}

\usepackage{multirow}
\usepackage{comment}
\usepackage{amsmath}
\usepackage{natbib}

\def\sec#1{Section~\ref{sec:#1}}

\def\Table#1{Table~\ref{tab:#1}}

\newcommand{\Chap}[1]{{\protect\hyperref[ch:#1]{Chapter~\ref*{ch:#1}}}}

\newcommand{\Sec}[1]{{\protect\hyperref[sec:#1]{Section~\ref*{sec:#1}}}}
\newcommand{\Secs}[2]{{\protect\hyperref[sec:#1]{Sections~\ref*{sec:#1}}~and~\ref{sec:#2}}}
\newcommand{\Fig}[1]{{\protect\hyperref[fig:#1]{Fig.~\ref*{fig:#1}}}}

\newcommand{\subFig}[2]{{\protect\hyperref[fig:#1]{Fig.~\ref*{fig:#1}~#2}}}
\newcommand{\Eq}[1]{{\protect\hyperref[eq:#1]{Eq.~\ref*{eq:#1}}}}
\newcommand{\Eqs}[2]{{\protect\hyperref[eq:#1]{Eqs.~\ref*{eq:#1}}~and~\ref{eq:#2}}}
\newcommand{\Tab}[1]{{\protect\hyperref[tab:#1]{Table~\ref*{tab:#1}}}}

\newcommand{\App}[1]{{\protect\hyperref[app:#1]{Appendix~\ref*{app:#1}}}}

\newcommand{\kms}{\ifmmode\,{\rm km}\,{\rm s}^{-1}\else km$\,$s$^{-1}$\fi}
\newcommand{\Rd}{\ifmmode\,R_{\rm d}\else $R_{\rm d}$\fi}
\newcommand{\be}{\begin{equation}}
\newcommand{\ee}{\end{equation}}
\newcommand\ltsima{$\; \buildrel < \over \sim \;$}
\newcommand\ltsim{\lower.5ex\hbox{\ltsima}}
\newcommand\gtsima{$\; \buildrel > \over \sim \;$}
\newcommand\gtsim{\lower.5ex\hbox{\gtsima}}

\newcommand{\magss}{\ifmmode {{{{\rm mag}~{\rm arcsec}}^{-2}}}
             \else {{{mag}$~${arcsec}$^{-2}$}}
             \fi}

\def \ion#1#2{#1{\footnotesize{#2}}\relax}

\def \littleprime{\ifmmode{\scriptscriptstyle \prime }
     \else{\hbox{$\scriptscriptstyle \prime$ }}\fi}
\def \arcsec{\raise .9ex \hbox{\littleprime\hskip-3pt\littleprime}}

\begin{document}

\title{The Nearly Universal Disk Galaxy Rotation Curve}

\correspondingauthor{Stephane Courteau}
\email{stephanecourteau07@gmail.com}

\shorttitle{The URC}
\shortauthors{Patel et al.}

\author[0000-0000-0000-0000]{Raj Patel}
\affiliation{Department of Physics, Engineering Physics \& Astronomy, Queen's University, Kingston, ON K7L 3N6, Canada}
\author[0000-0002-3929-9316]{Nikhil Arora}
\affiliation{Department of Physics, Engineering Physics \& Astronomy, Queen's University, Kingston, ON K7L 3N6, Canada}
\author[0000-0002-8597-6277]{Ste\'phane Courteau}
\affiliation{Department of Physics, Engineering Physics \& Astronomy, Queen's University, Kingston, ON K7L 3N6, Canada}
 \author[0000-0002-9086-6398]{Connor Stone}
 \affiliation{Department of Physics, Universit{\'e} de Montr{\'e}al, Montr{\'e}al, Qu{\' e}bec, Canada}
 \affiliation{Mila - Qu{\' e}bec Artificial Intelligence Institute, Montr{\'e}al, Qu{\' e}bec, Canada}
 \affiliation{Ciela - Montr{\'e}al Institute for Astrophysical Data Analysis and Machine Learning, Montr{\'e}al, Qu{\' e}bec, Canada}
 \author[0000-0000-0000-0000]{Matthew Frosst}
 \affiliation{ICRAR, M468, University of Western Australia, Crawley WA 6009, Australia}
 \author[0000-0000-0000-0000]{Lawrence Widrow}
 \affiliation{Department of Physics, Engineering Physics \& Astronomy, Queen's University, Kingston, ON K7L 3N6, Canada}



\begin{abstract}

The Universal Rotation Curve (URC) of disk galaxies was originally proposed to predict the shape and amplitude of any rotation curve (RC) based solely on photometric data.  
Here, the URC is investigated with an extensive set of spatially-resolved rotation curvesdrawn from the PROBES-I, PROBES-II, and MaNGA data bases with matching multi-band surface brightness profiles from the DESI-LIS and WISE surveys for 3,846 disk galaxies.
Common URC formulations fail to achieve an adequate level of accuracy to qualify as truly universal over fully sampled RCs.
We develop neural network (NN) equivalents for the proposed URCs which predict RCs with higher accuracy, showing that URC inaccuracies are not due to insufficient data but rather non-optimal formulations or sampling effects.
This conclusion remains even if the total RC sample is pruned for symmetry. 
The latest URC prescriptions and their NN equivalents trained on our sub-sample of 579 disk galaxies with symmetric RCs perform similarly to the URC/NN trained on the complete data sample.
We conclude that a URC with an acceptable level of accuracy ($\Delta V_{\rm circ} \lesssim15$ per cent) at all radii would require a detailed modelling of a galaxy's central regions and outskirts (e.g., for baryonic effects leading to contraction or expansion of any dark-matter-only halo).

\end{abstract}

\keywords{Dark matter distribution, Galaxy disks, Galaxy dynamics, Galaxy kinematics, Galaxy mass distribution, Galaxy properties, Galaxy rotation curves, Galaxy structure}


\section{Introduction}
\label{sec:intro}

The Universal Rotation Curve (URC) hypothesis is the proposition that a disk galaxy's rotation curve (RC) can be predicted from a few key photometric properties.
This concept is motivated by the remarkable similarity of RCs between disk galaxies of varying Hubble type, first noted by \cite{Rubin+1985}.
While peak amplitudes differ by scaling factors between Hubble types, these authors noted that the form and shape of RCs between bulge-dominant (Sa) and nearly bulge-less (Sc) systems are comparable. 
Higher galaxy luminosity would scale with higher velocity and shallower outer velocity gradients.
\citet[][hereafter PS91]{PS91} and \citet[][hereafter PSS96]{PSS96} quantified these observations in the first URC formulations, proposing that a galaxy's RC may be determined by its luminosity and radial scale length when scaled by a velocity input\footnote{This assertion holds over the range $1/3 \leq R/R_{\rm opt} \leq 1.1$, where $R_{\rm opt}$
is the ``optical'' radius which encloses 83 per cent of the total light.}.

\cite{Salucci+2007}, \cite{Karukes+Salucci2017}, and \cite{Paolo+2019} built upon the URC foundation set by PSS96, which was largely based on a sample dominated by bright galaxies, to establish similar URCs for dwarf and low surface brightness (LSB) galaxies.
A new compactness parameter, similar to a central concentration parameter, was proposed by \cite{Karukes+Salucci2017} to improve the performance of the classic PSS96 URC. 

The confirmation of a true URC parametrization for disk galaxies would have far-reaching implications for the development and understanding of disk galaxy formation and evolution models in the universe.
In what follows, we shall take advantage of high-quality, spatially-resolved photometric and RC data, spanning a broad morphological and stellar mass range 
to test the various URC candidates above. 

We wish to determine the feasibility and accuracy of any proposed URC given our available data and ultimately identify areas of improvements as needed.  
To assess the true universality of a URC candidate, we must determine if the model:
(a) is the best possible mathematical representation given the input galaxy parameters,
(b) produces reasonably accurate velocity predictions and, 
(c) applies to the broadest range of disk galaxy morphologies over a specified radial extent.
Universality based on the three criteria above has yet to be established.

The testing of literature URCs requires a benchmark against which to gauge their accuracy.
Due to their impressive expressive ability, we can use neural networks 
(NNs)
to quantify the maximum potential accuracy of a URC given the set of inputs it requires.
This provides a baseline within which to frame our results.
See \cite{SmithGeach2023} for a thorough review of NNs and their applications in astrophysics.

We aim to take advantage of a powerful strength of NNs: their ability to approximate any continuous function.
Referred to as a universal approximation theorem (UAT), this well-established capability of NNs has been studied since the early works of \cite{Hornik+1989} and \cite{Cybenko1989}.

By providing the same input properties to both the NN and literature URCs, we can compare their relative performances. 
If the literature and NN URCs perform similarly, we can confirm that formulation is likely the best possible one for that set of inputs.
Conversely, if the NN URC is significantly more accurate than the literature URC, we can conclude that a better formulation exists.

To assess criterion (a) above, we leveraged the universal approximation ability of NNs to determine a benchmark accuracy level that a URC candidate must approach.
For each URC candidate, we designed a NN architecture that works best for that set of input parameters and optimized it with the same training data as the URC candidate.
We can then fairly compare URC and NN velocity predictions and establish if the URC is formulated optimally.

For criterion (b), we used the accuracy of velocity predictions obtained via a stellar Tully-Fisher (TF) relation \citep{Reyes+2011,Hall2012,Arora+2019} as another benchmark for the success of a URC candidate.
For instance, \cite{Reyes+2011} reported a measured scatter of 0.06 dex
in their stellar TF study, which corresponds to inaccuracies at the level of 20 per cent for velocity, similar to those quoted by PSS96. 
Aiming to improve upon this, we adopt predicting velocities $\lesssim\,$15 per cent as another benchmark.

For criterion (c), the fullest range of galaxy morphologies was sought by exploiting the largest sample of spatially-resolved extended disk galaxy RCs with matching photometry available to us. 

Our paper is organized as follows:
\sec{datamethods} provides a broad description of our data and the methods applied to extract bulge-disk decompositions and stellar mass estimates. 
A more detailed presentation of the data is given in Appendix~\ref{app:data_app}. 
We outline our URC fitting and NN development procedures in 
\Secs{URC_fitting}{NN_arch} respectively.
The performance (accuracy) of our URCs and NNs is then assessed in \sec{URC},
followed in \sec{discussion} by a discussion of URC shortcomings and suggested improvements.
We conclude with views towards future related investigations. 

\section{Data Base and Processing} 
\label{sec:datamethods}

The Photometry and Rotation Curve Observations 
from Extragalactic Surveys \citep[][hereafter PROBES-I]{PROBES-I}, 
its extension PROBES-II \citep{frosst+2022},
and the catalogues of galaxy structural parameters based on the survey ``Mapping Nearby Galaxies at Apache Point Observatory (MaNGA)'' \citep{MaNGA-I, MaNGA-II} provided us with spatially resolved photometric and kinematic data for 3846 galaxies 
(PROBES-I = 1442, PROBES-II=171, MaNGA=2233). 
These form the basis of our URC investigations. 
While photometric data are widely available through large-scale surveys \citep[e.g.,][]{DESI}, the key strength of our combined data set consists of the matching RCs for each galaxy.
This powerful combination of information, along with suitable stellar mass-to-light ratios (\sec{stellarmasscalcs}), allows us to separate the luminous and dark components of a vast number of galaxies. 
Details on RC and photometric observations for each catalogue are presented in Appendix~\ref{app:data_app}.
This Section specifies the various data processing steps taken to prepare the data for our analysis.  
The reader interested in URC modeling and analysis may skip to \Sec{URC_fitting}.

\subsection{RC smoothing}
\label{sec:RCfits}

The directly measured RCs presented in Appendix~\ref{app:data_app} must be processed and cleaned before use in our URC analysis.
Firstly, the number of data points per galaxy and their location in the galaxy are not uniform across the sample.
For example, some galaxies have many ($>50$) measurements ranging from their centers to the faint outskirts, while others only have a few ($<20$) that terminate where the stellar distribution ends.
If left as is, galaxies with more data points would have a greater influence on the optimization procedure of URC parameters.
Secondly, these measured RCs contain fluctuations caused by local non-axisymmetric perturbations (e.g., spiral arms, bars).
As we aim to model the underlying global mass distribution with a URC, the presence of small-scale fluctuations will inhibit the URC optimization procedure's performance.
To address these issues, we use RC smoothing functions for each of the three data sets.
We stress that the fitting functions discussed below are not URCs.
Their parameters are optimized individually for each galaxy and require a full set of RC measurements, whereas a URC should be able to \emph{predict} a galaxy's RC.
These fitting functions effectively produce a smoothed (or an average) version of the RC measurements they fit\footnote{PSS96 excluded asymmetric RCs between the receding and outgoing sides, and poorly sampled RCs. Our RC smoothing functions aim to alleviate these concerns.}.

PROBES-I and -II include parameters for the empirical RC fitting function proposed in \citet[][hereafter C97]{C97}:
\be
\label{eq:C97_smoothing}
    V(R) = V_{\rm sys} + V_{\rm c} \frac{(1+x)^\beta}{(1+x^\gamma)^{1/\gamma} } ,
\ee
where $V_{\rm sys}$ is the systematic velocity of the galaxy (from cosmological expansion and peculiar motion), $V_{\rm c}$ is an asymptotic velocity (roughly analogous to maximum velocity, $V_{\rm max}$), $x=R_{\rm t}/(R-R_0)$, $R_{\rm t}$ is a transition radius, and $R_0$ is the spatial center of the galaxy.
While not strictly true, since these fitting parameters are somewhat degenerate between each other, $\gamma$ tunes the ``sharpness'' of turnover for galaxies which have a bump and falling shape, and $\beta$ adjusts the rise or fall of outer RCs.
This function is excellent at modelling one-dimensional RCs and the variety of shapes they exhibit \citep{PROBES-I}.

\cite{MaNGA-II} provide parameters for the simpler fitting function (see also C97):
\be
    V(R) = V_{\rm max} \tanh{(R/R_{\rm t})} ,
\ee
where $V_{\rm max}$ is the asymptotic (maximum) velocity of the function and $R_{\rm t}$ is a transition radius.
This function is effective at modelling RCs that rise and become asymptotically flat (within the range of observations). 
Care must be taken not to extrapolate RCs too far, where they may show a drop-off.

The C97 fit parameters are used to construct inferred RCs for galaxies in the PROBES-I and -II samples.
These inferred RCs begin $1''$ (matching the typical resolution of our RCs) outwards from the center of the RC and extend to the larger of $R_{23.5}$ (in the r-band\footnote{We use the $r$-band as a reference band throughout this work due to its minimal dust extinction and high signal-to-noise ratio, 
and to remain consistent with the PROBES-I, -II, and MaNGA data sources.
Unless otherwise noted, we refer to $r$-band isophotes when defining radii like $R_{23.5}$.
}) 
or the last measured RC point at $R_{\rm last}$.
We set the outer bound to at least $R_{23.5}$ since our RCs often extend close to this point and reasonable extrapolations beyond $R_{\rm last}$ to $R_{23.5}$ are possible.  
Where available, RC data beyond $R_{23.5}$ were always used and we could set the upper bound to $R_{\rm last}$.
We linearly sampled 100 data points using the C97 fit for each galaxy between these lower and upper bounds.

We used the C97 fits of PROBES-I and -II to confirm that RCs are generally flat or rising until $R_{23.5}$.
Only 17 per cent have a falling shape, and these few still have a slope close to zero.
This motivated our choice of inferred RC bounds for the $\tanh$ model adopted for the MaNGA sample.
As in the PROBES-I and -II case, the lower bound for MaNGA is set to $1''$.
The upper bound is fixed at $R_{23.5}$ here however, since the $\tanh$ function cannot model RCs that may fall beyond this point.
Again, 100 data points were linearly sampled for each galaxy between these bounds using the $\tanh$ RC fit.

We have assumed that velocities generated by these C97 and $\tanh$ functions represent the true underlying RC of a galaxy, and we do not assign uncertainties to them.

The inferred RCs produced by the $\tanh$ function for MaNGA galaxies are inherently corrected for inclination due to the two-dimensional velocity map fitting procedure.
For C97 fits of PROBES-I and -II RCs, 
we have corrected for projection effects (and ignored redshift broadening) with:
\be
    V_{\rm corr}(R) = \frac{V_{\rm LOS} - V_{\rm sys}}{\sin{i}} ,
\ee
where $V_{\rm corr}(R)$ is the inclination-corrected velocity, $V_{\rm LOS}$ is the raw line-of-sight velocity from the C97 model, and $i$ is the galaxy's inclination (see \sec{corrections} for inclination calculation).
Both sources include inclination values based on the ellipticity of photometric isophotes.

\subsection{Surface brightness profile truncation}
\label{sec:SB_truncation}

The deep SB levels reached by the DESI-LIS imaging ($\sim$28-29 $r$-\magss) used in the PROBES-I, PROBES-II, and MaNGA catalogues are greatly beneficial for our analysis as they probe the faint outer disk features.
However, the identification of reliable isophotes at such low intensities becomes erratic and the \emph{AutoProf} sky subtraction procedure is not always accurate. 
The impact of an uncertain sky background rises at fainter light levels as does the contamination from line-of-sight interlopers whether physically connected (e.g., interacting neighbour) or not (foreground/background objects). 
These potential sources of contamination must be carefully accounted for. 
We take several steps to mitigate their impact and clean our galaxy SB profiles in each photometric band.

First, SB profiles were truncated beyond the point where measurement uncertainty exceeds 0.2 \magss \citep{MaNGA-I,PROBES-I}.
We use a moving mean of uncertainty values instead of simply truncating beyond the first data point that exceeds 0.2 \magss, to eliminate the potential of a single high uncertainty data point prematurely terminating a profile.
The rolling window had a size equal to 7.5 per cent of the total number of data points in a profile.
This value is a compromise between too fine a resolution, where in the extreme case there is only one data point per window,
and too coarse a resolution, where too many data points are averaged and structural information is lost. 
Profiles are truncated from the central point of the rolling window.

Second, SB profiles were truncated at the point where they are indistinguishable from the sky background and the target galaxy is no longer the dominant source of light.
Noting that this would be indicated by a flat and faint surface brightness profile, we calculated the first and second derivatives of the surface brightness profile to determine the best truncation radii.
Using a rolling mean for the derivative profiles again, we truncated profiles where: 
\begin{enumerate}
    \item The profile is flat, defined numerically as where the first derivative is less than 0.05 \magss kpc$^{-1}$.
    \item The flatness is not a transient feature of the profile, defined as where the second derivative is less than 0.005 \magss kpc$^{-2}$.
    \item The profile is faint enough that the flatness is likely not a feature of the galaxy, defined as where the SB is greater than 26.5 \magss.
\end{enumerate}
The profile is truncated once all three criteria are satisfied; with the first removed data point defined as the beginning of the rolling window.
A number of techniques to accomplish a similar goal exist \citep{Gilhuly2018, MaNGA-I, PROBES-I}, but our method above worked best for our full data set.

Finally, all data points with SB error greater than 0.3 \magss \citep{MaNGA-I,PROBES-I} were removed.
This handled any remaining erratic observations that the previous steps may have missed.

The specifics of our truncation procedure do not significantly affect the non-parametric measures that we used.
Most of our galaxy parameters (stellar mass, size) were defined relative to an isophotal radius, typically $R_{23.5}$, located well within any truncation radius. 
While concentration does depend on total light, it involves a ratio between radii (see \Eq{stellar_c28}) so its reliance on the exact truncation radius is minimal.

\subsection{Corrections}
\label{sec:corrections}

SB profiles must be corrected for several effects that alter the observed flux. 
The corrections described in this section were applied to both SB and integrated magnitude outputs from \emph{AutoProf}, and take the form:
\be
    A_{\rm corr,\lambda} = A_{\rm obs,\lambda} - A_{\rm g,\lambda} - A_{\rm K,\lambda} - A_{\gamma,\lambda},
\label{eq:correct}
\ee
where $A_{\rm corr}$ is the corrected SB or magnitude value, $A_{\rm obs}$ is the observed SB or magnitude value, and the remaining three terms are the corrections described below.
The $\lambda$ subscript denotes the band dependency of each correction.

First, dust from within our Galaxy causes extinction of incoming light from a target galaxy ($ A_{\rm g}$).
All three catalogues provide Galactic extinction correction values, ranging from 0.007 - 1.0 mag (higher values in bluer bands).
Extinction data for PROBES-I were taken from \cite{SchlaflyFinkbeiner2011}, while PROBES-II and MaNGA used \cite{Schlegel+1998}.
This Galactic extinction correction is typically the largest of the three in \Eq{correct}. 

Second, shifting of spectral energy distributions (SEDs) of galaxies due to cosmological redshifting causes the flux received by a filter to be different than the flux received if the galaxy was at rest relative to the observer (K-correction, $ A_{\rm K}$).
This is important for mass-to-light color relations in \Sec{stellarmasscalcs}, which are calibrated for rest frame SEDs.
PROBES-II and MaNGA used K-corrections from \cite{BlantonRoweis2007}.
PROBES-I used instead the method of \cite{ChilingarianZolotukhin2012} to calculate K-corrections using redshift and color information.
Since all galaxies in our samples are relatively close, K-corrections have the lowest magnitudes ($<0.16$ mag).

Lastly, we corrected for the diminished flux observed for an edge-on galaxy relative to face-on due to dust extinction within the galaxy ($A_{\gamma}$).
We applied the procedure of \cite{PROBES-I} for all galaxies, where the inclination correction takes the form:
\be
\label{eq:incl_corr}
    A_{\gamma,\lambda} = - \gamma_\lambda \log{(\cos{i})} ,
\ee
where $\gamma_\lambda$ is a band-dependent constant ($\gamma_g = 0.48$, $\gamma_r = 0.30$, $\gamma_z = 0.01$, $\gamma_{W1} = 0.04$)  and $i$ is the galaxy inclination.
Inclination is determined using the 25.5 $r$-\magss isophote:
\be
    \cos{i^2} = \frac{q^2-q_0^2}{1-q_0^2} ,
\ee
where $q=b/a$ is the semi-minor to semi-major axis ratio of the isophote and $q_0$ is intrinsic thickness. 
An average value of $q_0=0.13$ was adopted for each galaxy \citep{Hall2012}.
Inclination corrections range from small to high values (0.006 - 0.55 mag) for face-on to edge-on galaxies, respectively.
At their highest, these are comparable to Galactic extinction corrections, and at their lowest, comparable to K-corrections.

\subsection{Non-parametric measures}
Using the corrected SB and integrated magnitude profiles from \emph{AutoProf}, some key galaxy properties were computed.
Firstly, we determined $R_{23.5}$ in units of both arcsec and kpc by linearly interpolating for the radius value in a profile where SB = 23.5 $r$-\magss.
Using $R_{23.5}$ as a size metric standardizes comparisons between galaxies and provides a normalization radius for our analysis.
We then interpolated for the rotational velocity at $R_{23.5}$, which we call $V_{23.5}$.

Next, we converted apparent magnitudes ($m_\lambda$) to absolute magnitudes ($M_\lambda$) and luminosities ($L_{\lambda}$).
All three catalogues have distance estimates and their uncertainties, typically obtained through Hubble flow ($H_0 = 73$ \kms Mpc$^{-1}$), tip of the red giant branch, or surface brightness fluctuation methods.

Lastly, we calculated a concentration metric that traces how much light or mass is concentrated in the central regions of galaxies.
We adopt \citep[][see references therein]{Courteau1996}: 
\be
\label{eq:stellar_c28}
    C_{28} = 5 \log{\frac{R_{80}}{R_{20}}} ,
\ee
where $R_{80}$ ($R_{20}$) is the radius containing 80 (20) per cent of the galaxy's total luminosity.
Since we apply a constant stellar mass-to-light ratio across a luminosity profile, $C_{28}$ is equivalent to stellar mass concentration.

\subsection{Stellar mass estimates}
\label{sec:stellarmasscalcs}

Assessing accurate stellar mass-to-light ratios for galaxies is inherently challenging.
If observed galaxy spectra or SEDs are available, they can be matched with template galaxy spectra or SEDs from stellar population synthesis (SPS) models to infer a stellar mass-to-light ratio \citep{Conroy2013,Courteau2014}.
Of all available methods, full spectrum fitting is generally thought to be the most accurate.
\cite{Roediger2015} advocated the use of stellar mass-to-light color relations (MLCRs), a simplified version of SED fitting that achieves similarly accurate stellar mass estimates much more rapidly.
They found that SED fitting and MLCRs stellar mass estimates agree within 0.05 dex,  and recover true masses of mock galaxies within 0.2 dex.
Applied to real galaxies, systematic discrepancies can rise to 0.3 dex for both methods.
We have adopted the MLCR method to estimate stellar mass-to-light ratios ($\Upsilon_*$) for our galaxies.

\begin{table}
\singlespace
\centering
\begin{tabular}{ccccc}
\hline
\textbf{Source}                    & \textbf{Color}              & \textbf{Luminosity}      & \textbf{$m$}              & \textbf{$b$}              \\ \hline
\textbf{}                          & $g-z$                       & $g$                      & 1.116                     & -1.132                    \\
\citet{2003MNRAS.344.1000B} & $g-r$                       & $r$                      & 1.629                     & -0.792                    \\
\textbf{}                          & $r-z$                       & $z$                      & 1.483                     & -0.935                    \\ \hline
\textbf{}                          & $g-z$                       & $g$                      & 0.942                     & -0.764                    \\
\citet{2009ApJ...699..486C} & $g-r$                       & $r$                      & 1.497                     & -0.647                    \\
\textbf{}                          & $r-z$                       & $z$                      & 1.021                     & -0.487                    \\ \hline
\textbf{}                          & $g-z$                       & $g$                      & 1.13                      & -1.07                     \\
\citet{garciabenito+2019}     &     $g-r$                       & $r$                      & 1.54                      & -0.68                     \\
\textbf{}                          & $r-z$                       & $z$                      & 1.46                      & -0.88                     \\ \hline
\citet{2014ApJ...782...90C} & $W1-W2$ & $W1$ & -1.93 & -0.04 \\ \hline
\end{tabular}
\caption{\label{tab:MLCRs}
MLCR coefficients for stellar mass estimates, with
$m$ and $b$ as in Eq.~\ref{eq:MLCR}.
From each source, we used coefficients that best applied for our sample (i.e., late-type, star-forming).
}
\end{table}

With this procedure, the log of the stellar mass-to-light ratio is modelled to have a linear relation with galaxy color:
\be
\label{eq:MLCR}
    \log{\Upsilon_*} = b_\lambda + m_\lambda (color) ,
\ee
where we used a set of colors ($g - r$, $g - z$, $r - z$, and where available $W1 - W2$) and corresponding luminosities from each band ($g$, $r$, $z$, and $W1$) to create a set of $\Upsilon_*$ estimates.
Our $b$ and $m$ values are shown in Table~\ref{tab:MLCRs}.
We applied these $\Upsilon_*$ values to galaxy total luminosities (defined by the last point in the truncated SB profile) and luminosity within $R_{23.5}$ (linearly interpolated from SB profile), creating a set of stellar mass estimates for both $M_{*,\rm last}$ and $M_{*,23.5}$ for each galaxy.
The median of these was taken as our best value to limit the influence of outlier estimates, and the standard deviation served as a measure of the uncertainty in the best value.

\begin{figure}
\singlespace
\centering
\includegraphics[width=\textwidth]{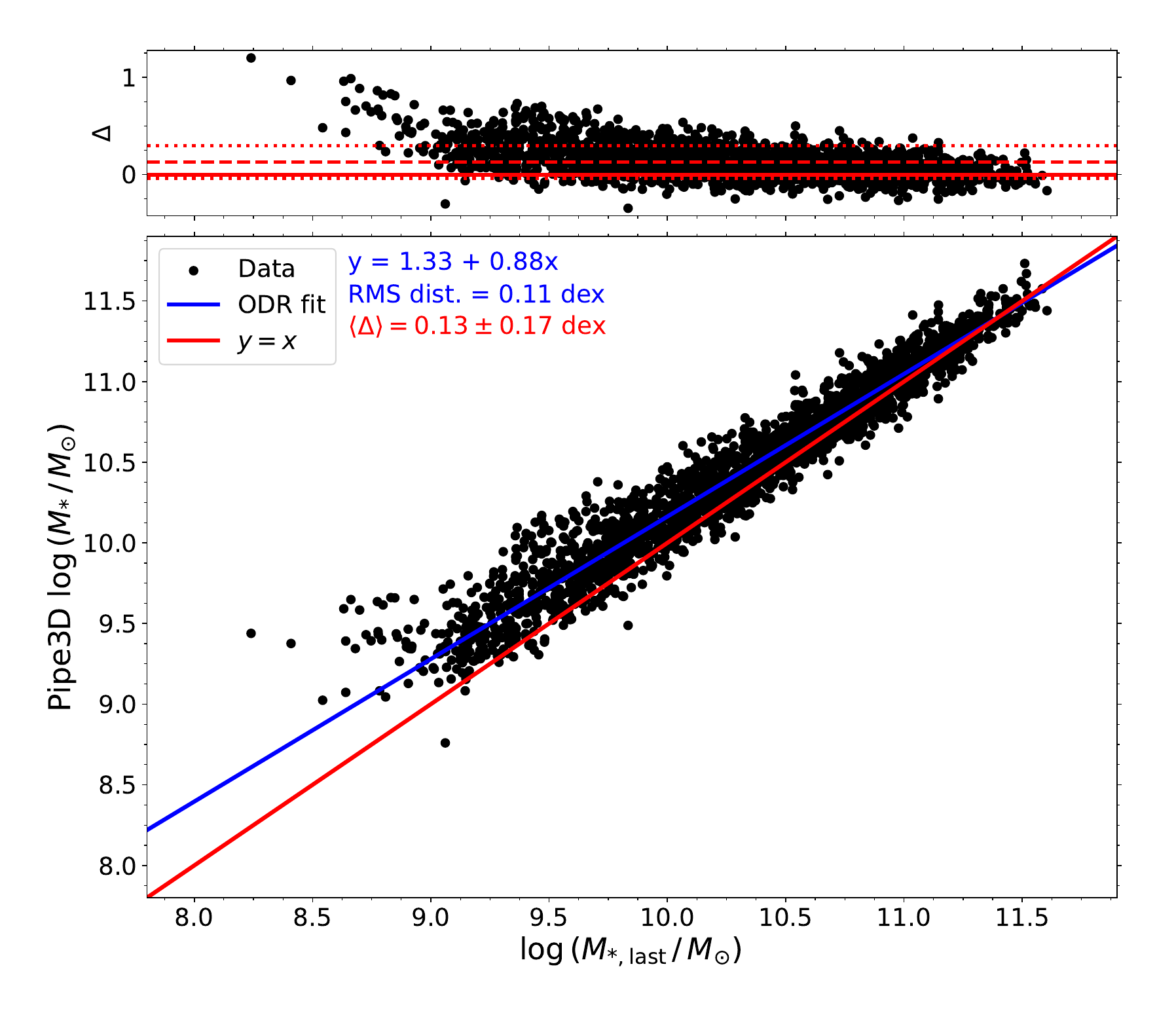}
\caption{Comparison of our total stellar masses ($x$-axis) versus those reported by \cite{Pipe3D} ($y$-axis).
Only MaNGA galaxies were used for this exercise.
The red line shows a 1:1 correspondence between our respective values, and the blue line shows an ODR fit of the actual correlation.
ODR fit coefficients and RMS orthogonal distance of each point to the ODR line are also shown in blue.
The top panel shows residuals between the 1:1 line and data, while the dashed line is the mean of the residuals and the dotted lines are standard deviation levels.
Mean and standard deviation of residuals are shown in the red inset text.
}
\label{fig:pipe3d_comp}
\end{figure}

The MaNGA Pipe3D catalogue \citep{Pipe3D} provides stellar mass estimates with $\Upsilon_*$ determined from spectrum fitting for our MaNGA galaxies.
Spectral analysis yields best possible $\Upsilon_*$ values, and their spatially resolved pixel summation method for determining galaxy luminosity is more accurate than our global approach, so we used their stellar mass calculations as a validation of our procedure.
\Fig{pipe3d_comp} shows a comparison between Pipe3D stellar masses and our stellar masses for each of our MaNGA galaxies.
Note that the spectral analysis assumed a different initial mass function for their SPS model than our MLCRs.
To account for this, we applied the 0.28 dex correction to our stellar mass values \citep{Gonzalez+2016}.

Pipe3D masses are systemically higher by typically 0.13 dex, though there is a stellar mass dependency and discrepancies are higher at the lower mass end.
This could be caused by the superior accuracy of pixel flux summation utilized by \cite{Pipe3D}, which affects faint lower stellar mass galaxies more than higher mass ones, or systematics in $\Upsilon_*$ calculations for MLCRs that under-predict $\Upsilon_*$ at low stellar mass (and their corresponding colors), as gauged by \Fig{pipe3d_comp}. 
Regardless, most galaxies fall within the 0.3 dex systematic uncertainty for stellar mass calculations, and we surmise that our stellar mass estimates are reasonable.

We also applied a quality criterion since some outlier stellar mass estimates exceeded $> 10^{12} M_\odot$, an unlikely limit for disk galaxies.
These extreme cases were caused by atypically high inclinations ($N=22$) or K-corrections ($N=2$), and were removed from our analysis.
The inclination correction issue is due to the asymptotic behavior of \Eq{incl_corr} for the edge-on configuration (as $i$ approaches 90$^{\circ}$). 

\subsection{Bulge-disk decompositions}
\label{sec:BD_decomps}

Disk galaxy SB profiles have commonly been decomposed into pre-determined parametric functions to describe the bulge, bar, disk, and even stellar halo of a galaxy \citep[e.g.,][]{MacArthur2003, PyMorph}.
Bulge-disk ($B/D$) decompositions provide galactic parameters for each component such as the effective radius of the bulge ($R_{\rm e}$),  intensity at the effective radius ($I_{\rm e}$), S{\'e}rsic index of the bulge ($n$), central intensity of the disk ($I_{0}$), and the scale length of the disk ($R_{\rm d}$).
The latter ($R_{\rm d}$) is an essential ingredient for the parameterizations of the literature URCs tested below.
To this end, we decomposed our one-dimensional SB profiles into the sum of a bulge and disk component to obtain reasonable disk scale lengths.
We followed the guidelines of \cite{MacArthur2003}, with some necessary modifications for our data set.
Appendix~\ref{app:BD_decomp} describes our procedure in detail.
Our results are validated in this section via comparison with literature sources.

\cite{PyMorph-DR17} used the \emph{PyMorph} \citep{PyMorphPackage} and \emph{GALFIT} \citep{galfitpackage} software packages to build a database of two-dimensional bulge-disk decomposition properties for MaNGA galaxies.
Their reported values for apparent magnitude of the bulge and disk components were used to calculate total bulge + disk luminosities.
This allowed for a direct comparison of our respective $B/D$ values for each MaNGA galaxy.
Even though we performed one-dimensional decompositions, which differ from their two-dimensional counterparts \citep{GaoHo2017}, an excellent 1:1 agreement is found between our luminosities and those of \cite{PyMorph-DR17} with a residual mean consistent with zero 
($\left< \Delta L \right> = -0.02 \pm 0.16$ dex). 

\begin{figure}
\singlespace
\centering
\includegraphics[width=\textwidth]{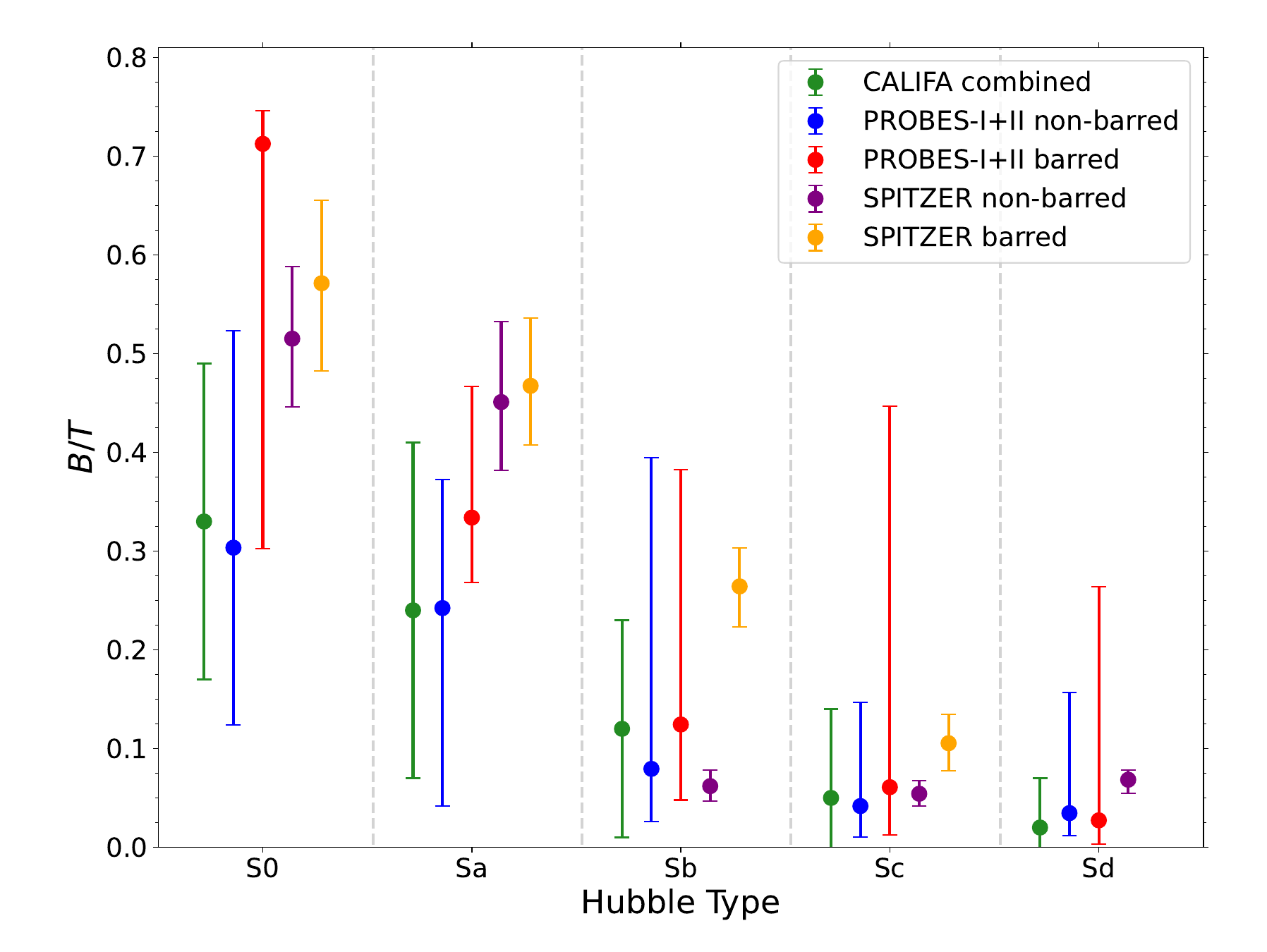}
\caption{Distribution of $B/T$ ratio as a function of galaxy morphology and barredness for PROBES-I and -II data, and comparison with literature sources.
\cite{Spitzer_BD_decomp} reported similar parameters for SPITZER galaxies, 
as did \cite{CALIFA_BD_decomp} for CALIFA galaxies, though not separated by bar presence.
The median and 16th/84th percentiles were computed for each bin (see inset).
}
\label{fig:BD_morph_comp}
\end{figure}

Next, we compared bulge-to-total light ratios ($B/T$) for our PROBES-I and -II galaxies\footnote{
Only PROBES-I and -II data were used since we do not have barredness information for MaNGA galaxies.}
with those reported in \cite{Spitzer_BD_decomp} with galaxies from the Spitzer survey \citep{2010PASP..122.1397S} and \cite{CALIFA_BD_decomp} for CALIFA survey galaxies \citep{2012A&A...538A...8S}.
Both of these works performed two-dimensional decompositions; \cite{Spitzer_BD_decomp} used the software GALFIT and \cite{CALIFA_BD_decomp} used GASP2D \citep{2008A&A...478..353M}.
\cite{Spitzer_BD_decomp} reported $B/T$ values for two-component (bulge + disk) fits, similar to ours, while \cite{CALIFA_BD_decomp} included a bar component in their models. 

\Fig{BD_morph_comp} shows that our $B/T$ values for non-barred galaxies (in blue) match well those from \cite{CALIFA_BD_decomp} (in green). 
Our $B/T$ values are lower than those of \cite{Spitzer_BD_decomp} non-barred galaxies (in purple) for S0 and Sa types, but are a good match for Sb, Sc, and Sd galaxies.
Overall, our decomposition procedure works well for non-barred galaxies.

However, the presence of a bar raises $B/T$ relative to galaxies without bars for all bins in \Fig{BD_morph_comp}.
This is true here and for \cite{Spitzer_BD_decomp}, whereby light attributed to a bulge component should be modelled as a bar component instead
\citep[e.g.,][]{Laurikainen+2005,2008MNRAS.384..420G,Spitzer_BD_decomp}.
Short of proper two-dimensional modeling of images, we acknowledge that our $B/T$ ratios may be over-estimated for barred galaxies. 
Given our good match with literature sources for non-barred galaxies, 
we conclude that our $B/D$ decompositions are satisfactory.

To easily distinguish between different galaxy morphologies, we divided our data into three subsets: disk-dominated ($B/D \; \mathrm{ratio}< 1/3$, N = 2804), two-component ($1/3 \leq B/D \; \mathrm{ratio} < 3$, N = 907), and bulge-dominated ($3 \leq B/D \; \mathrm{ratio} < 10$, N = 135).
Galaxies with $B/D \; \mathrm{ratio} > 10$ have such low disk mass that the disk scale length does not meaningfully represent the galaxy's mass distribution. 
Since the disk scale length, $R_{\rm d}$, is a key variable of the tested URCs, we did not include the latter systems in our analysis (N = 24).

\subsection{A sub-sample of symmetric RCs}
\label{sec:symmetric_RCs}

The performance of a URC can be limited by structural asymmetries that arise from interactions or merger events with nearby galaxies, as well as internal processes such as non-axisymmetric bar-like perturbations or localised self-propagating stochastic star formation. 
Current URC formulations do not take these effects into account and may therefore fail to characterise the full range of disk galaxy RCs. 
To assess URC performance in a more idealized context, we can also examine isolated PROBES I and II galaxies on the basis of RC symmetry.

The RC smoothing procedure described in \Sec{RCfits} provides the spatial center and systematic velocity for each RC. 
These can be used to divide a given RC into its red-shifted and blue-shifted portions.
A RC can be deemed symmetric if it meets the following two (arbitrary) criteria:
(a) the red- and blue-shifted portions of the RC extend out to comparable radii from the center of the RC;
(b) the red- and blue-shifted portions reach a similar asymptotic velocity relative to each other.

For criterion (a), we required that the absolute difference between the radial extents of the red- and blue-shifted portions of the RC be less than 30 per cent of the larger value of the two radial extents.
For criterion (b), the velocity amplitude was measured relative to the smoothed RC and the actual data.  
If the mean absolute percent difference between the smoothed RC and observed velocity values differed by more than 30 per cent for a galaxy, the system was rejected. 
This procedure yielded a subsample of 579 symmetric RCs from the PROBES I-II samples. 
Most of these ($\sim$80 per cent) fall into our disk-dominated category.
Another 17 per cent belong to the two-component category, and the remainder (3 per cent) are bulge-dominated.

\subsection{Normalized RCs}
\label{sec:normalized_RCs}

\begin{figure}
\singlespace
\centering
\includegraphics[width=\textwidth]{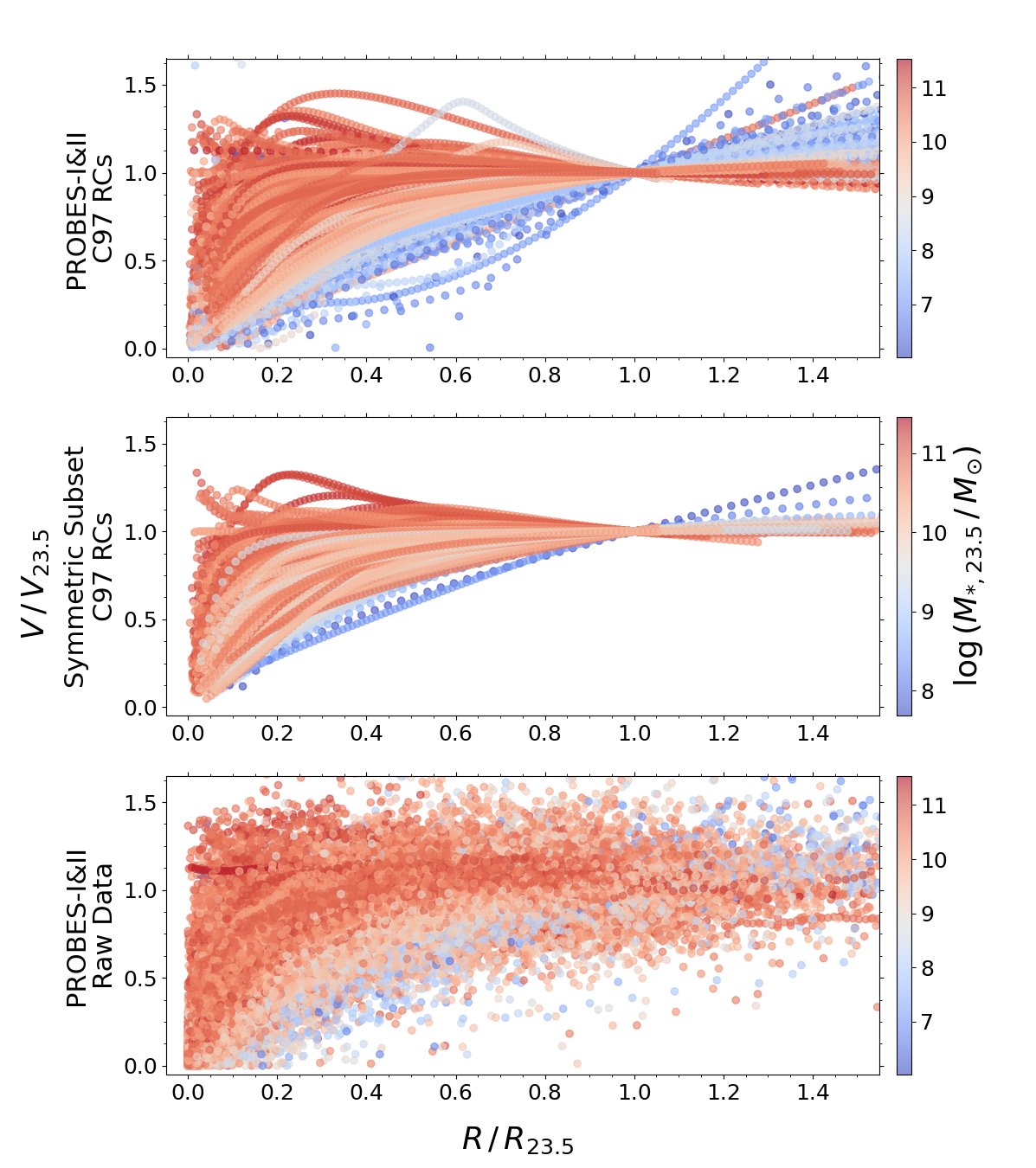}
\caption{Normalized RCs for the PROBES-I \& II data set.
Velocities and radii are normalized by their value at $R_{23.5}$.
Points are colored according to the stellar mass enclosed within $R_{23.5}$ of the corresponding galaxy.
The top panel shows the C97 fit RCs for PROBES-I \& II galaxies (1613 galaxies),
the middle panel shows the C97 fit RCs for the symmetric subset (579 galaxies),
and the bottom panel shows the raw data (1613 galaxies).}
\label{fig:normalized_RCs}
\end{figure}

PSS96 presented RCs normalized in velocity and radial units, as do we for PROBES-I\&II galaxies in \Fig{normalized_RCs}. 
Our normalization uses the velocity and size at $R_{23.5}$, a choice that differs slightly from PSS96 as discussed in \Sec{PSS96_URC_formulation}.
Note that MaNGA RCs are not presented in \Fig{normalized_RCs} since one-dimensional observed RCs are lacking for them.

The compilation of normalized fitted and raw RCs is shown in \Fig{normalized_RCs}.
It features a broad range of RC shapes for both the full (top panel) and symmetric (middle panel) samples fitted with the C97 functions. 
The raw data in the bottom panel show particularly significant breadth. 
The normalizations are meant to enhance any putative universality of RC shapes,
which in \Fig{normalized_RCs} does not seem to emerge. 
A multi-parameter URC may still exist if the scatter of these distributions can be captured by a third parameter such as luminosity or stellar mass \citep{PS91,PSS96,Salucci+2007,Karukes+Salucci2017,Paolo+2019}, as possibly suggested by \Fig{normalized_RCs}.
This possibility is explored below. 

Later in \sec{symmetric_URC}, a detailed URC analysis is applied to both the full sample of 3,846 disk galaxies drawn from our collective PROBES I-II and MaNGA samples and the symmetric sub-sample {\bf of 579 disk galaxies} identified here.
It will be shown that the URC and NN performances remain largely the same whether the full sample or smaller symmetric sub-sample is considered.

\section{URC Fitting}
\label{sec:URC_fitting}

This section addresses literature URCs and the fitting procedures used to determine their parameters.
\Secs{PSS96_URC_formulation}{LSBD_formulation} provide overviews of the original PSS96 URC \citep{PSS96}, and the modifications suggested by \cite{Karukes+Salucci2017} and \cite{Paolo+2019} to improve the performance over the PSS96 formulation.

Throughout this work, it is assumed that observed rotational velocities trace circular velocities (i.e., the total mass).
While this is a reasonable assumption at large radii, 
stellar orbits near galaxian centers may show significant non-circular motion driven by perturbations such as bars or spiral arms, 
causing notable deviations between rotational and circular velocities \citep{Trachternach+2008, Sellwood2015}.
Velocity predictions from mass models will therefore, like all URCs tested in this work, show inaccuracies at innermost radii since a mass model converts to circular speed.
We can test the upper limit of prediction accuracy with our NN approach, yielding reasonable expectations for URCs without data that probe non-circular motions (e.g., bar strength, velocity dispersion).
Note that gas distribution measurements are absent from our analysis.
A bulge velocity contribution component is also not included in the PSS96 formulation.

\subsection{PSS96 formulation}\label{sec:PSS96_URC_formulation}

\citet[][hereafter PSS96]{PSS96} modeled 1100 RCs largely from the data collection of \cite{Mathewson+1992} with: 
\be
\label{eqn:V_comb}
    V^2(R) = V_{\rm d}^2(R) + V_{\rm h}^2(R) ,
\ee
where $V_{\rm d}^2$ represents the circular velocity contribution from the luminous disk and $V_{\rm h}^2$ represents the dark halo.
PSS96 assumed that the luminous part of the galaxy is described by an 
infinitesimally thin exponential disk, which we model as: 
\be
\label{eq:PSS96_disk_final}
    V_{\rm d}^2(x) = V^2(R_{23.5}) \, \beta \, \frac{g(x)}{g(x_{23.5})},
\ee
where the function $g(x)$ is:
\be
    g(x) = \frac{x^2}{2} \, (I_0 K_0 - I_1 K_1) ,
\ee
$\beta$ is the disk mass fraction, $\beta \equiv M_{\rm d} / M_{\rm tot, 23.5}$, 
$M_{\rm d}$ is the disk mass,
$x=R/R_{\rm d}$, 
$x_{23.5}=R_{23.5}/R_{\rm d}$,
and $I_i$, $K_i$ are modified Bessel functions evaluated at $x/2$.
By definition, $\beta$ is bound between 0 and 1.

PSS96 proposed that $\beta$ is solely a function of luminosity.
However, since luminosity is the observable property resulting from a galaxy's stellar mass, 
the former is here converted into the latter.
This is motivated by the fact that stellar mass is the underlying physically relevant galaxy property when considering mass models.
The drawback of this operation, however, is that stellar mass estimates carry an intrinsic uncertainty of $\sim$0.3 dex \citep{Conroy2013,Roediger2015}.

$\beta$ was fitted as:
\be
\label{eq:PSS96_beta}
    \beta = k_0 + k_1 \log{\left(\frac{M_{*,23.5}}{M_{\rm ref}}\right)} ,
\ee
where $k_i$ are fitting parameters, $M_{*,23.5}$ is the inferred stellar mass enclosed within $R_{23.5}$, and $M_{\rm ref}$ is a reference value used to scale all stellar masses.
$M_{\rm ref}$ can be converted from the luminosity reference value given in PSS96, $L_{\rm B,ref} = 10^{10.4} \, L_{\rm B,\odot}$, with a approximate mass-to-light ratio of 1.2 $M_\odot/L_{\rm B,\odot}$ \citep{McGaugh+Schombert2014}.
Since the $k_i$ parameters are fitted for, the choice of $M_{\rm ref}$ does not affect the URC performance.  
The reference stellar mass value is $ M_{\rm ref} = 10^{10.48} \, M_\odot $.

Since mass-to-light ratios and disk fit parameters are available for our galaxies (\Secs{stellarmasscalcs}{BD_decomps}), we can calculate total disk stellar masses with $M_{\rm d} = 2 \pi \Sigma_{0} R_{\rm d}^2$,
where $\Sigma_{0} = I_0 \Upsilon_*$ is the disk central surface mass density.
With $M_{\rm d}$ and the total (baryons + DM) combined mass enclosed within $R_{23.5}$, $M_{\rm tot,23.5}$, 
we can determine $\beta$ for each galaxy. 
Using those $\beta$ values, we fitted for $k_i$ in \Eq{PSS96_beta} through ordinary least squares (OLS) regression for each of our data sets.

We note a disconnect between $M_{\rm d}$ and $M_{\rm tot,23.5}$ in the definition of $\beta$.
$M_{\rm d}$ is the disk mass integrated to the limit of $R$ approaching infinity, while $M_{\rm tot,23.5}$ is measured within $R_{23.5}$.
The total mass of a halo-galaxy system is certainly not contained entirely within $R_{23.5}$ and so for galaxies where the dark halo does not play a significant role in these inner regions, it may be that $\beta > 1$.
Galaxies with $\beta > 5$ were deemed to have errant B/D decompositions, RC fits, and/or mass-to-light ratios, and were removed from analysis (N = 16).
Galaxies with $1 \leq \beta \leq 5$ are effectively stellar mass dominated within $R_{23.5}$ and the physically-motivated limit of $\beta = 1$ is applied for those systems.

We could have also used disk mass values directly, eliminating the need for \Eq{PSS96_beta}.
However, the latter is an integral part of the PSS96 formulation and removing it would significantly alter the URC. 
Therefore, we have kept \Eq{PSS96_beta} for consistency with the original PSS96 formulation.

Following again PSS96's methods, we have modelled the dark halo velocity contribution as:
\be
\label{eq:PSS96_DM}
    V_{\rm h}^2(z) = V^2(R_{23.5}) (1-\beta)(1+a^2) \frac{z^2}{z^2+a^2} ,
\ee
where $z=R/R_{23.5}$ and $a$ is the halo core radius (in units of $R_{23.5}$) free parameter which is fit for in the form:
\be
\label{eq:PSS96_a}
    a = w_0 \left(\frac{M_{*,23.5}}{M_{\rm ref}}\right)^{w_1} ,
\ee
where $w_i$ are optimization parameters.

This is identical to the original PSS96 form, except using stellar mass instead of luminosity.
The scale radius was also changed from $R_{\rm opt}$ (in PSS96) to $R_{23.5}$ as a model-independent standardized radius is preferred.
$R_{23.5}$ also marks the end point of many of our RCs (\Sec{RCfits}).
The combined circular velocity is calculated via Eq.~\ref{eqn:V_comb}.

As opposed to the definition for $\beta$, the halo core radius $a$ in \Eq{PSS96_a} is not an observable and $w_0$ and $w_1$ must be left as optimization parameters for fitting RC data.
$k_0$ and $k_1$ are determined from fitting \Eq{PSS96_beta} with OLS regression, which is possible since we know both variables in the equation.
$w_0$ and $w_1$ are optimized to fit RC data.

A randomly sampled testing set consisting of 20 per cent of all galaxies (N= 769 galaxies) is set aside at the beginning of the fitting procedure.
Of the remaining galaxies, 75 per cent (N= 2308 galaxies) are randomly sampled to use as training data and the other 25 per cent (N= 769 galaxies) as validation data.
The testing set is used to gauge the performance of a model for new data, 
replicating the end use case of applying a URC for any disk galaxy.

A combination of OLS regression and a cost function minimization algorithm from the Python package \emph{SciPy} \citep{SciPy} was used to fit for $k_i$ and $w_i$.
We used a mean squared error (MSE) cost function to optimize for $w_i$ by minimizing the difference between observed and predicted velocity values.
The initial values for $w_0$ and $w_1$ were chosen randomly from a normal distribution centered on zero with a standard deviation of two. 
We required that $w_0$ be positive, as a negative $w_0$ would result in a non-physical result of a negative core radius value.

To ensure that the optimized parameters are independent of the random training set used and initial estimates, the training and validation data were re-sampled from the data set (excluding testing data) and the optimization routine was recalculated.
This was repeated 100 times and the performance of each trained model was judged by its cost function value when applied to the validation set.
Resampling like this also yields distributions for the best-fit values for each optimization parameter, allowing us to determine errors via 16th and 84th percentiles.

\begin{table}
\singlespace
\centering
\begin{tabular}{cccr|cccc}
\hline
\textbf{Data Set}                         & \textbf{N}           & \multicolumn{1}{l}{\textbf{Vali. Cost}} & \textbf{}     & \textbf{$k_0$}  & \textbf{$k_1$}  & \textbf{$w_0$}  & \textbf{$w_1$}  \\ \hline
\multirow{3}{*}{\textbf{All}}             & \multirow{3}{*}{64}  & \multirow{3}{*}{478}                    & Low           & 0.4048          & 0.0865          & 0.203          & -0.312          \\
                                          &                      &                                         & \textbf{Best} & \textbf{0.4078} & \textbf{0.0889} & \textbf{0.208} & \textbf{-0.299} \\
                                          &                      &                                         & High          & 0.4101          & 0.0920          & 0.210          & -0.291    \\ \hline      
\end{tabular}
\caption{Best values for each parameter of the PSS96 URC model. 
The model was trained 100 times, but some were removed due to our fit quality criterion (see text).
The number of models accepted and mean validation cost of the accepted models is shown in the second and third columns.
The median of each optimized set of $w_i$ is shown here as the best value, and the 16th and 84th percentiles show the lower and upper end of the distributions.
Precision of each parameter is determined by the standard error of the distribution.}
\label{tab:our_PSS96_params}
\end{table}

Some trained models reached a local minimum of the cost function, as opposed to the preferred global minimum.
Two distinct groups of models stood out, with a majority clustered below a validation cost of MSE = 750, and a minority residing at local minima above MSE = 750.
Using validation cost as a discriminant for models having reached a local minimum, any models with a validation cost greater than 750 were removed.
Table~\ref{tab:our_PSS96_params} shows our results.

\subsection{LSB-D formulation}
\label{sec:LSBD_formulation}
\cite{Karukes+Salucci2017} and \cite{Paolo+2019} both improved the PSS96 URC to account for dwarf and LSB spirals, respectively. 
Given their close resemblance, we have treated them as a single URC that we refer to as the ``LSB-D'' URC below.
This is just a naming convention; we apply this URC to all galaxies, not just LSB and dwarf spirals.
We also used $k_i$ and $w_i$ as notations for optimization parameters for the LSB-D URC as well as the PSS96 URC.
While we used the same variable name, the values will be different for each URC.

\begin{figure}
\singlespace
\centering
\includegraphics[width=\textwidth]{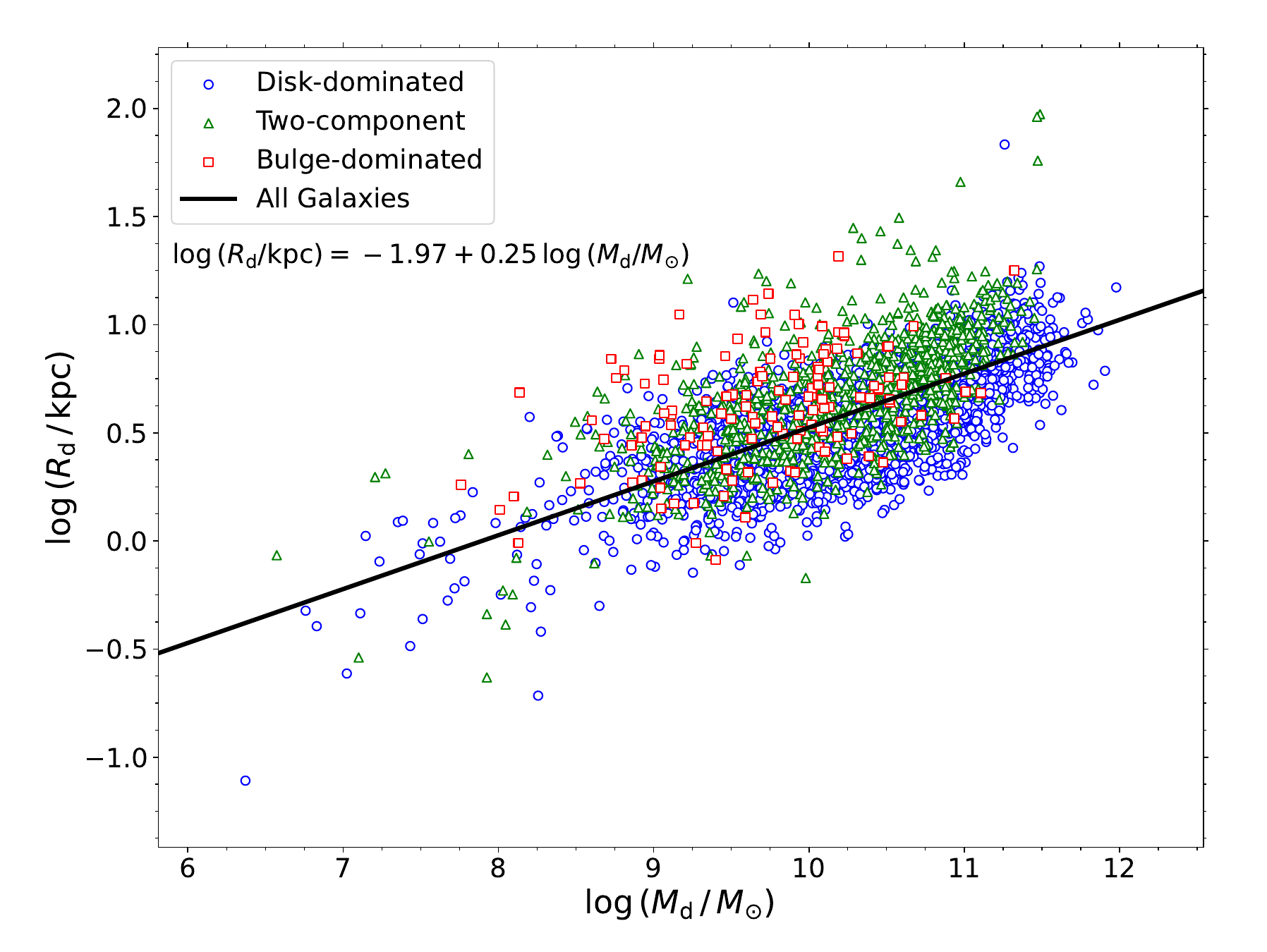}
\caption{Correlation of the disk stellar mass ($M_{\rm d}$) and disk scale length ($R_{\rm d}$) colored by morphology subset. 
The OLS fit (\Eq{LSBD_RdMd}) is shown in black, and the inset text shows the equation of the line.
}
\label{fig:Rd_Md}
\end{figure}

Both works updated the PSS96 formulation by including the compactness quantity in their analysis:
\be
\label{eq:LSBD_C*}
    C_* = \frac{10^{k_0 + k_1\log{M_{\rm d}}}}{R_{\rm d}} .
\ee
The definition of the compactness parameter was motivated by the large scatter observed in the $R_{\rm d} - M_{\rm d}$ relation (see \Fig{Rd_Md}). 
The compactness parameter is calculated according to \Eq{LSBD_C*} for each set of training data, with the coefficients $k_0$ and $k_1$ determined through OLS regression. 
$C_*$ is essentially a measure of the deviation of the expected $R_{\rm d}$ value relative to the observed, if the expected $\log{R_{\rm d}}$ value is defined as a linear relation only dependent on $\log{M_{\rm d}}$ (as in the exponent of the numerator in Eq.~\ref{eq:LSBD_C*}):
\be
\label{eq:LSBD_RdMd}
    \log{R_{\rm d}} = k_0 + k_1 \log{M_{\rm d}} .
\ee
We will verify if this updated formulation including compactness is a viable URC for any disk galaxy, not only dwarfs and LSBs.
In addition, we fitted this URC for the symmetric RC subset separately from the collective data set to determine if RC symmetry is an important factor for URC accuracy.
 
Following PSS96, the exponential thin disk model is adopted; however, 
a bulge component is still missing. 
\cite{Karukes+Salucci2017} found that the commonly used NFW halo profile performed poorly for this URC formulation, but the Burkert profile \citep{Burkert1995} works well.
We have also adopted the Burkert profile.
The circular velocity contribution at radius $R$ for this halo is then given by:
\be
\label{eq:LSBD_DM}
    V^2_{\rm h}(R) = 2\pi \rho_0 R_{\rm c}^3 \left[ \ln{\left(1+\frac{R}{R_{\rm c}}\right)} - \arctan{\left(\frac{R}{R_{\rm c}}\right)} + 0.5 \ln{\left(1+\left(\frac{R}{R_{\rm c}}\right)^2\right)} \right] \frac{G}{R},
\ee
where $\rho_0$ is the halo central mass density and $R_{\rm c}$ is the halo core radius.
The velocity contribution from the dark halo is combined with the disk contribution via Eq.~\ref{eqn:V_comb}.

The free parameters for this URC are then $M_{\rm d}$, $\rho_0$, and $R_{\rm c}$. 
They are modelled as:
\be
\label{eq:LSBD_Md}
    \log{M_{\rm d}} = k_2 + k_3 \log{V_{23.5}} + k_4 \log{C_*} ,
\ee

\be
\label{eq:LSBD_Rc}
    \log{R_{\rm c}} = w_0 + w_1 \log{V_{23.5}} + w_2 \log{C_*} ,
\ee

\be
\label{eq:LSBD_rho0} 
    \log{\rho_0} = w_3 + w_4 \log{V_{23.5}} + w_5 \log{C_*} .
\ee

The model training procedure here is similar to the one described in 
\Sec{PSS96_URC_formulation}.
Disk mass estimates are available for each galaxy, so $k_2, k_3, k_4$ were determined through minimization of a MSE cost function defined from the difference between modeled $M_d$ values via \Eq{LSBD_Md} and inferred disk mass values from the B/D decompositions.
Analogous to the $\beta$ parameter in PSS96, we used modelled $M_{\rm d}$ values for the disk velocity contribution equation when testing the URC instead of the directly inferred values to recreate the results of \cite{Karukes+Salucci2017} and \cite{Paolo+2019}.
$\rho_0$ and $R_{\rm c}$ are parameters associated with the dark halo and are not observable, so their $w_i$ were determined through minimization of a MSE cost function defined from the difference between URC modelled and observed RCs.

The model training procedure differs from PSS96 in one aspect for this URC: the initialization of the optimization parameters.
Instead of a completely random initialization, $w_i$ were instead initialized as the parameters given in \cite{Paolo+2019} with a random noise factor added in.
Due to the increased number of $w_i$, initializing $w_i$ completely randomly would result in excessively long training times and/or the cost function reaching local minima.

Again, 100 models were trained and those with a validation cost below 750 were retained.
Medians and 16th/84th percentiles of each set of $k_i$ and $w_i$ are presented in Table~\ref{tab:LSBD_params_0} and \ref{tab:LSBD_params_1}.

\begin{table}
\singlespace
\centering
\begin{tabular}{cccr|ccccc}
\hline
\textbf{Data Set} & \textbf{N} & \textbf{Vali. Cost} & \textbf{} & $k_0$           & $k_1$           & \textbf{$k_2$} & \textbf{$k_3$} & \textbf{$k_4$} \\ \hline
                                            &                                      &                                               & Low                                 & -1.779          & 0.2245          & 3.583                                    & 3.008                                    & 0.061                                    \\
\textbf{All Data}               & 69                                   & 427                                           & \textbf{Best}      & \textbf{-1.752} & \textbf{0.2282} & \textbf{3.637}          & \textbf{3.029}          & \textbf{0.085}          \\
                                            &                                      &                                               & High                                & -1.714          & 0.2308          & 3.685                                    & 3.052                                    & 0.107         \\ \hline
                                            &                                      &                                               & Low                                 & -1.82           & 0.216           & 4.77                                     & 2.519                                     & 0.000                                   \\
\textbf{Symmetric RCs}        & 87                                   & 563                                           & \textbf{Best}      & \textbf{-1.71}  & \textbf{0.227}  & \textbf{4.92}           & \textbf{2.574}           & \textbf{0.030}         \\
                                            &                                      &                                               & High                                & -1.60           & 0.237           & 5.03                                     & 2.639                                     & 0.063                                    \\ \hline
                                            
\end{tabular}
\caption{Best values for \Eqs{LSBD_RdMd}{LSBD_Md} of the LSB-D URC model (parameters are continued in Table~\ref{tab:LSBD_params_1}) . 
The model was trained 100 times for each of our data sets, but some are removed due to our fit quality criterion of validation cost $<750$.
The number of models accepted and mean validation cost of the accepted models is shown in the second and third columns.
The median of each optimized set of $k_i$ is shown here as the best value, and the 16th and 84th percentiles show the lower and upper end of the distributions.
Precision of each parameter is determined by the standard error of the distribution.}
\label{tab:LSBD_params_0}
\end{table}

\begin{table}
\singlespace
\centering
\begin{tabular}{cr|cccccc}
\hline
\textbf{Data Set}   & \textbf{}     & \textbf{$w_0$} & \textbf{$w_1$} & \textbf{$w_2$}  & \textbf{$w_3$}  & \multicolumn{1}{c}{\textbf{$w_4$}} & \textbf{$w_5$} \\ \hline
                    & Low           & -0.54          & 0.45           & 0.58            & -25.33          & 0.80                                & -1.61          \\
\textbf{All Data}        & \textbf{Best} & \textbf{-0.44} & \textbf{0.48}  & \textbf{0.61}   & \textbf{-25.16} & \textbf{0.88}                       & \textbf{-1.54} \\
                    & High          & -0.36          & 0.53           & 0.65            & -24.99          & 0.95                                & -1.48          \\ \hline
                   & Low           & -1.17          & 0.40           & 1.21          & -26.2          & 0.47                                & -3.41           \\
\textbf{Symmetric RCs} & \textbf{Best} & \textbf{-0.86} & \textbf{0.67}  & \textbf{1.28} & \textbf{-24.9} & \textbf{0.73}                       & \textbf{-3.19}  \\
                    & High          & -0.23          & 0.80           & 1.35          & -24.3          & 1.26                                & -3.02          \\ \hline
\end{tabular}
\caption{Best values for Eq.~\ref{eq:LSBD_Rc} \& Eq.~\ref{eq:LSBD_rho0} of the LSB-D URC model (continued from Table~\ref{tab:LSBD_params_0}).
The median of each optimized set of $w_i$ is shown here as the best value, and the 16th and 84th percentiles show the lower and upper end of the distributions.
Precision of each parameter is determined by the standard error of the distribution.}
\label{tab:LSBD_params_1}
\end{table}

\section{NN Architecture}
\label{sec:NN_arch}

By leveraging the exceptional expressiveness capability of NNs, we can characterize the best possible performance a URC for a given set of inputs.
We applied this concept in two ways: (1) to directly compare against the PSS96 and LSB-D URCs and determine if their performance is optimal given their inputs (\Sec{NN_URC_comp}), and (2) to define upper limits on performance for any URC candidate (\Sec{defining_upper_limits}).

\subsection{Hyperparameter tuning}

Our application of NNs is predicated on an appropriate set of NN hyperparameters for a given set of input variables.
Hyperparameters are chosen to generate NN architectures (e.g., number of nodes $N$, number of layers $K$, optimizer choice).
The ideal hyperparameters vary for each problem and corresponding input set, and testing all possible combinations is not resource-efficient.
Fortunately, we can minimize the hyperparameter search space by following the UAT guidelines of \cite{Lu+2017}.
While the classic work of \cite{Cybenko1989} (and others) chose a fixed depth (number of hidden layers) for a NN and varied its width (number of nodes) to approximate a function, 
\cite{Lu+2017} verified the common belief that increasing depth is more efficient than increasing width when attempting to increase the expressive ability of a NN.
In their formalism, the number of nodes is fixed at $N = x + 4$, where $x$ is the number of input variables, and NN depth is arbitrarily increased until a choice for $K$ is found that best minimizes the cost function.

We used the \emph{TensorFlow} library \citep{tensorflow2015} with the \emph{Keras} API \citep{Keras} to build and optimize our NNs.
We set our cost function to MSE for all NNs to align with choices made in \Sec{URC_fitting}.
While not strictly true, choice of optimizer and batch size generally only affects the speed at which the learning procedure reaches a cost function minimum, not the exact minimum value.
Faster training speeds are of course beneficial, but not necessary, so we do not individually optimize our NNs for these hyperparameters.
We set our optimizer to the ``Nadam'' algorithm, a state-of-the-art implementation of gradient descent with an adaptive learning rate and Nesterov momentum \citep{Nadam}, and our batch size to 512.

\begin{table}
\singlespace
    \centering
    \begin{tabular}{c|c|c}
        \hline 
         \textbf{Input Variable Set} & \textbf{Input Variables} & \textbf{Num. of Layers} \\
         \hline 
         PSS96& $R$, $R_{\rm d}$, $R_{23.5}$, $x=R/R_{\rm d}$, $z=R/R_{23.5}$, $M_{*,23.5}$, $V_{23.5}$ & 14\\
         LSB-D& $R$, $R_{\rm d}$, $x=R/R_{\rm d}$, $M_{\rm d}$, $C_*$, $V_{23.5}$ & 18\\
         Necessary& $R$, $R_{23.5}$, $z=R/R_{23.5}$, $V_{\rm 23.5}$ & 12\\
         $B/D$ parameters& $I_0$, $R_{\rm d}$, $I_{\rm e}$, $R_{\rm e}$, $n$, $x=R/R_{\rm d}$, $B/D$, $\Upsilon_*$, $V_{*}(R)$, $V_{23.5}$ & 4\\
 Non-parametric&$M_{*,23.5}$, $C_{28}$, $\Upsilon_*$, $V_{23.5}$& 14\\
 All variables&$B/D$ parameters + Non-parametric + Necessary  & 4\\
 \hline
 \end{tabular}
    \caption{Input variable sets and final choice for number of layers ($K$) for each of our NNs. Variable names are: 
    galactocentric radius ($R$), isophotal radius ($R_{23.5}$), 
    disk central intensity ($I_0$), disk scale length ($R_{\rm d}$), 
    bulge effective intensity ($I_{\rm e}$), bulge effective radius ($R_{\rm e}$), S\'ersic index for the bulge ($n$),
    bulge-disk mass ratio ($B/D$), mass-to-light ratio ($\Upsilon_*$), 
    stellar mass within $R_{23.5}$ ($M_{*,23.5}$),
    stellar concentration ($C_{28}$),
    compactness ($C_*$),
    stellar velocity contribution as a function of radius ($V_*(R)$),
    and rotational velocity at $R_{23.5}$ ($V_{23.5}$).}
    \label{tab:NN_variable_sets}
\end{table}

\subsection{URC comparison}
\label{sec:NN_URC_comp}

The fair comparison of a NN against a URC requires identical inputs for each.
Table~\ref{tab:NN_variable_sets} summarizes input variables for both the PSS96 and LSB-D NN equivalents.
These input variables represent the data needed for either URC to predict a test galaxy's RC.
Following a similar logic, we use the same training/testing data for the NN equivalents of the PSS96/LSB-D URCs as well.

Tuning for $K$ also dictates how many epochs should be used to train the NN, which we set to 1000.
We avoided over-fitting the data by: (1) rejecting models with validation cost over 750 (as in \Sec{URC_fitting}), and (2) saving the model parameters that produced the minimum validation cost, not simply the model parameters at the end of the 1000 epoch training cycle.
As described in \Sec{URC_fitting}, we re-sampled the training and validation data multiple times in order to reduce any bias from training data choice.
We reduced the number of re-sampling times to 25 however, as NN training is more time-consuming than parameter fitting (\Sec{URC_fitting}).
The mean of the velocity predictions made by the ensemble of 25 NNs was used as the final velocity prediction.

\subsection{URC performance upper limits}\label{sec:defining_upper_limits}

An obvious test is to provide the NN with a large number of variables to set a standard for ``best-possible'' performance.
We can then limit the input variables and compare prediction results to determine which variables are most impactful for predicting RCs.
In the extreme case, we can input only the necessary variables for RC predictions.
While the NN cannot provide a URC in mathematical form, we used this method as a first step to select the best variables for a future URC.

Our variables can be grouped into three broad categories: $B/D$ decomposition parameters, non-parametric galaxy properties, and necessary inputs.
Table~\ref{tab:NN_variable_sets} shows the variables in each set.
The mass-to-light ratio ($\Upsilon_*$) is included in both the $B/D$ parameter and non-parametric group.
Necessary inputs are the radii at which RCs are evaluated, a scaling velocity, and the radius at which the scaling velocity is provided ($R_{23.5}$).
All of the NNs trained must include the necessary input group.

We are motivated to train the NNs with $V_{23.5}$ to compare against the PSS96 and LSB-D URCs and determine if their performance can be improved upon by a different URC formulation that also has a direct velocity input.
Importantly, we can also define the best possible performance over inner radii where non-circular motions exist.

Ultimately, we have four NNs trained with a direct velocity input ($V_{23.5}$). 
Each of the four corresponds to one of the combinations: necessary + $B/D$ parameters + non-parametric properties (``All Variables''), necessary + $B/D$ parameters, necessary + non-parametric, or only necessary variables.
We refer to the NN for the necessary + $B/D$ parameters variables set simply as the ``$B/D$ parameters'' NN for short.
A similar naming convention applies for the necessary + non-parametric variables set.
All four NNs used the same training/testing data. 
Our final choices for number of layers for each NN are found in \Table{NN_variable_sets}.
As in the NNs for direct URC comparison, we re-sampled training/validation data to create an ensemble of 25 sets of predictions for testing data, and used the mean as our final predictions.

\section{URC Performance}
\label{sec:URC}

The URC parameters and NNs obtained through the procedures described in \Secs{URC_fitting}{NN_arch} enable testing the accuracy (performance) of each URC and NN.
\Secs{PSS96_URC_accuracy}{LSBD_URC} address the accuracy of the PSS96 and LSB-D URCs and their NN equivalents, while \Sec{presenting_upper_limits} presents the NN-only approach of determining best possible URC performance with a given set of input variables.
\Sec{symmetric_URC} compares LSB-D URC performance for the symmetric RC subset with the combined data set.

We have quantified model accuracy using two metrics: (i) ``data - model'' velocity difference and (ii) absolute percent difference.
These are calculated as $V_{\rm observed} - V_{\rm model}$, and $|V_{\rm observed} - V_{\rm model}| / V_{\rm observed}$, respectively.
We determined their values by separating our data into normalized radial bins (in the manner of PSS96) and calculating both metrics for data points in each bin.
Binning with normalized radius allows us to fairly evaluate model accuracy for all galaxies in the testing set as a collective, as opposed to data - model comparisons for each galaxy individually.
Bins with less than 25 data points were removed as their population statistics (median, 16th and 84th percentiles) behaved erratically.
Observed velocity values were extrapolated from our RC fits at least to $1.05 R_{23.5}$, or further if the observed RC data points extended beyond $1.05R_{23.5}$ (see \Sec{RCfits}).
Few galaxy RCs extended beyond this point, explaining the relative abundance of data in the histograms in the top panels of each figure below.

\subsection{Accuracy of the PSS96 URC}
\label{sec:PSS96_URC_accuracy}

\begin{figure}
\singlespace
\centering
\includegraphics[width=\textwidth]{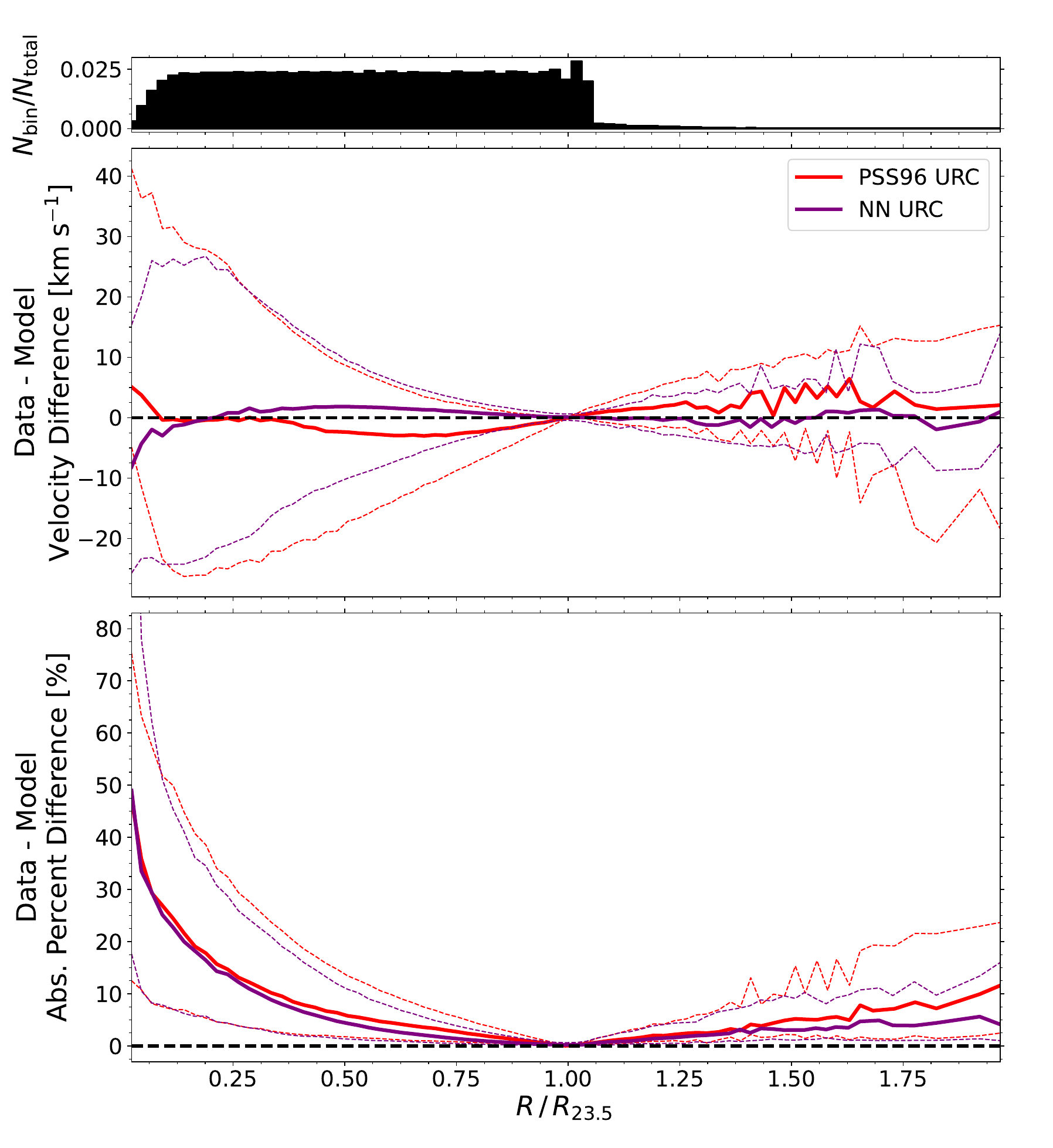}
\caption{Velocity difference and absolute percent difference relative to testing data for the PSS96 URC and its NN equivalent as a function of normalized radius.
Calculated as $V_{\rm observed} - V_{\rm model}$ and $|V_{\rm observed} - V_{\rm model}| / V_{\rm observed}$ for the middle and bottom panels respectively. 
The solid line traces the median and the dashed lines represent the 16th and 84th percentiles of data in each radial bin. 
The histogram in the top panel shows the distribution of data in each bin.}
\label{fig:PSS_combined_errors}
\end{figure}

\Fig{PSS_combined_errors} presents the performance of the PSS96 URC.
In terms of velocity difference, the PSS96 URC works best near $R_{23.5}$ as expected, since $V_{23.5}$ is a direct input.
Beyond $R_{23.5}$, the model consistently under-estimates velocity.
For $R < R_{23.5}$, there is significant scatter but the median tends to over-estimation at intermediate radii ($0.1 < R/R_{23.5} < 0.9)$, transitioning to typically under-predictions at the innermost radii. 
In general, while the median difference at any given radius is similar (between $-10$ and $10$ \kms), the spread in differences becomes significant for $R/R_{23.5} < 0.8$.
These inner regions of galaxies are complex, since: 
(a) morphological differences are most conspicuous, 
(b) galaxies transition from light to dark matter domination, and, 
(c) RC data are not well defined due to non-circular motions.

The issue of unreliable inner RC data is further exacerbated by rapid variations in orbit inclination in the inner regions of galaxies.
A single inclination value is typically used to deproject line-of-sight velocities. 
However, this relies on the assumption that the galaxy is a thin disk and does not hold for the inner regions of galaxies with bulges, which extend well above the galactic plane.
This presents another barrier for the model, as the velocity data used for training purposes at inner radii may be poorly defined.

Under-predicted velocities are more often encountered at outermost radii, which may be indicative of the limited expressive power of the dark halo model for the PSS96 URC (see \Eq{PSS96_DM}).
The only free halo parameter for the halo, the scaled core radius $a$, accounts for the shape of the halo density profile and depends only on stellar mass.
Adjusting the magnitude of the halo contribution can only be achieved via the disk mass fraction parameter, $\beta$.
Without the flexibility afforded by a free central density parameter, the large core radii needed for URC accuracy at large radii cause over-predictions at intermediate radii due to an enhanced halo velocity contribution component throughout the RC.

\Fig{PSS_combined_errors} also shows the absolute percent difference between observed and modelled velocities as a function of radius.
The curve tracing the median reaches its lowest point near $R_{23.5}$, where velocity differences are artificially suppressed.
For the range $0.5 \leq R/R_{23.5} \leq 1.5$, the per cent differences are still fairly small, with median values below 10 per cent. 
The 84th percentile range is however much larger.
Percent differences often exceed 15 per cent at other radii. 

The NN URC is a marked improvement over PSS96, even though the 84th percentile line drops below 10 per cent for only a limited radii range.
Throughout the radial extents used in 
\Fig{PSS_combined_errors}, the NN URC shows tighter scatter, and therefore superior accuracy, than the PSS96 URC.
There are no systematic under- or over-predictions, and the scatter is symmetrical at all radii.
The superior performance of the NN URC shows that the PSS96 formulation does not maximally capture the information embedded in the input variables.
This indicates that, using the same input variables, a URC formulation exists that would perform better than the PSS96 URC.

\begin{figure}
\singlespace
\centering
\includegraphics[width=\textwidth]{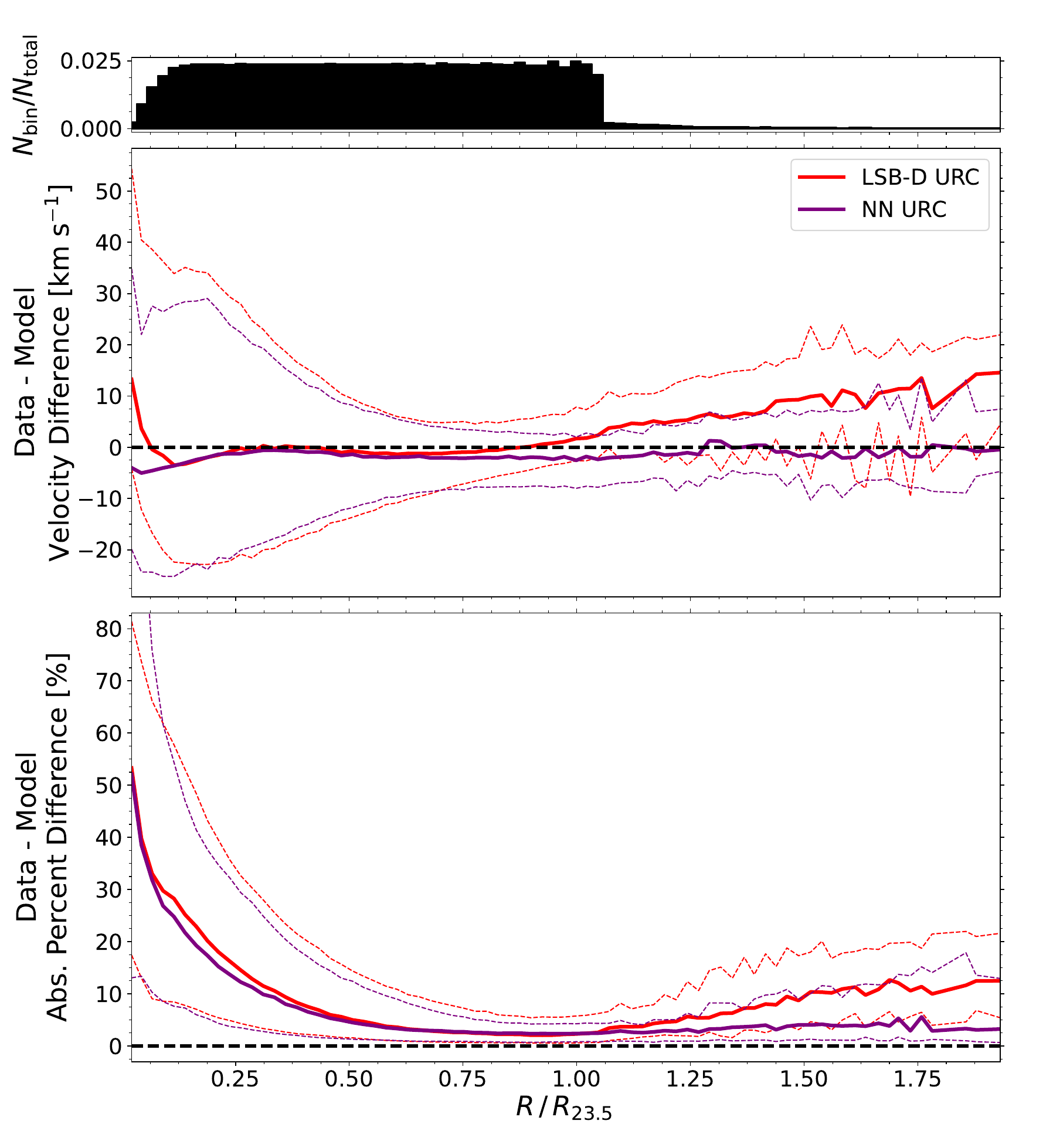}
\caption{Velocity difference and absolute percent difference relative to testing data for the LSB-D URC and its NN equivalent as a function of normalized radius.
Calculated as $V_{\rm observed} - V_{\rm model}$ and $|V_{\rm observed} - V_{\rm model}| / V_{\rm observed}$ for the top and bottom panels respectively. 
The solid line traces the median and the dashed lines represent the 16th and 84th percentiles of data in each radial bin. 
The histogram in the top panel shows the distribution of data in each bin.}
\label{fig:LSBD_combined_errors}
\end{figure}

\subsection{Accuracy of the LSB-D URC}
\label{sec:LSBD_URC}

The accuracy of the LSB-D URC is analysed in the same manner as the PSS96 URC - through radially binned velocity differences and absolute velocity differences.
\Fig{LSBD_combined_errors} shows the excellent agreement of the LSB-D URC and the NN over the range 0.3-1.0 $R_{23.5}$.
Both have large and symmetric under- and over-predictions, slightly skewed towards the latter.
Note that unlike the PSS96 URC which uses $V_{23.5}$ as a direct scaling input, $V_{23.5}$ only appears indirectly in the LSB-D URC via 
Equations \ref{eq:LSBD_Md}, \ref{eq:LSBD_Rc} and \ref{eq:LSBD_rho0}.
Since $V_{23.5}$ is not a direct scaling input, the tightening in velocity residuals near $R_{23.5}$ is no longer observed. 
Similarly, while $V_{23.5}$ serves as input to the NN, $R_{23.5}$ is not involved and there is no tightening near $R_{23.5}$ for its trend lines either.

Near $R_{23.5}$, both the LSB-D and NN URC trends diverge.
The LSB-D URC consistently under-predicts velocities at these large, dark-matter dominated radii while the NN URC maintains a median difference near zero. 
As was surmised for PSS96, this trend indicates that the dark halo parameterization for the LSB-D URC is still lacking.

The absolute percent difference plot of \Fig{LSBD_combined_errors} tells a similar story as the velocity differences.
The LBS-D URC performs as well as the NN within $R_{23.5}$ and under-predicts velocities beyond that point.
\cite{Paolo+2019} reported a typical uncertainty of 8 per cent for the LSB-D URC, which agrees with our results if we disregard the innermost high deviation regions ($<0.5 R_{23.5}$).

Overall, this URC and the PSS96 URC perform at the same level.
Despite adding a compactness parameter, the 84th percentile line in \Fig{LSBD_combined_errors} is above 20 per cent below $0.45 R_{23.5}$ and roughly between 5 and 20 per cent above $1.25 R_{23.5}$ for both.
The key distinction between the two formulations is how closely the LSB-D URC traces the NN URC relative to PSS96 within $0.75~R_{23.5}$.
The LSB-D formulation is close to the best-case NN scenario, at least over the inner luminous regions of galaxies.
We attributed the large deviations in velocity predictions for the PSS96 URC over inner/intermediate radii to the inadequacy of $\beta$ as a model parameter since the NN equivalent performed much better.
Here, the URC error trend matches the NN trend (over $0.3-1.0 R_{23.5}$) and we contend that \Eq{LSBD_Md} provides a fair modelling of the stellar velocity contribution.
While the scatter in \Fig{LSBD_combined_errors} is large at these radii, it cannot be improved without additional input variables.

\begin{figure}
\singlespace
\centering
\includegraphics[width=\textwidth]{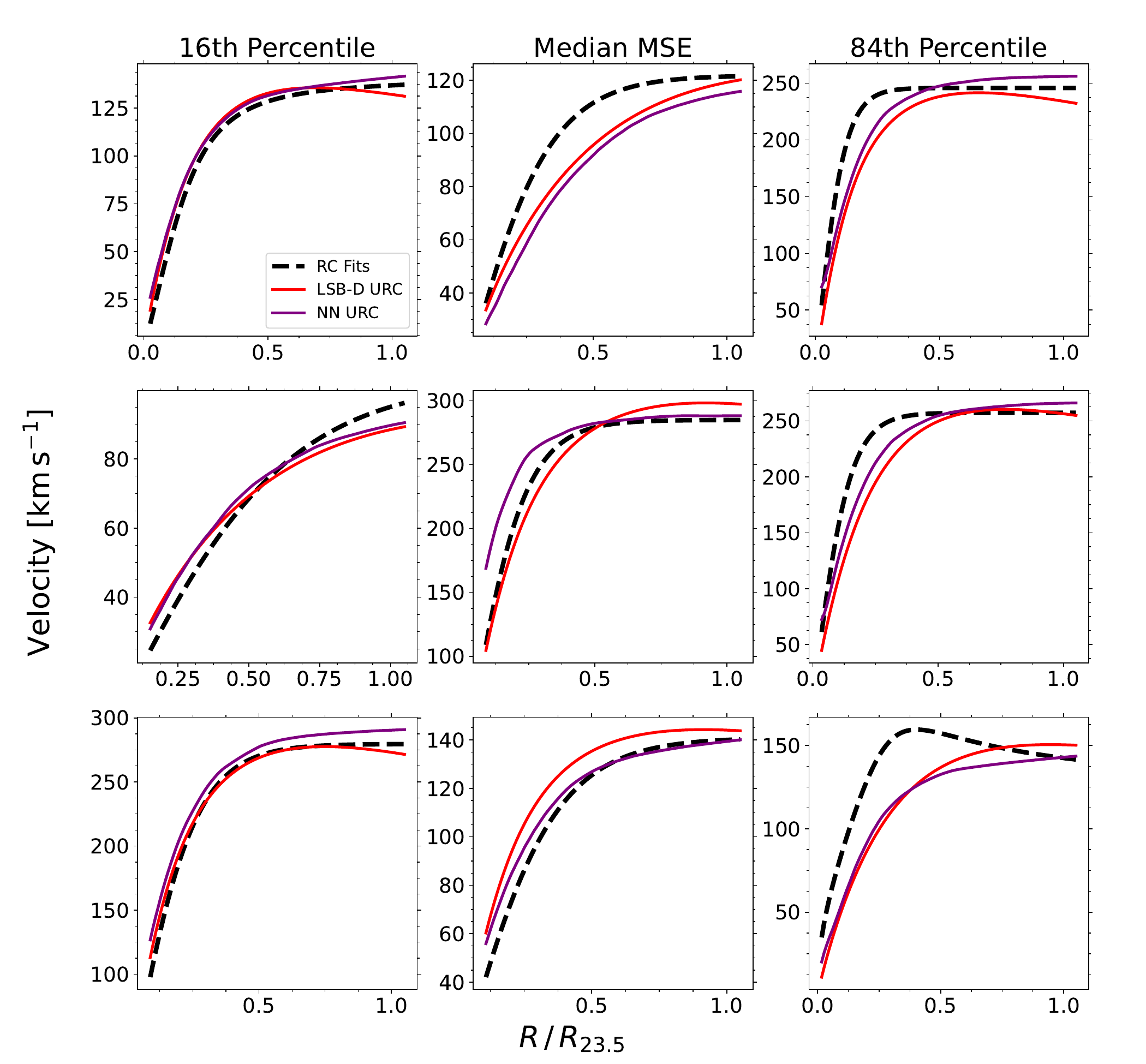}
\caption{LSB-D URC examples and matching NN predictions (solid lines for both) for typical smoothed RCs (thick dashed lines).
Each panel is a different sample galaxy.
Galaxies in each column have MSE costs (16th percentile, Median MSE, 84th percentile) that correspond to population statistics of the test distribution.
}
\label{fig:LSBD_examples}
\end{figure}

\Fig{LSBD_examples} shows examples of individual RC performance for the LSB-D URC.
The biggest concern noted here is that the LSB-D URC falls too quickly.
Since this falling behavior occurs at high radii, this is an issue caused by the dark halo profile.
As discussed, the LSB-D URC traces the NN closely at inner radii, indicating the luminous contribution parameterization extracts close to a maximal amount of information from the input variables.

Overall, the improvements of \cite{Karukes+Salucci2017} and \cite{Paolo+2019} (addition of $\rho_0$ and $C_*$; removal of $\beta$)
have yielded better velocity predictions at small radii but inaccuracies exceeding 15 per cent still remain at both innermost and outermost radii.

\subsection{NN upper limits}
\label{sec:presenting_upper_limits}

We now utilize NNs to determine the best-case performance for any URC given the available data.
By selectively including the various input variable sets introduced in \Sec{defining_upper_limits} as NN inputs, we can also pinpoint which variables are most useful for a URC formulation.
We present below velocity differences and absolute percent differences in the radially binned format for each input group.

\begin{figure}
\singlespace
\centering
\includegraphics[width=\textwidth]{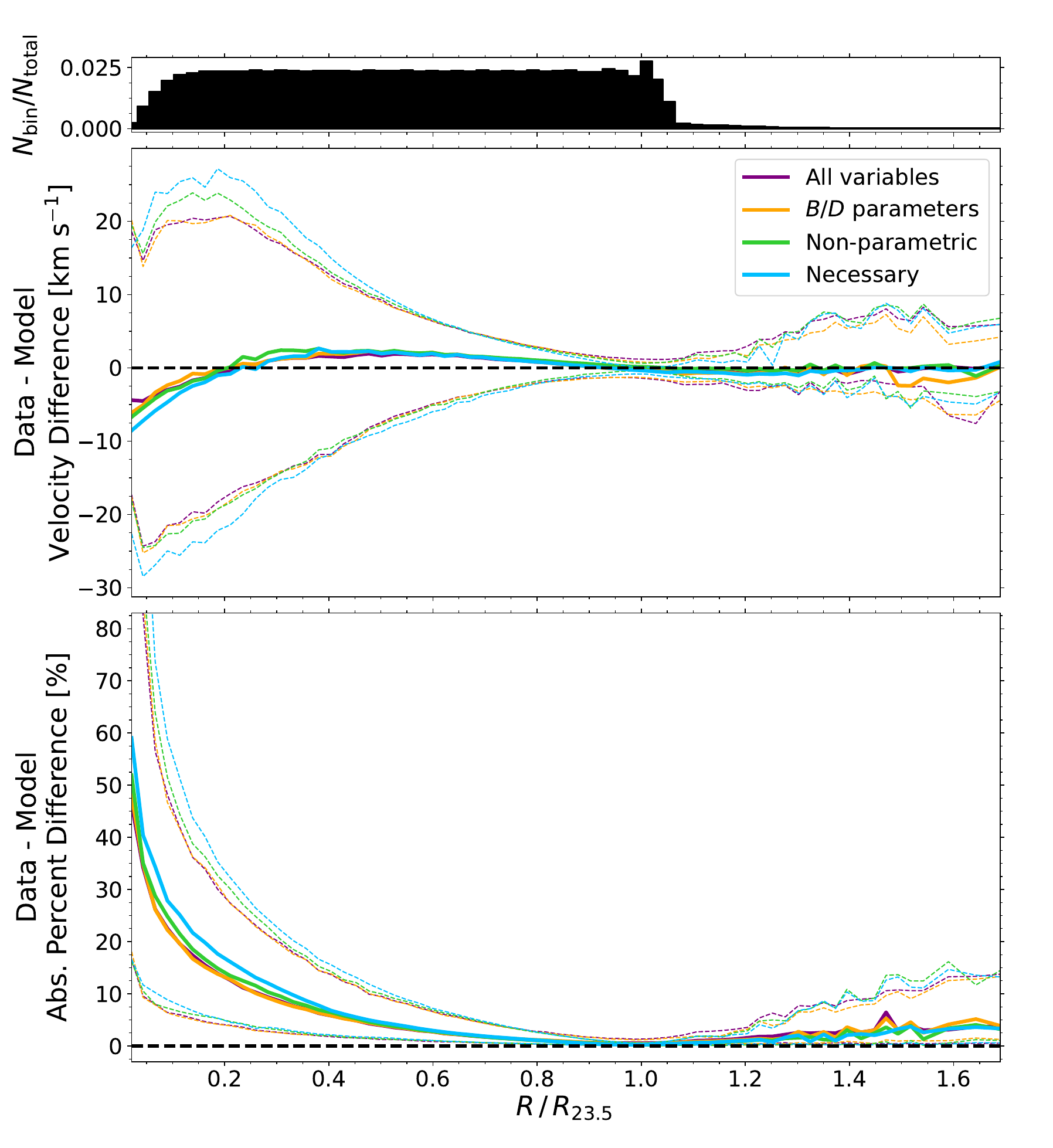}
\caption{Velocity difference and absolute percent difference relative to testing data for the $V_{23.5}$ scaled NNs as a function of normalized radius.
Calculated as $V_{\rm observed} - V_{\rm model}$ and $|V_{\rm observed} - V_{\rm model}| / V_{\rm observed}$ for the top and bottom panels respectively. 
The solid line traces the median and the dashed lines represent the 16th and 84th percentiles of data in each radial bin. 
The histogram in the top panel shows the distribution of data in each bin.}
\label{fig:custom_V235_combined_errors}
\end{figure}

\Fig{custom_V235_combined_errors} illustrates the performance of the $V_{23.5}$ scaled NNs.
As expected, both plots show a tight error spread near $R_{23.5}$ due to the $V_{23.5}$ input.
The all variables and $B/D$ parameters groups show slightly more spread than the necessary and non-parametric groups due to the increased complexity of their NNs.
With a small number of input variables, as in the necessary variables group, it is straightforward for the NN to adjust its weights to carry forward $V_{23.5}$ from the input layer to the output layer since there is no additional information to consider.
As input variables increase though, weights are tuned to produce the best possible predictions using this new information at all radii, not just $R_{23.5}$.
These weights produce more accurate predictions overall, but the NN performs slightly worse near $R_{23.5}$ due to this.

\begin{figure}
\singlespace
\centering
\includegraphics[width=\textwidth]{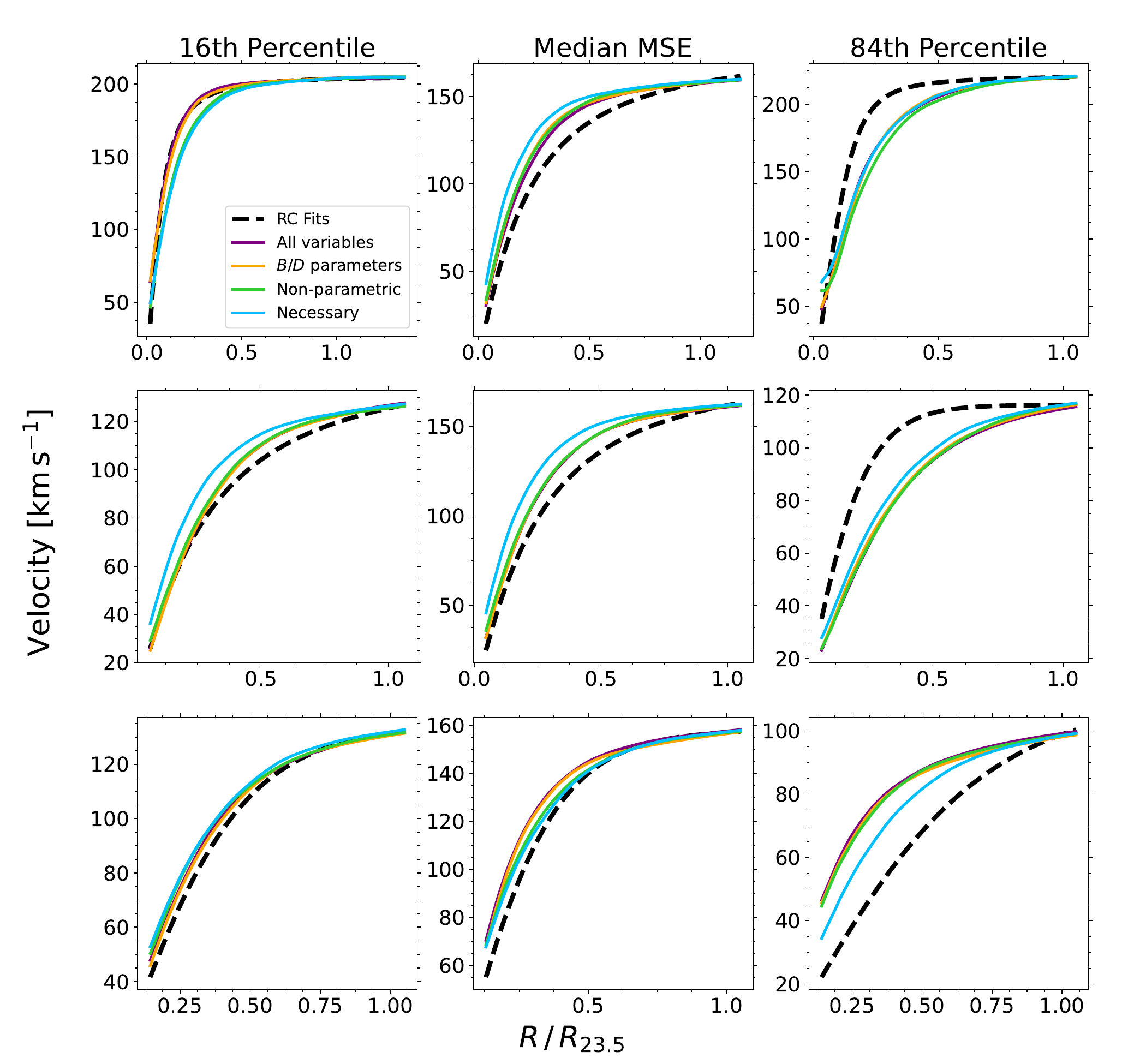}
\caption{Individual RCs showing the performance of each $V_{23.5}$ scaled NN.
See \Fig{LSBD_examples} caption for details.
}
\label{fig:custom_V235_examples}
\end{figure}

All groups performed similarly over the dark halo dominated outer regions of galaxies.
Data are sparse there, so the minor differences and erratic behaviors are due to small sample size and remain rather insignificant.
Accurate performance at high radii across all groups is not surprising, as most RCs are nearly flat beyond $R_{23.5}$.
With an input of $V_{23.5}$ and $R_{23.5}$, the NNs can derive a $V_{23.5}$ - $R_{23.5}$ - outer RC slope relation to simply predict a line beyond $R_{23.5}$ and, on average, be reasonably accurate.
Indeed, even a flat line prediction at $V_{23.5}$ would be a good estimate.
However, we see in the middle column, bottom row example in \Fig{custom_V235_examples} that the NNs predict a slightly rising RC at these high radii, so they are affecting more than a simple flat line prediction.
This $V_{23.5}$ - $R_{23.5}$ - outer RC slope relation dependence is confirmed by the performance of the necessary variables group.
This NN only has $V_{23.5}$ and $R_{23.5}$ information available, so this is the only relation that it could possibly derive.
Since the other groups show the same error trends, they are using a similar relation to predict velocity at outer radii and do not use the additional information provided to them.

The impact of adding stellar distribution information is most noticeable over the inner luminous matter dominated regions of galaxies.
Variable groups that include stellar properties all show less spread than the baseline necessary variables set.
The $B/D$ parameter group performs better than the non-parametric group, indicating that while galactic scale measures of stellar distribution like $M_{*,23.5}$ and $C_{28}$ offer some improvement on the necessary input group, an accurate $B/D$ breakdown is ideal for a URC.
The $B/D$ parameter trend closely parallels the all variables group trend in 
\Fig{custom_V235_combined_errors}, implying that the non-parametric measures are inconsequential and the NN is largely utilizing $B/D$ parameters.

Individual sample RCs are presented in \Fig{custom_V235_examples} with their NN RC predictions.
With $V_{23.5}$ serving as a fixed point to ensure correct scaling, all NNs can reproduce a variety of RC shapes well.
As expected, inclusion of stellar distribution parameters produces RC adjustments that align better with the actual shape.

\begin{figure}
\singlespace
\centering
\includegraphics[width=\textwidth]{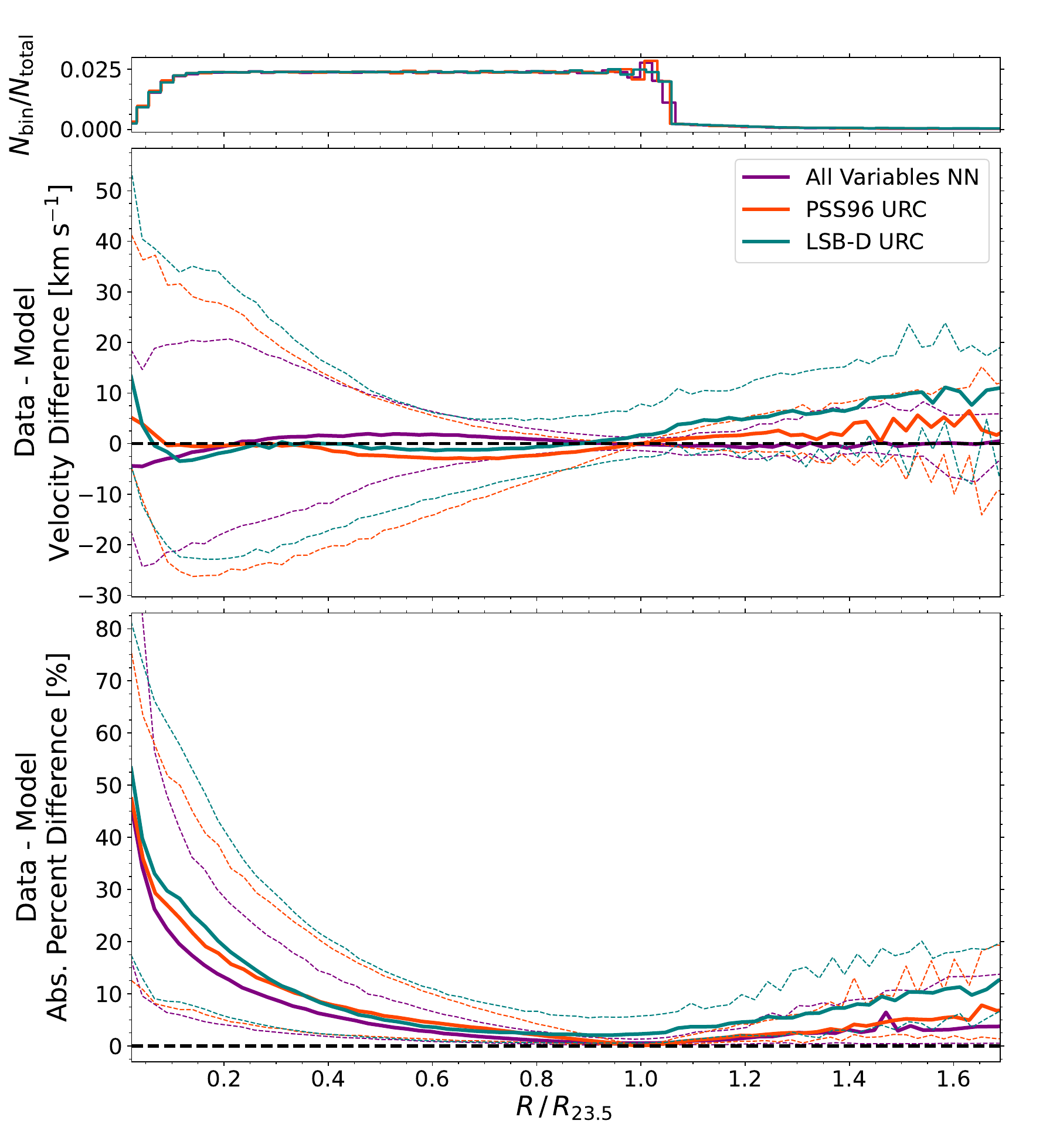}
\caption{Velocity difference and absolute percent difference relative to testing data for the all variables group $V_{23.5}$ scaled NN, the PSS96 URC, and the LSB-D URC as a function of normalized radius.
Calculated as $V_{\rm observed} - V_{\rm model}$ and $|V_{\rm observed} - V_{\rm model}| / V_{\rm observed}$ for the top and bottom panels respectively. 
The solid line traces the median and the dashed lines represent the 16th and 84th percentiles of data in each radial bin. 
The histogram in the top panel shows the distribution of data in each bin.}
\label{fig:custom_URC_comp_combined_errors}
\end{figure}

\begin{figure}
\singlespace
\centering
\includegraphics[width=\textwidth]{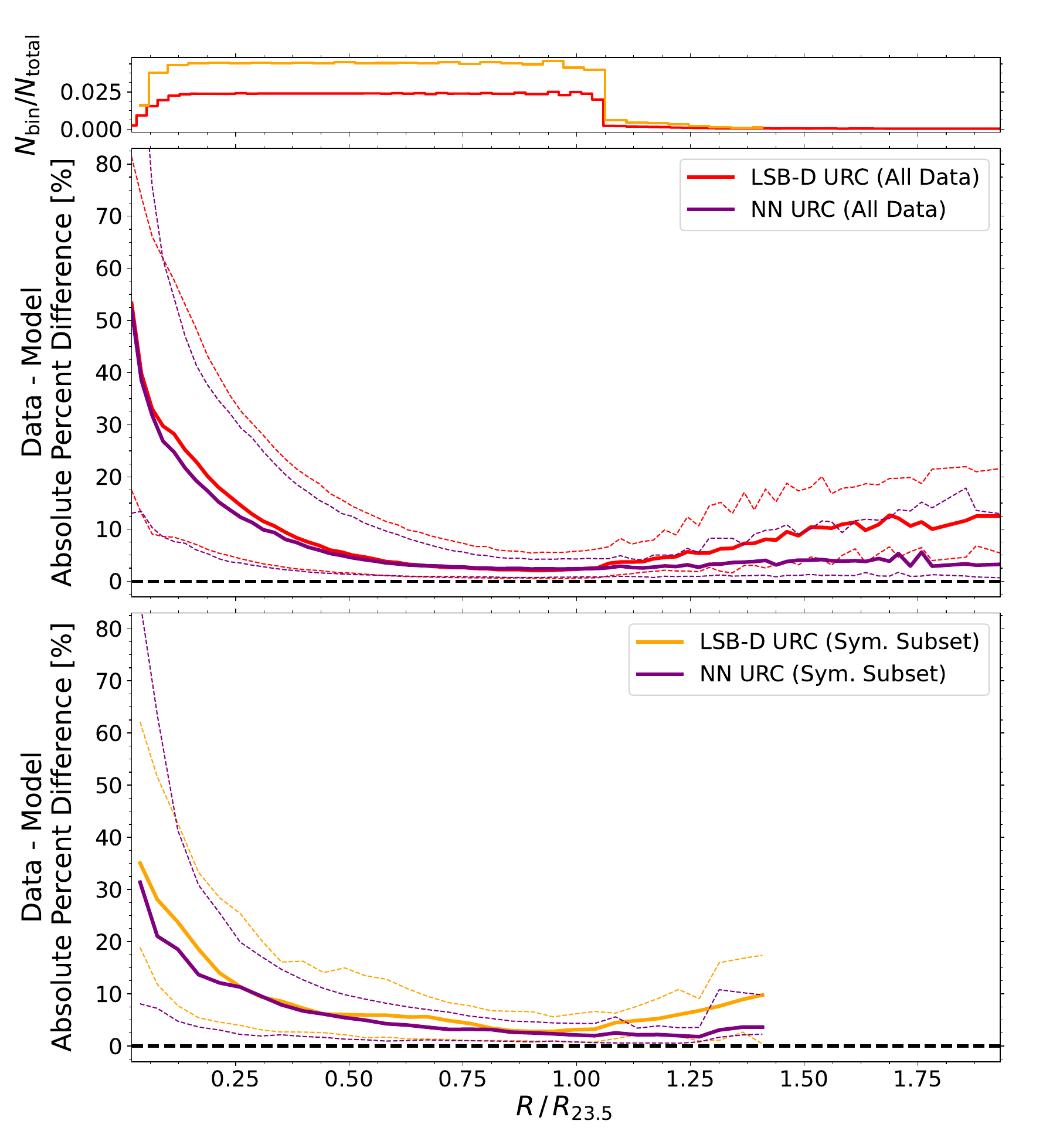}
\caption{Absolute percent difference relative to testing data for the full data set (middle panel) and the symmetric RC sub-sample (bottom panel) for the LSB-D URC and its NN equivalent as a function of normalized radius.
RC data for galaxies in the ``symmetric'' subset do not extend as far as the full data set, which results in the shorter radial cutoff seen here.
Calculated as $|V_{\rm observed} - V_{\rm model}| / V_{\rm observed}$. 
The solid line traces the median and the dashed lines represent the 16th and 84th percentiles of data in each radial bin. 
The histogram in the top panel shows the distribution of data in each bin for each data set.}
\label{fig:symmetric_URC}
\end{figure}

Next, we use the all variables group NN as a performance benchmark for the PSS96 and LSB-D URCs.
Given the considerable improvement the NN provides relative to the PSS96 and LSB-D URC (\Fig{custom_URC_comp_combined_errors}), we surmise that a superior URC formulation is possible.
As discussed in previous sections, both the PSS96 and LSB-D URCs are ineffective at modelling velocities over the dark halo dominated regions of RCs.
The NN performs significantly better at those large radii. 
Furthermore, the all variables NN also performs significantly better at inner radii, suggesting that the URCs are not limited by non-circular motions or inclination correction concerns over this region.
With additional variables, they could perform better than they do currently.
While the all variables NN includes many more variables than the URCs, parameters with a bulge velocity contribution are more likely to have greater impact.

\subsection{URC performance for symmetric RCs}
\label{sec:symmetric_URC}

In \Sec{symmetric_RCs}, we devised a sub-sample of 579 symmetric RC galaxies.
This sub-sample is now examined for the performance of our LSB-D URC and its NN equivalent. 
\Fig{symmetric_URC} reveals that the LSB-D URC and its NN equivalent perform marginally better for the symmetric RC subset relative to the full data set.
The 84th percentile curve drops below 20 per cent at $\sim$0.45 $R_{\rm 23.5}$ for the URC trained on all the data, while for the symmetric RC subset, this value is reached at $\sim$0.3 $R_{\rm 23.5}$.
This relative performance improvement in the central galactic regions likely stems from the dearth of prominent bulges in the symmetric RC subset, where $\sim$80 per cent of galaxies are disk-dominated.
The disk velocity component of the URC may better represent the baryonic mass distribution of these symmetric systems relative to the URC trained on the full data set, which includes more galaxies with prominent bulges.

The URC trained on the symmetric RC subset also shows a trend of increasing errors at higher radii, as seen for the URC trained on all data. 
However, because the symmetric RC subset galaxies lack data at large galactocentric radii, a definitive assessment of their relative URC behavior in the halo regions is not currently possible.

Similar to the full data set trained URC, the symmetric subset trained URC shows increased percent differences relative to its NN equivalents at all radii.
Given the above observations, we conclude that even for the symmetric RC data subset, the LSB-D URC does not reach a sufficient level of accuracy at all radii to qualify as a truly universal. 

\subsection{Is there a URC?}
\label{sec:is_there_URC}

In this paper, we have tested various URC formulations and their NN equivalents with a most extensive compilations of RC based on our PROBES I-II and MaNGA data sets. 
We have also reduced the latter to a sub-sample of galaxies with symmetric RCs. 
We have verified that the LSB-D URC performs relatively well for our full and smaller symmetric samples within $R_{23.5}$, in agreement with \cite{Karukes+Salucci2017} and \cite{Paolo+2019}. 
However, in no circumstances can these formulations be deemed universal since 
modifications would be needed to improve their utility over the full radial range from center to edge. 
This statement is further supported by the superior performance of the URC NN equivalents.
These NNs provide a glimpse at the level of accuracy that could be achieved by future URC improvements.

In the general case, disregarding exact URC formulations, our results for the $V_{23.5}$ scaled NN suggest that a universal shape of RCs may exist when a measure of dark matter content (via $V_{23.5}$) is considered in the formulation.
Individual RCs in \Fig{custom_V235_examples} show that the NNs can predict RC shapes quite well, even at the 84th percentile level.
Effects from non-circular motions become prominent at small radii ($<0.3 R_{23.5}$), as seen in \Fig{custom_V235_combined_errors}, though the NNs perform well (with discrepancies below 30 per cent) beyond $\sim0.3 R_{23.5}$) . 
Most of the RC predictions are well within $\pm10$ \kms relative to the smoothed RCs. 
While the inclusion of stellar properties is beneficial for prediction accuracy, the necessary group NN with only $R_{23.5}$ and $V_{23.5}$ still achieves accuracy levels fairly close to the all variables group.
If radial information alone produces this level of accuracy, it suggests that a truly universal RC formulation could be achieved.

The practicality of a URC that includes $V_{23.5}$ is limited however, primarily useful for spatially unresolved HI or high-redshift observations\footnote{If $V_{23.5}$ is known, data for the entire RC are likely available anyways, defeating the purpose of a URC as a predictive tool.}.
Such a URC could be useful to draw connections between galactic luminous and dark components and their elusive interplay in galaxies, but improvements to the PSS96 and LSB-D URCs are needed before this method can be exploited.

\section{Discussion and Conclusions}
\label{sec:discussion}

We have highlighted various successes and shortcomings of the PSS96 and LSB-D URC formulations above.  
Two majors areas for improving URC parametrizations were identified, pertaining respectively to the modeling of the galactic bulge and dark halo. 

Including a bulge velocity component would be a natural extension to the disk velocity contribution currently used in the PSS96 and LSB-D URCs, leading to a total stellar velocity contribution in the form of \Eq{stellar_V}.
We showed in \Sec{BD_decomps}, and see also \cite{Spitzer_BD_decomp}, that the one-dimensional photometric bulge component of barred galaxies is systematically over-estimated in photometric decompositions if a bar component is not included.
Therefore, the bulge velocity contribution based on our derived one-dimensional photometric bulge parameters would over-estimate the stellar mass in central regions of galaxies.
The improved two-dimensional modeling of galactic centers (with or without bars), and ideally using tilted ring models, could constrain non-circular motions yielding more accurate stellar masses and mass models \citep{Courteau2014,Sellwood2015}.
The proper two-dimensional modelling of a bar and bulge for our sample galaxies will be explored elsewhere.
AstroPhot \citep{astrophot} is an ideal photometric decomposition package for such a task. 

We also found that current URCs typically under-predict velocities in the galaxy's outskirts, calling for an improved dark matter halo parameterization.
The URC NN equivalents also performed significantly better than the original URC formulations, implying that both URC input variable sets have enough information encoded within them to improve prediction accuracy at large radii, but the latter is not being fully utilized with current parameterizations.
Current URCs may be amended by using a different halo profile (only the NFW and Burkert profiles were tested) and/or by changing the relations used to define $R_{\rm c}$ and $\rho_0$.

Equally important, pure dark matter profiles must be modified to account for any adiabatic contraction and/or expansion due to feedback of the baryonic matter on the dark halo. 
The primary tools for investigating halo expansion and contraction are cosmological simulations.
Much work along those lines has supported the halo contraction argument 
\citep[e.g.,][]{Blumenthal+1986,Abadi+2010,Sawala+2016}, 
the halo expansion argument \citep[e.g.,][]{Dutton+2007,Governato+2010,Macciò+2012,Chan+2015}, 
or both occurring at different epochs \citep[e.g.,][]{Tollet+2016}.
Generally, halo response is a function of mass.
Lower mass disk galaxies ($M_* \lesssim 10^{10} M_\odot$) typically encounter halo expansion, while higher mass disk galaxies ($M_* \gtrsim 10^{10}$) experience halo contraction, though not as strong as predicted by adiabatic evolution.
Not all simulations agree and some do not show expansion for lower mass galaxies \citep{Sawala+2016,Grand+2016}.

\cite{Li+2022} developed and applied a methodology of rotation curve fitting that explicitly accounts for adiabatic contraction to a subset of our sample \citep[from][]{Lelli+2016}.
In their model, galaxies should undergo a strong level of adiabatic contraction which cannot be reconciled with the flatter profiles of present-epoch galaxies without artificially reducing mass-to-light ratios or elevated levels of feedback.
However, their model does not include mergers and assumes that halos have completed their growth before baryonic compression begins.
While they acknowledged this shortcoming, the resulting over-estimation of adiabatic contraction resulting from this assumption is not addressed. 

Likewise, \cite{DelPopolo2009} and \cite{DelPopolo+2014} also investigated the impact of a bulge on dark halo profiles using a semi-analytic halo-galaxy evolution model, and found that:
(a) the presence of a bulge causes steeper inner density profiles and, 
(b) lower mass galaxies have flatter profiles than their higher mass counterparts.
Statement (a) implies that bulge-dominant galaxies have smaller core radii and larger central densities. 
Statement (b) corresponds to decreasing normalized radii with increasing stellar mass for PSS96 URC normalized core radii.
Their model does not include mergers, the mechanism thought to be the primary deterrent of adiabatic contraction.

It has also been shown that simulated shapes of dark halo profiles depend directly on the choice of gas density threshold for star formation \citep{PontzenGovernato2012,Benítez-Llambay+2019,Dutton+2020}.
Such a strong dependence on a single simulation parameter explains why similar studies may yield contradicting results. 
For instance, \cite{Chan+2015} and \cite{Sawala+2016} found that the shape of their simulated halo profiles was controlled by their choice of star formation threshold.
Short of a definitive choice for such a value, any conclusions from simulations on dark halo profiles and their evolution based on RC shapes alone is premature.
The search for a URC based on numerical simulations requires the latest gastrophysical implementations.

Extensive accurate data go a long way towards guiding model developments.  
For the purpose of detailed RC modeling over all galactocentric radii, more neutral gas data for improved bulge/disk models and the most extended RCs to constrain dark halo models are sorely needed. 
The broadest morphological sampling is also required to establish the existence of a truly universal rotation curve. 

Our PROBES I-II and MaNGA database of spatially-resolved rotation curves and matching multi-band photometry is available upon request to the authors. 

\section*{Acknowledgements}

NA, SC, and LW acknowledge generous support from the Natural Sciences and Engineering Research Council of Canada and Queen's University through various scholarships and grants. 
CS acknowledges the support of the Natural Sciences and Engineering Research Council of Canada (NSERC). 
Cette recherche a été financée par le Conseil de recherches en sciences naturelles et en génie du Canada (CRSNG). 
CS also acknowledges support from the Canadian Institute for Theoretical Astrophysics (CITA) National Fellowship program.
NA and RP also gratefully acknowledge support from Queen’s University.
MF acknowledges support from the Australian Government Research Training Program (RTP) Scholarship.
SC wishes to thank Paolo Salucci for his hospitality at SISSA where some of this work was completed. 
Paolo Salucci and Kristine Spekkens are thanked for insightful discussions and references.

\bibliographystyle{aasjournal}
\bibliography{citations_fixed}{}

\begin{thebibliography}{}
\expandafter\ifx\csname natexlab\endcsname\relax\def\natexlab#1{#1}\fi
\providecommand{\url}[1]{\href{#1}{#1}}
\providecommand{\dodoi}[1]{doi:~\href{http://doi.org/#1}{\nolinkurl{#1}}}
\providecommand{\doeprint}[1]{\href{http://ascl.net/#1}{\nolinkurl{http://ascl.net/#1}}}
\providecommand{\doarXiv}[1]{\href{https://arxiv.org/abs/#1}{\nolinkurl{https://arxiv.org/abs/#1}}}

\bibitem[{Abadi {et~al.}(2015)Abadi, Agarwal, Barham, Brevdo, Chen, Citro,
  Corrado, Davis, Dean, Devin, Ghemawat, Goodfellow, Harp, Irving, Isard, Jia,
  Jozefowicz, Kaiser, Kudlur, Levenberg, Man\'{e}, Monga, Moore, Murray, Olah,
  Schuster, Shlens, Steiner, Sutskever, Talwar, Tucker, Vanhoucke, Vasudevan,
  Vi\'{e}gas, Vinyals, Warden, Wattenberg, Wicke, Yu, \&
  Zheng}]{tensorflow2015}
Abadi, M., Agarwal, A., Barham, P., {et~al.} 2015, {TensorFlow}: Large-Scale
  Machine Learning on Heterogeneous Systems.
\newblock \url{https://www.tensorflow.org/}

\bibitem[{{Abadi} {et~al.}(2010){Abadi}, {Navarro}, {Fardal}, {Babul}, \&
  {Steinmetz}}]{Abadi+2010}
{Abadi}, M.~G., {Navarro}, J.~F., {Fardal}, M., {Babul}, A., \& {Steinmetz}, M.
  2010, \mnras, 407, 435, \dodoi{10.1111/j.1365-2966.2010.16912.x}

\bibitem[{{Abdurro'uf} {et~al.}(2022){Abdurro'uf}, {Accetta}, {Aerts}, {Silva
  Aguirre}, {Ahumada}, {Ajgaonkar}, {Filiz Ak}, {Alam}, {Allende Prieto},
  {Almeida}, {Anders}, {Anderson}, {Andrews}, {Anguiano}, {Aquino-Ort{\'\i}z},
  {Arag{\'o}n-Salamanca}, {Argudo-Fern{\'a}ndez}, {Ata}, {Aubert},
  {Avila-Reese}, {Badenes}, {Barb{\'a}}, {Barger}, {Barrera-Ballesteros},
  {Beaton}, {Beers}, {Belfiore}, {Bender}, {Bernardi}, {Bershady}, {Beutler},
  {Bidin}, {Bird}, {Bizyaev}, {Blanc}, {Blanton}, {Boardman}, {Bolton},
  {Boquien}, {Borissova}, {Bovy}, {Brandt}, {Brown}, {Brownstein}, {Brusa},
  {Buchner}, {Bundy}, {Burchett}, {Bureau}, {Burgasser}, {Cabang}, {Campbell},
  {Cappellari}, {Carlberg}, {Wanderley}, {Carrera}, {Cash}, {Chen}, {Chen},
  {Cherinka}, {Chiappini}, {Choi}, {Chojnowski}, {Chung}, {Clerc}, {Cohen},
  {Comerford}, {Comparat}, {da Costa}, {Covey}, {Crane}, {Cruz-Gonzalez},
  {Culhane}, {Cunha}, {Dai}, {Damke}, {Darling}, {Davidson}, {Davies},
  {Dawson}, {De Lee}, {Diamond-Stanic}, {Cano-D{\'\i}az}, {S{\'a}nchez},
  {Donor}, {Duckworth}, {Dwelly}, {Eisenstein}, {Elsworth}, {Emsellem},
  {Eracleous}, {Escoffier}, {Fan}, {Farr}, {Feng}, {Fern{\'a}ndez-Trincado},
  {Feuillet}, {Filipp}, {Fillingham}, {Frinchaboy}, {Fromenteau}, {Galbany},
  {Garc{\'\i}a}, {Garc{\'\i}a-Hern{\'a}ndez}, {Ge}, {Geisler}, {Gelfand},
  {G{\'e}ron}, {Gibson}, {Goddy}, {Godoy-Rivera}, {Grabowski}, {Green},
  {Greener}, {Grier}, {Griffith}, {Guo}, {Guy}, {Hadjara}, {Harding},
  {Hasselquist}, {Hayes}, {Hearty}, {Hern{\'a}ndez}, {Hill}, {Hogg},
  {Holtzman}, {Horta}, {Hsieh}, {Hsu}, {Hsu}, {Huber}, {Huertas-Company},
  {Hutchinson}, {Hwang}, {Ibarra-Medel}, {Chitham}, {Ilha}, {Imig}, {Jaekle},
  {Jayasinghe}, {Ji}, {Johnson}, {Jones}, {J{\"o}nsson}, {Katkov}, {Khalatyan},
  {Kinemuchi}, {Kisku}, {Knapen}, {Kneib}, {Kollmeier}, {Kong}, {Kounkel},
  {Kreckel}, {Krishnarao}, {Lacerna}, {Lane}, {Langgin}, {Lavender}, {Law},
  {Lazarz}, {Leung}, {Leung}, {Lewis}, {Li}, {Li}, {Lian}, {Liang}, {Lin},
  {Lin}, {Lin}, {Lintott}, {Long}, {Longa-Pe{\~n}a}, {L{\'o}pez-Cob{\'a}},
  {Lu}, {Lundgren}, {Luo}, {Mackereth}, {de la Macorra}, {Mahadevan},
  {Majewski}, {Manchado}, {Mandeville}, {Maraston}, {Margalef-Bentabol},
  {Masseron}, {Masters}, {Mathur}, {McDermid}, {Mckay}, {Merloni},
  {Merrifield}, {Meszaros}, {Miglio}, {Di Mille}, {Minniti}, {Minsley},
  {Monachesi}, {Moon}, {Mosser}, {Mulchaey}, {Muna}, {Mu{\~n}oz}, {Myers},
  {Myers}, {Nadathur}, {Nair}, {Nandra}, {Neumann}, {Newman}, {Nidever},
  {Nikakhtar}, {Nitschelm}, {O'Connell}, {Garma-Oehmichen}, {Luan Souza de
  Oliveira}, {Olney}, {Oravetz}, {Ortigoza-Urdaneta}, {Osorio}, {Otter},
  {Pace}, {Padilla}, {Pan}, {Pan}, {Parikh}, {Parker}, {Peirani}, {Pe{\~n}a
  Ram{\'\i}rez}, {Penny}, {Percival}, {Perez-Fournon}, {Pinsonneault},
  {Poidevin}, {Poovelil}, {Price-Whelan}, {B{\'a}rbara de Andrade Queiroz},
  {Raddick}, {Ray}, {Rembold}, {Riddle}, {Riffel}, {Riffel}, {Rix}, {Robin},
  {Rodr{\'\i}guez-Puebla}, {Roman-Lopes}, {Rom{\'a}n-Z{\'u}{\~n}iga}, {Rose},
  {Ross}, {Rossi}, {Rubin}, {Salvato}, {S{\'a}nchez}, {S{\'a}nchez-Gallego},
  {Sanderson}, {Santana Rojas}, {Sarceno}, {Sarmiento}, {Sayres}, {Sazonova},
  {Schaefer}, {Schiavon}, {Schlegel}, {Schneider}, {Schultheis}, {Schwope},
  {Serenelli}, {Serna}, {Shao}, {Shapiro}, {Sharma}, {Shen}, {Shetrone}, {Shu},
  {Simon}, {Skrutskie}, {Smethurst}, {Smith}, {Sobeck}, {Spoo}, {Sprague},
  {Stark}, {Stassun}, {Steinmetz}, {Stello}, {Stone-Martinez},
  {Storchi-Bergmann}, {Stringfellow}, {Stutz}, {Su}, {Taghizadeh-Popp},
  {Talbot}, {Tayar}, {Telles}, {Teske}, {Thakar}, {Theissen}, {Tkachenko},
  {Thomas}, {Tojeiro}, {Hernandez Toledo}, {Troup}, {Trump}, {Trussler},
  {Turner}, {Tuttle}, {Unda-Sanzana}, {V{\'a}zquez-Mata}, {Valentini},
  {Valenzuela}, {Vargas-Gonz{\'a}lez}, {Vargas-Maga{\~n}a}, {Alfaro},
  {Villanova}, {Vincenzo}, {Wake}, {Warfield}, {Washington}, {Weaver},
  {Weijmans}, {Weinberg}, {Weiss}, {Westfall}, {Wild}, {Wilde}, {Wilson},
  {Wilson}, {Wilson}, {Wolf}, {Wood-Vasey}, {Yan}, {Zamora}, {Zasowski},
  {Zhang}, {Zhao}, {Zheng}, {Zheng}, \& {Zhu}}]{2022ApJS..259...35A}
{Abdurro'uf}, {Accetta}, K., {Aerts}, C., {et~al.} 2022, \apjs, 259, 35,
  \dodoi{10.3847/1538-4365/ac4414}

\bibitem[{{Adams} {et~al.}(2014){Adams}, {Simon}, {Fabricius}, {van den Bosch},
  {Barentine}, {Bender}, {Gebhardt}, {Hill}, {Murphy}, {Swaters}, {Thomas}, \&
  {van de Ven}}]{2014ApJ...789...63A}
{Adams}, J.~J., {Simon}, J.~D., {Fabricius}, M.~H., {et~al.} 2014, \apj, 789,
  63, \dodoi{10.1088/0004-637X/789/1/63}

\bibitem[{{Arora} {et~al.}(2023){Arora}, {Courteau}, {Stone}, \&
  {Macci{\`o}}}]{MaNGA-II}
{Arora}, N., {Courteau}, S., {Stone}, C., \& {Macci{\`o}}, A.~V. 2023, \mnras,
  522, 1208, \dodoi{10.1093/mnras/stad1023}

\bibitem[{{Arora} {et~al.}(2019){Arora}, {Fossati}, {Fontanot}, {Hirschmann},
  \& {Wilman}}]{Arora+2019}
{Arora}, N., {Fossati}, M., {Fontanot}, F., {Hirschmann}, M., \& {Wilman},
  D.~J. 2019, \mnras, 489, 1606, \dodoi{10.1093/mnras/stz2266}

\bibitem[{{Arora} {et~al.}(2021){Arora}, {Stone}, {Courteau}, \&
  {Jarrett}}]{MaNGA-I}
{Arora}, N., {Stone}, C., {Courteau}, S., \& {Jarrett}, T.~H. 2021, \mnras,
  505, 3135, \dodoi{10.1093/mnras/stab1430}

\bibitem[{{Begum} {et~al.}(2003){Begum}, {Chengalur}, \&
  {Hopp}}]{2003NewA....8..267B}
{Begum}, A., {Chengalur}, J.~N., \& {Hopp}, U. 2003, \na, 8, 267,
  \dodoi{10.1016/S1384-1076(02)00238-5}

\bibitem[{Benítez-Llambay {et~al.}(2019)Benítez-Llambay, Frenk, Ludlow, \&
  Navarro}]{Benítez-Llambay+2019}
Benítez-Llambay, A., Frenk, C.~S., Ludlow, A.~D., \& Navarro, J.~F. 2019,
  Monthly Notices of the Royal Astronomical Society, 488, 2387,
  \dodoi{10.1093/mnras/stz1890}

\bibitem[{{Blanton} \& {Roweis}(2007)}]{BlantonRoweis2007}
{Blanton}, M.~R., \& {Roweis}, S. 2007, \aj, 133, 734, \dodoi{10.1086/510127}

\bibitem[{{Blumenthal} {et~al.}(1986){Blumenthal}, {Faber}, {Flores}, \&
  {Primack}}]{Blumenthal+1986}
{Blumenthal}, G.~R., {Faber}, S.~M., {Flores}, R., \& {Primack}, J.~R. 1986,
  \apj, 301, 27, \dodoi{10.1086/163867}

\bibitem[{{Broeils}(1992)}]{broeils92}
{Broeils}, A.~H. 1992, PhD thesis, University of Groningen, Netherlands

\bibitem[{{Bruzual} \& {Charlot}(2003)}]{2003MNRAS.344.1000B}
{Bruzual}, G., \& {Charlot}, S. 2003, \mnras, 344, 1000,
  \dodoi{10.1046/j.1365-8711.2003.06897.x}

\bibitem[{Bundy {et~al.}(2014)Bundy, Bershady, Law, Yan, Drory, MacDonald,
  Wake, Cherinka, Sánchez-Gallego, Weijmans, Thomas, Tremonti, Masters,
  Coccato, Diamond-Stanic, Aragón-Salamanca, Avila-Reese, Badenes,
  Falcón-Barroso, Belfiore, Bizyaev, Blanc, Bland-Hawthorn, Blanton,
  Brownstein, Byler, Cappellari, Conroy, Dutton, Emsellem, Etherington,
  Frinchaboy, Fu, Gunn, Harding, Johnston, Kauffmann, Kinemuchi, Klaene,
  Knapen, Leauthaud, Li, Lin, Maiolino, Malanushenko, Malanushenko, Mao,
  Maraston, McDermid, Merrifield, Nichol, Oravetz, Pan, Parejko, Sanchez,
  Schlegel, Simmons, Steele, Steinmetz, Thanjavur, Thompson, Tinker, van~den
  Bosch, Westfall, Wilkinson, Wright, Xiao, \& Zhang}]{Bundy+2015}
Bundy, K., Bershady, M.~A., Law, D.~R., {et~al.} 2014, The Astrophysical
  Journal, 798, 7, \dodoi{10.1088/0004-637X/798/1/7}

\bibitem[{{Burkert}(1995)}]{Burkert1995}
{Burkert}, A. 1995, \apjl, 447, L25, \dodoi{10.1086/309560}

\bibitem[{{Chan} {et~al.}(2015){Chan}, {Kere{\v{s}}}, {O{\~n}orbe}, {Hopkins},
  {Muratov}, {Faucher-Gigu{\`e}re}, \& {Quataert}}]{Chan+2015}
{Chan}, T.~K., {Kere{\v{s}}}, D., {O{\~n}orbe}, J., {et~al.} 2015, \mnras, 454,
  2981, \dodoi{10.1093/mnras/stv2165}

\bibitem[{{Chilingarian} \& {Zolotukhin}(2012)}]{ChilingarianZolotukhin2012}
{Chilingarian}, I.~V., \& {Zolotukhin}, I.~Y. 2012, \mnras, 419, 1727,
  \dodoi{10.1111/j.1365-2966.2011.19837.x}

\bibitem[{Chollet {et~al.}(2015)}]{Keras}
Chollet, F., {et~al.} 2015, Keras, \url{https://keras.io}

\bibitem[{{Ciotti} \& {Bertin}(1999)}]{CiottiBertin1999}
{Ciotti}, L., \& {Bertin}, G. 1999, \aap, 352, 447,
  \dodoi{10.48550/arXiv.astro-ph/9911078}

\bibitem[{{Cluver} {et~al.}(2014){Cluver}, {Jarrett}, {Hopkins}, {Driver},
  {Liske}, {Gunawardhana}, {Taylor}, {Robotham}, {Alpaslan}, {Baldry}, {Brown},
  {Peacock}, {Popescu}, {Tuffs}, {Bauer}, {Bland-Hawthorn}, {Colless},
  {Holwerda}, {Lara-L{\'o}pez}, {Leschinski}, {L{\'o}pez-S{\'a}nchez},
  {Norberg}, {Owers}, {Wang}, \& {Wilkins}}]{2014ApJ...782...90C}
{Cluver}, M.~E., {Jarrett}, T.~H., {Hopkins}, A.~M., {et~al.} 2014, \apj, 782,
  90, \dodoi{10.1088/0004-637X/782/2/90}

\bibitem[{{Conroy}(2013)}]{Conroy2013}
{Conroy}, C. 2013, \araa, 51, 393, \dodoi{10.1146/annurev-astro-082812-141017}

\bibitem[{{Conroy} {et~al.}(2009){Conroy}, {Gunn}, \&
  {White}}]{2009ApJ...699..486C}
{Conroy}, C., {Gunn}, J.~E., \& {White}, M. 2009, \apj, 699, 486,
  \dodoi{10.1088/0004-637X/699/1/486}

\bibitem[{{Courteau}(1996)}]{Courteau1996}
{Courteau}, S. 1996, \apjs, 103, 363, \dodoi{10.1086/192281}

\bibitem[{{Courteau}(1997)}]{C97}
---. 1997, \aj, 114, 2402, \dodoi{10.1086/118656}

\bibitem[{{Courteau} {et~al.}(1996){Courteau}, {de Jong}, \&
  {Broeils}}]{CourteauBroeils1996}
{Courteau}, S., {de Jong}, R.~S., \& {Broeils}, A.~H. 1996, \apjl, 457, L73,
  \dodoi{10.1086/309906}

\bibitem[{{Courteau} {et~al.}(2000){Courteau}, {Willick}, {Strauss},
  {Schlegel}, \& {Postman}}]{2000ApJ...544..636C}
{Courteau}, S., {Willick}, J.~A., {Strauss}, M.~A., {Schlegel}, D., \&
  {Postman}, M. 2000, \apj, 544, 636, \dodoi{10.1086/317234}

\bibitem[{{Courteau} {et~al.}(2014){Courteau}, {Cappellari}, {de Jong},
  {Dutton}, {Emsellem}, {Hoekstra}, {Koopmans}, {Mamon}, {Maraston}, {Treu}, \&
  {Widrow}}]{Courteau2014}
{Courteau}, S., {Cappellari}, M., {de Jong}, R.~S., {et~al.} 2014, Reviews of
  Modern Physics, 86, 47, \dodoi{10.1103/RevModPhys.86.47}

\bibitem[{Cybenko(1989)}]{Cybenko1989}
Cybenko, G. 1989, Mathematics of Control, Signals, and Systems (MCSS), 2, 303,
  \dodoi{10.1007/BF02551274}

\bibitem[{{Dalcanton} {et~al.}(2004){Dalcanton}, {Yoachim}, \&
  {Bernstein}}]{2004ApJ...608..189D}
{Dalcanton}, J.~J., {Yoachim}, P., \& {Bernstein}, R.~A. 2004, \apj, 608, 189,
  \dodoi{10.1086/386358}

\bibitem[{{Dale} {et~al.}(1999){Dale}, {Giovanelli}, {Haynes}, {Campusano}, \&
  {Hardy}}]{1999AJ....118.1489D}
{Dale}, D.~A., {Giovanelli}, R., {Haynes}, M.~P., {Campusano}, L.~E., \&
  {Hardy}, E. 1999, \aj, 118, 1489, \dodoi{10.1086/301048}

\bibitem[{{de Blok} \& {Bosma}(2002)}]{db2002}
{de Blok}, W.~J.~G., \& {Bosma}, A. 2002, \aap, 385, 816,
  \dodoi{10.1051/0004-6361:20020080}

\bibitem[{{de Blok} {et~al.}(2008){de Blok}, {Walter}, {Brinks},
  {Trachternach}, {Oh}, \& {Kennicutt}}]{2008AJ....136.2648D}
{de Blok}, W.~J.~G., {Walter}, F., {Brinks}, E., {et~al.} 2008, \aj, 136, 2648,
  \dodoi{10.1088/0004-6256/136/6/2648}

\bibitem[{{Del Popolo}(2009)}]{DelPopolo2009}
{Del Popolo}, A. 2009, \apj, 698, 2093, \dodoi{10.1088/0004-637X/698/2/2093}

\bibitem[{{Del Popolo} \& {Hiotelis}(2014)}]{DelPopolo+2014}
{Del Popolo}, A., \& {Hiotelis}, N. 2014, \jcap, 2014, 047,
  \dodoi{10.1088/1475-7516/2014/01/047}

\bibitem[{{Dey} {et~al.}(2019){Dey}, {Schlegel}, {Lang}, {Blum}, {Burleigh},
  {Fan}, {Findlay}, {Finkbeiner}, {Herrera}, {Juneau}, {Landriau}, {Levi},
  {McGreer}, {Meisner}, {Myers}, {Moustakas}, {Nugent}, {Patej}, {Schlafly},
  {Walker}, {Valdes}, {Weaver}, {Y{\`e}che}, {Zou}, {Zhou}, {Abareshi},
  {Abbott}, {Abolfathi}, {Aguilera}, {Alam}, {Allen}, {Alvarez}, {Annis},
  {Ansarinejad}, {Aubert}, {Beechert}, {Bell}, {BenZvi}, {Beutler}, {Bielby},
  {Bolton}, {Brice{\~n}o}, {Buckley-Geer}, {Butler}, {Calamida}, {Carlberg},
  {Carter}, {Casas}, {Castander}, {Choi}, {Comparat}, {Cukanovaite}, {Delubac},
  {DeVries}, {Dey}, {Dhungana}, {Dickinson}, {Ding}, {Donaldson}, {Duan},
  {Duckworth}, {Eftekharzadeh}, {Eisenstein}, {Etourneau}, {Fagrelius},
  {Farihi}, {Fitzpatrick}, {Font-Ribera}, {Fulmer}, {G{\"a}nsicke},
  {Gaztanaga}, {George}, {Gerdes}, {Gontcho}, {Gorgoni}, {Green}, {Guy},
  {Harmer}, {Hernandez}, {Honscheid}, {Huang}, {James}, {Jannuzi}, {Jiang},
  {Joyce}, {Karcher}, {Karkar}, {Kehoe}, {Kneib}, {Kueter-Young}, {Lan},
  {Lauer}, {Le Guillou}, {Le Van Suu}, {Lee}, {Lesser}, {Perreault Levasseur},
  {Li}, {Mann}, {Marshall}, {Mart{\'\i}nez-V{\'a}zquez}, {Martini}, {du Mas des
  Bourboux}, {McManus}, {Meier}, {M{\'e}nard}, {Metcalfe},
  {Mu{\~n}oz-Guti{\'e}rrez}, {Najita}, {Napier}, {Narayan}, {Newman}, {Nie},
  {Nord}, {Norman}, {Olsen}, {Paat}, {Palanque-Delabrouille}, {Peng},
  {Poppett}, {Poremba}, {Prakash}, {Rabinowitz}, {Raichoor}, {Rezaie},
  {Robertson}, {Roe}, {Ross}, {Ross}, {Rudnick}, {Safonova}, {Saha},
  {S{\'a}nchez}, {Savary}, {Schweiker}, {Scott}, {Seo}, {Shan}, {Silva},
  {Slepian}, {Soto}, {Sprayberry}, {Staten}, {Stillman}, {Stupak}, {Summers},
  {Sien Tie}, {Tirado}, {Vargas-Maga{\~n}a}, {Vivas}, {Wechsler}, {Williams},
  {Yang}, {Yang}, {Yapici}, {Zaritsky}, {Zenteno}, {Zhang}, {Zhang}, {Zhou}, \&
  {Zhou}}]{DESI}
{Dey}, A., {Schlegel}, D.~J., {Lang}, D., {et~al.} 2019, \aj, 157, 168,
  \dodoi{10.3847/1538-3881/ab089d}

\bibitem[{{Di Paolo} {et~al.}(2019){Di Paolo}, {Salucci}, \&
  {Erkurt}}]{Paolo+2019}
{Di Paolo}, C., {Salucci}, P., \& {Erkurt}, A. 2019, \mnras, 490, 5451,
  \dodoi{10.1093/mnras/stz2700}

\bibitem[{{Dom{\'\i}nguez S{\'a}nchez} {et~al.}(2022){Dom{\'\i}nguez
  S{\'a}nchez}, {Margalef}, {Bernardi}, \& {Huertas-Company}}]{PyMorph-DR17}
{Dom{\'\i}nguez S{\'a}nchez}, H., {Margalef}, B., {Bernardi}, M., \&
  {Huertas-Company}, M. 2022, \mnras, 509, 4024, \dodoi{10.1093/mnras/stab3089}

\bibitem[{Dozat(2016)}]{Nadam}
Dozat, T. 2016, in Proceedings of the 4th International Conference on Learning
  Representations, 1--4

\bibitem[{{Drory} {et~al.}(2015){Drory}, {MacDonald}, {Bershady}, {Bundy},
  {Gunn}, {Law}, {Smith}, {Stoll}, {Tremonti}, {Wake}, {Yan}, {Weijmans},
  {Byler}, {Cherinka}, {Cope}, {Eigenbrot}, {Harding}, {Holder}, {Huehnerhoff},
  {Jaehnig}, {Jansen}, {Klaene}, {Paat}, {Percival}, \&
  {Sayres}}]{2015AJ....149...77D}
{Drory}, N., {MacDonald}, N., {Bershady}, M.~A., {et~al.} 2015, \aj, 149, 77,
  \dodoi{10.1088/0004-6256/149/2/77}

\bibitem[{{Dutton} {et~al.}(2020){Dutton}, {Buck}, {Macci{\`o}}, {Dixon},
  {Blank}, \& {Obreja}}]{Dutton+2020}
{Dutton}, A.~A., {Buck}, T., {Macci{\`o}}, A.~V., {et~al.} 2020, \mnras, 499,
  2648, \dodoi{10.1093/mnras/staa3028}

\bibitem[{{Dutton} {et~al.}(2007){Dutton}, {van den Bosch}, {Dekel}, \&
  {Courteau}}]{Dutton+2007}
{Dutton}, A.~A., {van den Bosch}, F.~C., {Dekel}, A., \& {Courteau}, S. 2007,
  \apj, 654, 27, \dodoi{10.1086/509314}

\bibitem[{{Epinat} {et~al.}(2008){Epinat}, {Amram}, \&
  {Marcelin}}]{2008MNRAS.390..466E}
{Epinat}, B., {Amram}, P., \& {Marcelin}, M. 2008, \mnras, 390, 466,
  \dodoi{10.1111/j.1365-2966.2008.13796.x}

\bibitem[{Fathi {et~al.}(2010)Fathi, Allen, Boch, Hatziminaoglou, \&
  Peletier}]{Fathi+2010}
Fathi, K., Allen, M., Boch, T., Hatziminaoglou, E., \& Peletier, R.~F. 2010,
  Monthly Notices of the Royal Astronomical Society, 406, 1595,
  \dodoi{10.1111/j.1365-2966.2010.16812.x}

\bibitem[{{Fischer} {et~al.}(2019){Fischer}, {Dom{\'\i}nguez S{\'a}nchez}, \&
  {Bernardi}}]{PyMorph}
{Fischer}, J.~L., {Dom{\'\i}nguez S{\'a}nchez}, H., \& {Bernardi}, M. 2019,
  \mnras, 483, 2057, \dodoi{10.1093/mnras/sty3135}

\bibitem[{{Freeman}(1970)}]{Freeman1970}
{Freeman}, K.~C. 1970, \apj, 160, 811, \dodoi{10.1086/150474}

\bibitem[{{Frosst} {et~al.}(2022){Frosst}, {Courteau}, {Arora}, {Stone},
  {Macci{\`o}}, \& {Blank}}]{frosst+2022}
{Frosst}, M., {Courteau}, S., {Arora}, N., {et~al.} 2022, \mnras, 514, 3510,
  \dodoi{10.1093/mnras/stac1497}

\bibitem[{{Gadotti}(2008)}]{2008MNRAS.384..420G}
{Gadotti}, D.~A. 2008, \mnras, 384, 420,
  \dodoi{10.1111/j.1365-2966.2007.12723.x}

\bibitem[{Gao \& Ho(2017)}]{GaoHo2017}
Gao, H., \& Ho, L.~C. 2017, The Astrophysical Journal, 845, 114,
  \dodoi{10.3847/1538-4357/aa7da4}

\bibitem[{{Garc{\'\i}a-Benito} {et~al.}(2019){Garc{\'\i}a-Benito},
  {Gonz{\'a}lez Delgado}, {P{\'e}rez}, {Cid Fernandes}, {S{\'a}nchez}, \& {de
  Amorim}}]{garciabenito+2019}
{Garc{\'\i}a-Benito}, R., {Gonz{\'a}lez Delgado}, R.~M., {P{\'e}rez}, E.,
  {et~al.} 2019, \aap, 621, A120, \dodoi{10.1051/0004-6361/201833993}

\bibitem[{{Garrido} {et~al.}(2005){Garrido}, {Marcelin}, {Amram}, {Balkowski},
  {Gach}, \& {Boulesteix}}]{2005MNRAS.362..127G}
{Garrido}, O., {Marcelin}, M., {Amram}, P., {et~al.} 2005, \mnras, 362, 127,
  \dodoi{10.1111/j.1365-2966.2005.09274.x}

\bibitem[{{Gilhuly} \& {Courteau}(2018)}]{Gilhuly2018}
{Gilhuly}, C., \& {Courteau}, S. 2018, \mnras, 477, 845,
  \dodoi{10.1093/mnras/sty756}

\bibitem[{{Gonz{\'a}lez Delgado} {et~al.}(2016){Gonz{\'a}lez Delgado}, {Cid
  Fernandes}, {P{\'e}rez}, {Garc{\'\i}a-Benito}, {L{\'o}pez Fern{\'a}ndez},
  {Lacerda}, {Cortijo-Ferrero}, {de Amorim}, {Vale Asari}, {S{\'a}nchez},
  {Walcher}, {Wisotzki}, {Mast}, {Alves}, {Ascasibar}, {Bland-Hawthorn},
  {Galbany}, {Kennicutt}, {M{\'a}rquez}, {Masegosa}, {Moll{\'a}},
  {S{\'a}nchez-Bl{\'a}zquez}, \& {V{\'\i}lchez}}]{Gonzalez+2016}
{Gonz{\'a}lez Delgado}, R.~M., {Cid Fernandes}, R., {P{\'e}rez}, E., {et~al.}
  2016, \aap, 590, A44, \dodoi{10.1051/0004-6361/201628174}

\bibitem[{{Governato} {et~al.}(2010){Governato}, {Brook}, {Mayer}, {Brooks},
  {Rhee}, {Wadsley}, {Jonsson}, {Willman}, {Stinson}, {Quinn}, \&
  {Madau}}]{Governato+2010}
{Governato}, F., {Brook}, C., {Mayer}, L., {et~al.} 2010, \nat, 463, 203,
  \dodoi{10.1038/nature08640}

\bibitem[{{Grand} {et~al.}(2016){Grand}, {Springel}, {G{\'o}mez}, {Marinacci},
  {Pakmor}, {Campbell}, \& {Jenkins}}]{Grand+2016}
{Grand}, R. J.~J., {Springel}, V., {G{\'o}mez}, F.~A., {et~al.} 2016, \mnras,
  459, 199, \dodoi{10.1093/mnras/stw601}

\bibitem[{{Gunn} {et~al.}(2006){Gunn}, {Siegmund}, {Mannery}, {Owen}, {Hull},
  {Leger}, {Carey}, {Knapp}, {York}, {Boroski}, {Kent}, {Lupton}, {Rockosi},
  {Evans}, {Waddell}, {Anderson}, {Annis}, {Barentine}, {Bartoszek}, {Bastian},
  {Bracker}, {Brewington}, {Briegel}, {Brinkmann}, {Brown}, {Carr},
  {Czarapata}, {Drennan}, {Dombeck}, {Federwitz}, {Gillespie}, {Gonzales},
  {Hansen}, {Harvanek}, {Hayes}, {Jordan}, {Kinney}, {Klaene}, {Kleinman},
  {Kron}, {Kresinski}, {Lee}, {Limmongkol}, {Lindenmeyer}, {Long}, {Loomis},
  {McGehee}, {Mantsch}, {Neilsen}, {Neswold}, {Newman}, {Nitta}, {Peoples},
  {Pier}, {Prieto}, {Prosapio}, {Rivetta}, {Schneider}, {Snedden}, \&
  {Wang}}]{2006AJ....131.2332G}
{Gunn}, J.~E., {Siegmund}, W.~A., {Mannery}, E.~J., {et~al.} 2006, \aj, 131,
  2332, \dodoi{10.1086/500975}

\bibitem[{{Hall} {et~al.}(2012){Hall}, {Courteau}, {Dutton}, {McDonald}, \&
  {Zhu}}]{Hall2012}
{Hall}, M., {Courteau}, S., {Dutton}, A.~A., {McDonald}, M., \& {Zhu}, Y. 2012,
  \mnras, 425, 2741, \dodoi{10.1111/j.1365-2966.2012.21290.x}

\bibitem[{Hornik {et~al.}(1989)Hornik, Stinchcombe, \& White}]{Hornik+1989}
Hornik, K., Stinchcombe, M., \& White, H. 1989, Neural Networks, 2, 359,
  \dodoi{https://doi.org/10.1016/0893-6080(89)90020-8}

\bibitem[{{Karukes} \& {Salucci}(2017)}]{Karukes+Salucci2017}
{Karukes}, E.~V., \& {Salucci}, P. 2017, \mnras, 465, 4703,
  \dodoi{10.1093/mnras/stw3055}

\bibitem[{{Kauffmann} {et~al.}(2015){Kauffmann}, {Huang}, {Moran}, \&
  {Heckman}}]{kauff15}
{Kauffmann}, G., {Huang}, M.-L., {Moran}, S., \& {Heckman}, T.~M. 2015, \mnras,
  451, 878, \dodoi{10.1093/mnras/stv1014}

\bibitem[{{Kirby} {et~al.}(2012){Kirby}, {Koribalski}, {Jerjen}, \&
  {L{\'o}pez-S{\'a}nchez}}]{2012MNRAS.420.2924K}
{Kirby}, E.~M., {Koribalski}, B., {Jerjen}, H., \& {L{\'o}pez-S{\'a}nchez},
  {\'A}. 2012, \mnras, 420, 2924, \dodoi{10.1111/j.1365-2966.2011.20103.x}

\bibitem[{{Kuzio de Naray} {et~al.}(2008){Kuzio de Naray}, {McGaugh}, \& {de
  Blok}}]{2008ApJ...676..920K}
{Kuzio de Naray}, R., {McGaugh}, S.~S., \& {de Blok}, W.~J.~G. 2008, \apj, 676,
  920, \dodoi{10.1086/527543}

\bibitem[{{Lang}(2014)}]{unWISE}
{Lang}, D. 2014, \aj, 147, 108, \dodoi{10.1088/0004-6256/147/5/108}

\bibitem[{{Lange} {et~al.}(2016){Lange}, {Moffett}, {Driver}, {Robotham},
  {Lagos}, {Kelvin}, {Conselice}, {Margalef-Bentabol}, {Alpaslan}, {Baldry},
  {Bland-Hawthorn}, {Bremer}, {Brough}, {Cluver}, {Colless}, {Davies},
  {H{\"a}u{\ss}ler}, {Holwerda}, {Hopkins}, {Kafle}, {Kennedy}, {Liske},
  {Phillipps}, {Popescu}, {Taylor}, {Tuffs}, {van Kampen}, \&
  {Wright}}]{Lange2016}
{Lange}, R., {Moffett}, A.~J., {Driver}, S.~P., {et~al.} 2016, \mnras, 462,
  1470, \dodoi{10.1093/mnras/stw1495}

\bibitem[{{Laurikainen} {et~al.}(2005){Laurikainen}, {Salo}, \&
  {Buta}}]{Laurikainen+2005}
{Laurikainen}, E., {Salo}, H., \& {Buta}, R. 2005, \mnras, 362, 1319,
  \dodoi{10.1111/j.1365-2966.2005.09404.x}

\bibitem[{{Lelli} {et~al.}(2016){Lelli}, {McGaugh}, \&
  {Schombert}}]{Lelli+2016}
{Lelli}, F., {McGaugh}, S.~S., \& {Schombert}, J.~M. 2016, \aj, 152, 157,
  \dodoi{10.3847/0004-6256/152/6/157}

\bibitem[{{Li} {et~al.}(2022){Li}, {McGaugh}, {Lelli}, {Schombert}, \&
  {Pawlowski}}]{Li+2022}
{Li}, P., {McGaugh}, S.~S., {Lelli}, F., {Schombert}, J.~M., \& {Pawlowski},
  M.~S. 2022, \aap, 665, A143, \dodoi{10.1051/0004-6361/202243916}

\bibitem[{Lu {et~al.}(2017)Lu, Pu, Wang, Hu, \& Wang}]{Lu+2017}
Lu, Z., Pu, H., Wang, F., Hu, Z., \& Wang, L. 2017, in Proceedings of the 31st
  International Conference on Neural Information Processing Systems, NIPS'17
  (Red Hook, NY, USA: Curran Associates Inc.), 6232–6240

\bibitem[{{MacArthur} {et~al.}(2003){MacArthur}, {Courteau}, \&
  {Holtzman}}]{MacArthur2003}
{MacArthur}, L.~A., {Courteau}, S., \& {Holtzman}, J.~A. 2003, \apj, 582, 689,
  \dodoi{10.1086/344506}

\bibitem[{Macciò {et~al.}(2011)Macciò, Stinson, Brook, Wadsley, Couchman,
  Shen, Gibson, \& Quinn}]{Macciò+2012}
Macciò, A.~V., Stinson, G., Brook, C.~B., {et~al.} 2011, The Astrophysical
  Journal Letters, 744, L9, \dodoi{10.1088/2041-8205/744/1/L9}

\bibitem[{{Martinsson} {et~al.}(2013){Martinsson}, {Verheijen}, {Westfall},
  {Bershady}, {Schechtman-Rook}, {Andersen}, \& {Swaters}}]{martinsson2013}
{Martinsson}, T. P.~K., {Verheijen}, M. A.~W., {Westfall}, K.~B., {et~al.}
  2013, \aap, 557, A130, \dodoi{10.1051/0004-6361/201220515}

\bibitem[{{Mathewson} \& {Ford}(1996)}]{MathewsonFord1996}
{Mathewson}, D.~S., \& {Ford}, V.~L. 1996, \apjs, 107, 97,
  \dodoi{10.1086/192356}

\bibitem[{{Mathewson} {et~al.}(1992){Mathewson}, {Ford}, \&
  {Buchhorn}}]{Mathewson+1992}
{Mathewson}, D.~S., {Ford}, V.~L., \& {Buchhorn}, M. 1992, \apjs, 81, 413,
  \dodoi{10.1086/191700}

\bibitem[{{McGaugh} {et~al.}(2001){McGaugh}, {Rubin}, \& {de
  Blok}}]{2001AJ....122.2381M}
{McGaugh}, S.~S., {Rubin}, V.~C., \& {de Blok}, W.~J.~G. 2001, \aj, 122, 2381,
  \dodoi{10.1086/323448}

\bibitem[{{McGaugh} \& {Schombert}(2014)}]{McGaugh+Schombert2014}
{McGaugh}, S.~S., \& {Schombert}, J.~M. 2014, \aj, 148, 77,
  \dodoi{10.1088/0004-6256/148/5/77}

\bibitem[{{M{\'e}ndez-Abreu} {et~al.}(2008){M{\'e}ndez-Abreu}, {Aguerri},
  {Corsini}, \& {Simonneau}}]{2008A&A...478..353M}
{M{\'e}ndez-Abreu}, J., {Aguerri}, J.~A.~L., {Corsini}, E.~M., \& {Simonneau},
  E. 2008, \aap, 478, 353, \dodoi{10.1051/0004-6361:20078089}

\bibitem[{{M{\'e}ndez-Abreu} {et~al.}(2017){M{\'e}ndez-Abreu}, {Ruiz-Lara},
  {S{\'a}nchez-Menguiano}, {de Lorenzo-C{\'a}ceres}, {Costantin},
  {Catal{\'a}n-Torrecilla}, {Florido}, {Aguerri}, {Bland-Hawthorn}, {Corsini},
  {Dettmar}, {Galbany}, {Garc{\'\i}a-Benito}, {Marino}, {M{\'a}rquez},
  {Ortega-Minakata}, {Papaderos}, {S{\'a}nchez}, {S{\'a}nchez-Blazquez},
  {Spekkens}, {van de Ven}, {Wild}, \& {Ziegler}}]{CALIFA_BD_decomp}
{M{\'e}ndez-Abreu}, J., {Ruiz-Lara}, T., {S{\'a}nchez-Menguiano}, L., {et~al.}
  2017, \aap, 598, A32, \dodoi{10.1051/0004-6361/201629525}

\bibitem[{{Noordermeer} {et~al.}(2007){Noordermeer}, {van der Hulst},
  {Sancisi}, {Swaters}, \& {van Albada}}]{2007MNRAS.376.1513N}
{Noordermeer}, E., {van der Hulst}, J.~M., {Sancisi}, R., {Swaters}, R.~S., \&
  {van Albada}, T.~S. 2007, \mnras, 376, 1513,
  \dodoi{10.1111/j.1365-2966.2007.11533.x}

\bibitem[{{Oh} {et~al.}(2015{\natexlab{a}}){Oh}, {Hunter}, {Brinks},
  {Elmegreen}, {Schruba}, {Walter}, {Rupen}, {Young}, {Simpson}, {Johnson},
  {Herrmann}, {Ficut-Vicas}, {Cigan}, {Heesen}, {Ashley}, \& {Zhang}}]{oh11}
{Oh}, S.-H., {Hunter}, D.~A., {Brinks}, E., {et~al.} 2015{\natexlab{a}}, \aj,
  149, 180, \dodoi{10.1088/0004-6256/149/6/180}

\bibitem[{{Oh} {et~al.}(2015{\natexlab{b}}){Oh}, {Hunter}, {Brinks},
  {Elmegreen}, {Schruba}, {Walter}, {Rupen}, {Young}, {Simpson}, {Johnson},
  {Herrmann}, {Ficut-Vicas}, {Cigan}, {Heesen}, {Ashley}, \&
  {Zhang}}]{2015AJ....149..180O}
---. 2015{\natexlab{b}}, \aj, 149, 180, \dodoi{10.1088/0004-6256/149/6/180}

\bibitem[{{Ouellette} {et~al.}(2017){Ouellette}, {Courteau}, {Holtzman},
  {Dutton}, {Cappellari}, {Dalcanton}, {McDonald}, {Roediger}, {Taylor},
  {Tully}, {C{\^o}t{\'e}}, {Ferrarese}, \& {Peng}}]{2017ApJ...843...74O}
{Ouellette}, N. N.~Q., {Courteau}, S., {Holtzman}, J.~A., {et~al.} 2017, \apj,
  843, 74, \dodoi{10.3847/1538-4357/aa74b1}

\bibitem[{{Peng} {et~al.}(2010){Peng}, {Ho}, {Impey}, \& {Rix}}]{galfitpackage}
{Peng}, C.~Y., {Ho}, L.~C., {Impey}, C.~D., \& {Rix}, H.-W. 2010, \aj, 139,
  2097, \dodoi{10.1088/0004-6256/139/6/2097}

\bibitem[{{Persic} \& {Salucci}(1991)}]{PS91}
{Persic}, M., \& {Salucci}, P. 1991, \apj, 368, 60, \dodoi{10.1086/169670}

\bibitem[{{Persic} {et~al.}(1996){Persic}, {Salucci}, \& {Stel}}]{PSS96}
{Persic}, M., {Salucci}, P., \& {Stel}, F. 1996, \mnras, 281, 27,
  \dodoi{10.1093/mnras/278.1.27}

\bibitem[{Pontzen \& Governato(2012)}]{PontzenGovernato2012}
Pontzen, A., \& Governato, F. 2012, Monthly Notices of the Royal Astronomical
  Society, 421, 3464, \dodoi{10.1111/j.1365-2966.2012.20571.x}

\bibitem[{{Reyes} {et~al.}(2011){Reyes}, {Mandelbaum}, {Gunn}, {Pizagno}, \&
  {Lackner}}]{Reyes+2011}
{Reyes}, R., {Mandelbaum}, R., {Gunn}, J.~E., {Pizagno}, J., \& {Lackner},
  C.~N. 2011, \mnras, 417, 2347, \dodoi{10.1111/j.1365-2966.2011.19415.x}

\bibitem[{{Roediger} \& {Courteau}(2015)}]{Roediger2015}
{Roediger}, J.~C., \& {Courteau}, S. 2015, \mnras, 452, 3209,
  \dodoi{10.1093/mnras/stv1499}

\bibitem[{{Rubin} {et~al.}(1985){Rubin}, {Burstein}, {Ford}, \&
  {Thonnard}}]{Rubin+1985}
{Rubin}, V.~C., {Burstein}, D., {Ford}, W.~K., J., \& {Thonnard}, N. 1985,
  \apj, 289, 81, \dodoi{10.1086/162866}

\bibitem[{{Salo} {et~al.}(2015){Salo}, {Laurikainen}, {Laine}, {Comer{\'o}n},
  {Gadotti}, {Buta}, {Sheth}, {Zaritsky}, {Ho}, {Knapen}, {Athanassoula},
  {Bosma}, {Laine}, {Cisternas}, {Kim}, {Mu{\~n}oz-Mateos}, {Regan}, {Hinz},
  {Gil de Paz}, {Menendez-Delmestre}, {Mizusawa}, {Erroz-Ferrer}, {Meidt}, \&
  {Querejeta}}]{Spitzer_BD_decomp}
{Salo}, H., {Laurikainen}, E., {Laine}, J., {et~al.} 2015, \apjs, 219, 4,
  \dodoi{10.1088/0067-0049/219/1/4}

\bibitem[{{Salucci} {et~al.}(2007){Salucci}, {Lapi}, {Tonini}, {Gentile},
  {Yegorova}, \& {Klein}}]{Salucci+2007}
{Salucci}, P., {Lapi}, A., {Tonini}, C., {et~al.} 2007, \mnras, 378, 41,
  \dodoi{10.1111/j.1365-2966.2007.11696.x}

\bibitem[{{S{\'a}nchez} {et~al.}(2012){S{\'a}nchez}, {Kennicutt}, {Gil de Paz},
  {van de Ven}, {V{\'\i}lchez}, {Wisotzki}, {Walcher}, {Mast}, {Aguerri},
  {Albiol-P{\'e}rez}, {Alonso-Herrero}, {Alves}, {Bakos}, {Bart{\'a}kov{\'a}},
  {Bland-Hawthorn}, {Boselli}, {Bomans}, {Castillo-Morales}, {Cortijo-Ferrero},
  {de Lorenzo-C{\'a}ceres}, {Del Olmo}, {Dettmar}, {D{\'\i}az}, {Ellis},
  {Falc{\'o}n-Barroso}, {Flores}, {Gallazzi}, {Garc{\'\i}a-Lorenzo},
  {Gonz{\'a}lez Delgado}, {Gruel}, {Haines}, {Hao}, {Husemann},
  {Igl{\'e}sias-P{\'a}ramo}, {Jahnke}, {Johnson}, {Jungwiert}, {Kalinova},
  {Kehrig}, {Kupko}, {L{\'o}pez-S{\'a}nchez}, {Lyubenova}, {Marino},
  {M{\'a}rmol-Queralt{\'o}}, {M{\'a}rquez}, {Masegosa}, {Meidt},
  {Mendez-Abreu}, {Monreal-Ibero}, {Montijo}, {Mour{\~a}o}, {Palacios-Navarro},
  {Papaderos}, {Pasquali}, {Peletier}, {P{\'e}rez}, {P{\'e}rez}, {Quirrenbach},
  {Rela{\~n}o}, {Rosales-Ortega}, {Roth}, {Ruiz-Lara},
  {S{\'a}nchez-Bl{\'a}zquez}, {Sengupta}, {Singh}, {Stanishev}, {Trager},
  {Vazdekis}, {Viironen}, {Wild}, {Zibetti}, \&
  {Ziegler}}]{2012A&A...538A...8S}
{S{\'a}nchez}, S.~F., {Kennicutt}, R.~C., {Gil de Paz}, A., {et~al.} 2012,
  \aap, 538, A8, \dodoi{10.1051/0004-6361/201117353}

\bibitem[{Sawala {et~al.}(2016)Sawala, Frenk, Fattahi, Navarro, Bower, Crain,
  Vecchia, Furlong, Helly, Jenkins, Oman, Schaller, Schaye, Theuns, Trayford,
  \& White}]{Sawala+2016}
Sawala, T., Frenk, C.~S., Fattahi, A., {et~al.} 2016, Monthly Notices of the
  Royal Astronomical Society, 457, 1931, \dodoi{10.1093/mnras/stw145}

\bibitem[{{Schlafly} \& {Finkbeiner}(2011)}]{SchlaflyFinkbeiner2011}
{Schlafly}, E.~F., \& {Finkbeiner}, D.~P. 2011, \apj, 737, 103,
  \dodoi{10.1088/0004-637X/737/2/103}

\bibitem[{{Schlegel} {et~al.}(1998){Schlegel}, {Finkbeiner}, \&
  {Davis}}]{Schlegel+1998}
{Schlegel}, D.~J., {Finkbeiner}, D.~P., \& {Davis}, M. 1998, \apj, 500, 525,
  \dodoi{10.1086/305772}

\bibitem[{{Sellwood} \& {Spekkens}(2015)}]{Sellwood2015}
{Sellwood}, J.~A., \& {Spekkens}, K. 2015, arXiv e-prints, arXiv:1509.07120,
  \dodoi{10.48550/arXiv.1509.07120}

\bibitem[{{S\'ersic}(1968)}]{Sersic1968}
{S\'ersic}, J.~L. 1968, {Atlas de Galaxias Australes} (Observatorio
  Astronomico)

\bibitem[{{Sheth} {et~al.}(2010){Sheth}, {Regan}, {Hinz}, {Gil de Paz},
  {Men{\'e}ndez-Delmestre}, {Mu{\~n}oz-Mateos}, {Seibert}, {Kim},
  {Laurikainen}, {Salo}, {Gadotti}, {Laine}, {Mizusawa}, {Armus},
  {Athanassoula}, {Bosma}, {Buta}, {Capak}, {Jarrett}, {Elmegreen},
  {Elmegreen}, {Knapen}, {Koda}, {Helou}, {Ho}, {Madore}, {Masters},
  {Mobasher}, {Ogle}, {Peng}, {Schinnerer}, {Surace}, {Zaritsky},
  {Comer{\'o}n}, {de Swardt}, {Meidt}, {Kasliwal}, \&
  {Aravena}}]{2010PASP..122.1397S}
{Sheth}, K., {Regan}, M., {Hinz}, J.~L., {et~al.} 2010, \pasp, 122, 1397,
  \dodoi{10.1086/657638}

\bibitem[{{Smee} {et~al.}(2013){Smee}, {Gunn}, {Uomoto}, {Roe}, {Schlegel},
  {Rockosi}, {Carr}, {Leger}, {Dawson}, {Olmstead}, {Brinkmann}, {Owen},
  {Barkhouser}, {Honscheid}, {Harding}, {Long}, {Lupton}, {Loomis}, {Anderson},
  {Annis}, {Bernardi}, {Bhardwaj}, {Bizyaev}, {Bolton}, {Brewington}, {Briggs},
  {Burles}, {Burns}, {Castander}, {Connolly}, {Davenport}, {Ebelke}, {Epps},
  {Feldman}, {Friedman}, {Frieman}, {Heckman}, {Hull}, {Knapp}, {Lawrence},
  {Loveday}, {Mannery}, {Malanushenko}, {Malanushenko}, {Merrelli}, {Muna},
  {Newman}, {Nichol}, {Oravetz}, {Pan}, {Pope}, {Ricketts}, {Shelden},
  {Sandford}, {Siegmund}, {Simmons}, {Smith}, {Snedden}, {Schneider},
  {SubbaRao}, {Tremonti}, {Waddell}, \& {York}}]{2013AJ....146...32S}
{Smee}, S.~A., {Gunn}, J.~E., {Uomoto}, A., {et~al.} 2013, \aj, 146, 32,
  \dodoi{10.1088/0004-6256/146/2/32}

\bibitem[{Smith \& Geach(2023)}]{SmithGeach2023}
Smith, M.~J., \& Geach, J.~E. 2023, Royal Society Open Science, 10, 221454,
  \dodoi{10.1098/rsos.221454}

\bibitem[{{Sofue} {et~al.}(2003){Sofue}, {Koda}, {Nakanishi}, \&
  {Onodera}}]{2003PASJ...55...59S}
{Sofue}, Y., {Koda}, J., {Nakanishi}, H., \& {Onodera}, S. 2003, \pasj, 55, 59,
  \dodoi{10.1093/pasj/55.1.59}

\bibitem[{{Sofue} {et~al.}(1999){Sofue}, {Tutui}, {Honma}, {Tomita},
  {Takamiya}, {Koda}, \& {Takeda}}]{1999ApJ...523..136S}
{Sofue}, Y., {Tutui}, Y., {Honma}, M., {et~al.} 1999, \apj, 523, 136,
  \dodoi{10.1086/307731}

\bibitem[{{Stone} {et~al.}(2022){Stone}, {Courteau}, {Arora}, {Frosst}, \&
  {Jarrett}}]{PROBES-I}
{Stone}, C., {Courteau}, S., {Arora}, N., {Frosst}, M., \& {Jarrett}, T.~H.
  2022, \apjs, 262, 33, \dodoi{10.3847/1538-4365/ac83ad}

\bibitem[{{Stone} {et~al.}(2023){Stone}, {Courteau}, {Cuillandre}, {Hezaveh},
  {Perreault-Levasseur}, \& {Arora}}]{astrophot}
{Stone}, C.~J., {Courteau}, S., {Cuillandre}, J.-C., {et~al.} 2023, \mnras,
  525, 6377, \dodoi{10.1093/mnras/stad2477}

\bibitem[{{Swaters} {et~al.}(2000){Swaters}, {Madore}, \&
  {Trewhella}}]{2000ApJ...531L.107S}
{Swaters}, R.~A., {Madore}, B.~F., \& {Trewhella}, M. 2000, \apjl, 531, L107,
  \dodoi{10.1086/312540}

\bibitem[{{Swaters} {et~al.}(2003){Swaters}, {Madore}, {van den Bosch}, \&
  {Balcells}}]{swaters2003}
{Swaters}, R.~A., {Madore}, B.~F., {van den Bosch}, F.~C., \& {Balcells}, M.
  2003, \apj, 583, 732, \dodoi{10.1086/345426}

\bibitem[{{Swaters} {et~al.}(2009){Swaters}, {Sancisi}, {van Albada}, \& {van
  der Hulst}}]{swaters2009}
{Swaters}, R.~A., {Sancisi}, R., {van Albada}, T.~S., \& {van der Hulst}, J.~M.
  2009, \aap, 493, 871, \dodoi{10.1051/0004-6361:200810516}

\bibitem[{Sánchez {et~al.}(2022)Sánchez, Barrera-Ballesteros, Lacerda,
  Mejía-Narvaez, Camps-Fariña, Bruzual, Espinosa-Ponce, Rodríguez-Puebla,
  Calette, Ibarra-Medel, Avila-Reese, Hernandez-Toledo, Bershady, Cano-Diaz, \&
  Munguia-Cordova}]{Pipe3D}
Sánchez, S.~F., Barrera-Ballesteros, J.~K., Lacerda, E., {et~al.} 2022, The
  Astrophysical Journal Supplement Series, 262, 36,
  \dodoi{10.3847/1538-4365/ac7b8f}

\bibitem[{Tollet {et~al.}(2016)Tollet, Macciò, Dutton, Stinson, Wang, Penzo,
  Gutcke, Buck, Kang, Brook, Di~Cintio, Keller, \& Wadsley}]{Tollet+2016}
Tollet, E., Macciò, A.~V., Dutton, A.~A., {et~al.} 2016, Monthly Notices of
  the Royal Astronomical Society, 456, 3542, \dodoi{10.1093/mnras/stv2856}

\bibitem[{{Trachternach} {et~al.}(2008){Trachternach}, {de Blok}, {Walter},
  {Brinks}, \& {Kennicutt}}]{Trachternach+2008}
{Trachternach}, C., {de Blok}, W.~J.~G., {Walter}, F., {Brinks}, E., \&
  {Kennicutt}, R.~C., J. 2008, \aj, 136, 2720,
  \dodoi{10.1088/0004-6256/136/6/2720}

\bibitem[{{Trachternach, C.} {et~al.}(2009){Trachternach, C.}, {de Blok, W. J.
  G.}, {McGaugh, S. S.}, {van der Hulst, J. M.}, \& {Dettmar,
  R.-J.}}]{Trachternach+2009}
{Trachternach, C.}, {de Blok, W. J. G.}, {McGaugh, S. S.}, {van der Hulst, J.
  M.}, \& {Dettmar, R.-J.} 2009, A\&A, 505, 577,
  \dodoi{10.1051/0004-6361/200811136}

\bibitem[{Vikram {et~al.}(2010)Vikram, Wadadekar, Kembhavi, \&
  Vijayagovindan}]{PyMorphPackage}
Vikram, V., Wadadekar, Y., Kembhavi, A.~K., \& Vijayagovindan, G.~V. 2010,
  Monthly Notices of the Royal Astronomical Society, 409, 1379,
  \dodoi{10.1111/j.1365-2966.2010.17426.x}

\bibitem[{Virtanen {et~al.}(2020)Virtanen, Gommers, Oliphant, Haberland, Reddy,
  Cournapeau, Burovski, Peterson, Weckesser, Bright, {van der Walt}, Brett,
  Wilson, Millman, Mayorov, Nelson, Jones, Kern, Larson, Carey, Polat, Feng,
  Moore, {VanderPlas}, Laxalde, Perktold, Cimrman, Henriksen, Quintero, Harris,
  Archibald, Ribeiro, Pedregosa, {van Mulbregt}, \& {SciPy 1.0
  Contributors}}]{SciPy}
Virtanen, P., Gommers, R., Oliphant, T.~E., {et~al.} 2020, Nature Methods, 17,
  261, \dodoi{10.1038/s41592-019-0686-2}

\bibitem[{Wright {et~al.}(2010)Wright, Eisenhardt, Mainzer, Ressler, Cutri,
  Jarrett, Kirkpatrick, Padgett, McMillan, Skrutskie, Stanford, Cohen, Walker,
  Mather, Leisawitz, Gautier, McLean, Benford, Lonsdale, Blain, Mendez, Irace,
  Duval, Liu, Royer, Heinrichsen, Howard, Shannon, Kendall, Walsh, Larsen,
  Cardon, Schick, Schwalm, Abid, Fabinsky, Naes, \& Tsai}]{WISE}
Wright, E.~L., Eisenhardt, P. R.~M., Mainzer, A.~K., {et~al.} 2010, The
  Astronomical Journal, 140, 1868, \dodoi{10.1088/0004-6256/140/6/1868}

\end{thebibliography}

\appendix

\section{Data Sample}
\label{app:data_app}

This Appendix provides a broad overview of the rotation curves and photometry for the PROBES-I \citep{PROBES-I}, PROBES-II \citep{frosst+2022}, and MaNGA \citep{MaNGA-I,MaNGA-II} catalogues used here.
Refer to the original publications for greater details.

\begin{table*}
\singlespace
\centering
\begin{tabular}{ccccc}
\hline
\textbf{Source}             & \textbf{$N$} & \textbf{Obs. Type} & \textbf{log(M$_{\text{dyn}}$/M$_{\odot}$) Range} \\
\hline
\multicolumn{4}{c}{\textbf{PROBES-I}}                                                                                                                                                                                           \\ \hline
\citet{C97}                 & 296                         & H$\alpha$                                                                       & 8.5 - 10.0                                                        \\
\citet{2000ApJ...544..636C} & 171                         & H$\alpha$                                                                    & 10.5 - 12.1                                                       \\
\citet{1999AJ....118.1489D}            & 522                         & H$\alpha$                                                                     & 7.9 - 12.1                                                        \\
\citet{Lelli+2016}             & 175                         & HI + H$\alpha$                                           & 7.5 - 11.7                                                        \\
\citet{Mathewson+1992}       & 744                         & HI + H$\alpha$                                                 & 9.4 - 12.0                                                        \\
\citet{MathewsonFord1996}       & 1216                        & H$\alpha$                                                                    & 8.5 - 12.0                                                        \\
\citet{2017ApJ...843...74O}                 & 44                          & H$\alpha$                                                                       & 6.5 - 11.0 \\ \hline
\multicolumn{4}{c}{\textbf{PROBES-II}}                                                                                                                  \\ \hline
\citet{2014ApJ...789...63A} & 7                           & H$\beta$ + {[}OIII{]}                                                           & 9.5 - 10.2                                                        \\
\citet{2003NewA....8..267B} & 1                           & HI                                                                      & 8.0                                                               \\
\citet{broeils92}           & 12                          & HI                                                            & 9.3 - 11.1                                                        \\
\citet{db2002}              & 26                          & H$\alpha$                                                                       & 9.0 - 10.7                                                        \\
\citet{2004ApJ...608..189D} & 1                           & HI                                                                      & 9.5                                                               \\
\citet{2008AJ....136.2648D} & 19                          & HI                                                              & 10.0 - 11.1                                                       \\
\citet{2008MNRAS.390..466E} & 97                          & H$\alpha$                                                                      & 8.2 - 12.1                                                        \\
\citet{2005MNRAS.362..127G} & 24                          & H$\alpha$                                                                    & 9.9 - 11.3                                                        \\
\citet{kauff15}             & 106                         & H$\alpha$ + $\sigma_{\rm abs}$                                                      & 9.4 - 11.4                                                        \\
\citet{2012MNRAS.420.2924K} & 12                          & HI                                                              & 9.3 - 9.9                                                         \\
\citet{2008ApJ...676..920K} & 9                           & H$\alpha$                                                                     & 8.3 - 10.5                                                        \\
\citet{Lelli+2016}             & 175                         & HI + H$\alpha$                                                & 7.5 - 11.7                                                        \\
\citet{martinsson2013}      & 30                          &        [OIII]                                                                       & 9.8 - 11.4                                                        \\
\citet{2001AJ....122.2381M} & 36                          & HI + H$\alpha$                                                 & 9.6 - 10.6                                                        \\
\citet{2007MNRAS.376.1513N} & 1                           & HI                                                               & 10.4                                                              \\
\citet{oh11}                & 7                           & HI                                                             & 8.7 - 10.2                                                        \\
\citet{2015AJ....149..180O} & 26                          & HI                                                             & 9.5 - 11.1                                                        \\
\citet{1999ApJ...523..136S} & 40                          & H$\alpha$ + CO                                                                & 8.8 - 11.6                                                        \\
\citet{2003PASJ...55...59S} & 12                          & CO                                                                          & 8.8 - 11.6                                                        \\
\citet{2000ApJ...531L.107S} & 5                           & H$\alpha$                                                                     & 10.3 - 10.9                                                       \\
\citet{swaters2003}           & 15                          & H$\alpha$                                                                   & 10.5 - 10.7                                                       \\
\citet{swaters2009}         & 62                          & HI + CO                                                      & 8.8 - 10.4                                                        \\
\citet{Trachternach+2009}    & 11                          & HI                                                            & 8.5 - 10.0  \\ \hline
\end{tabular}
\caption{RC sources for the PROBES-I and PROBES-II data sets. 
Adapted from \cite{frosst+2022}. 
$N$ is the number of galaxies from each source, the method of observation is shown in the third column, and
the fourth column shows the dynamical mass range of each source.}
\label{tab:PROBES_sources}
\end{table*}

\subsection{PROBES-I}\label{sec:data_PROBESI}

Kinematic data were taken from seven different sources for PROBES-I (N=3168; see \Table{PROBES_sources}).
These galaxies populate the local universe, with only a few (15) having cosmological redshift $z>0.1$ ($z_{\rm max}=0.1867$).
Inclinations range from 20$^{\circ}$ to 90$^{\circ}$, though galaxies closest to edge-on will be removed due to improper photometric inclination corrections (see \Sec{corrections}) and transparency issues.
Most galaxies have Hubble types between Sb and Scd, a notable number of Sa spirals are also included. 
Most RCs were obtained from H$\alpha$ long-slit spectra, but some come from HI observations or integral field spectroscopy (IFS).

\subsection{PROBES-II}\label{sec:data_PROBESII}
The PROBES-II extension (N=356 galaxies; see \Table{PROBES_sources}) aims to expand the stellar mass and morphological coverage of the larger PROBES-I data set.
PROBES-II galaxies are still local ($z<0.05$), mostly of irregular types (Im, Sm) and low-mass bulge-less spirals (Sd), and with inclinations ranging from 20$^{\circ}$ to 85$^{\circ}$.
There is a small overlap between the PROBES-I and PROBES-II samples. 
For all duplicates, measurements from PROBES-I are used. 

The PROBES-II sample includes multiple \ion{H}{I} observations, though CO interferometry, H$\alpha$ long-slit spectroscopy, and H$\alpha$/H$\beta$/[OIII] IFS also exist.
PROBES-I and PROBES-II galaxies typically extend out to 0.9$R_{23.5}$ and 1.5$R_{23.5}$, respectively.

\subsection{MaNGA}\label{sec:data_MaNGA}

The sample of spiral galaxies extracted by \cite{MaNGA-I, MaNGA-II}, hereafter referred to as our ``MaNGA'' sample ($N=2368$ galaxies), is in fact a subset of the larger Sloan Digital Sky Survey (SDSS) Data Release 17 data set of the same name \citep{2006AJ....131.2332G,Bundy+2015,2013AJ....146...32S,2015AJ....149...77D,2022ApJS..259...35A}.
\cite{MaNGA-II} selected spiral galaxies from the SDSS data set and extracted one-dimensional RCs and homogeneous photometry (see \Sec{DESI_photo} for the latter). 
While still viewed as local, MaNGA galaxies cover a slightly higher redshift range than the other two samples ($z<0.15$).
Inclinations range from 30$^{\circ}$ to 80$^{\circ}$.
MaNGA RC data were not included in \Table{PROBES_sources} since they all come from the same source.
The MaNGA velocity cubes from IFS were modelled with a tanh RC fitting function to produce one-dimensional RCs.

\subsection{Photometry}
\label{sec:DESI_photo}
The Dark Energy Spectroscopic Instrument Legacy Imaging Surveys \citep[DESI-LIS;][]{DESI} was used to extract uniform photometry in the optical $g$, $r$, and $z$ bands for our three data sets (PROBES-I, PROBES-II, MaNGA).
For PROBES-I, we also used mid-infrared imaging from the Wide-Field Infrared Survey Explorer \citep[WISE;][]{WISE} in the $W1$ and $W2$ bands that were processed and ``unblurred'' by the unWISE project \citep{unWISE}.
The multi-band imaging enabled us to utilize several color-mass-to-light relations to build robust mass-to-light ratio estimates in \Sec{stellarmasscalcs}.

\section{Bulge-Disk Decompositions}
\label{app:BD_decomp}

This Appendix provides a description of our $B/D$ decomposition procedure, which builds upon \cite{MacArthur2003}.
Final distributions of key properties and comparison with the literature are shown in \Sec{BD_decomps}.

The radial bulge component is modelled as a S{\'e}rsic profile~\citep{Sersic1968}:
\be
\label{eq:BD_bulge}
    I_{\rm b}(R) = I_{\rm e} \, \exp{\left\{ -b_n \left[ \left( \frac{R}{R_{\rm e}} \right)^{1/n} - 1\right]\right\}} ,
\ee
where $R_{\rm e}$ is the projected radius which encloses half of the modelled light, $I_{\rm e}$ is the intensity at $R_{\rm e}$, $n$ is the S{\'e}rsic shape parameter, and $b_n$ is a normalization factor approximated as \citep{CiottiBertin1999}:
\be
\label{eq:bn_high}
    b_n \approx 2n - \frac{1}{3} + \frac{4}{405n} + \frac{46}{25525n^2} + \frac{131}{1148175n^3} - \frac{2194697}{30690717750n^4} , 
\ee
for $n>0.36$, and as \citep{MacArthur2003}:
\be
\label{eq:bn_low}
    b_n \approx 0.01945 - 0.8902n + 10.95n^2 - 19.67n^3 + 13.43n^4 ,
\ee
for $n\leq0.36$. 
The distinction between the two $n$ ranges avoids a divergence of \Eq{bn_high} at low $n$ values. 
Combined, these two approximations provide accurate values of $b_n$ for the typical range of fitted $n$ values
($0 \leq n \leq 4$). 

The radial disk component is modelled as a thin exponential disk~\citep{Freeman1970}:
\be
\label{eq:BD_disk}
    I_{\rm d}(R) = I_0 \, \exp{\left\{-\frac{R}{R_{\rm d}} \right\}} ,
\ee
where $I_0$ is the central intensity and $R_{\rm d}$ is the disk scale length.
The total radial light profile of a galaxy is then:
\be
\label{eq:BD_combined}
    I(R) = I_{\rm b}(R) + I_{\rm d}(R) .
\ee

While straightforward in theory, $B/D$ decompositions are nuanced and particularly sensitive to initial estimates and bounds placed on each free parameter \citep{MacArthur2003,Lange2016,Gilhuly2018}.
Performing a naive fit for all five free parameters ($I_0$, $R_{\rm d}$, $I_{\rm e}$, $R_{\rm e}$, $n$) may result in unrealistic values which have little correspondence to the physical definition of each parameter.
We instead use a two-pass fitting procedure designed to assert the known spiral galaxy structure of an inner bulge (if present) and outer disk. 
The first-pass independently fits the bulge and disk components to obtain reasonable initial estimates for the second-pass, which fits both components simultaneously. 
Throughout the fitting procedure, several choices and assumptions are made with the goal of limiting some of the common failure modes of only performing a single-pass five parameter fit.

Various B/D decomposition concerns include: 
\begin{itemize}
    \item A ``bulge'' being fit for most of the SB profile with Eq.~\ref{eq:BD_bulge}, while a stellar halo is interpreted as a ``disk'' via Eq.~\ref{eq:BD_disk}. 
    This is of course an improper characterization of each component by the fitting algorithm.
    In general, the disk component is especially sensitive to being artificially suppressed.
    \item Background contamination not being removed by our cleaning procedure (described in Section~\ref{sec:SB_truncation}), and the algorithm fitting the flat contaminated profile as a ``disk'' instead of the galaxy's actual disk.
    \item The presence of multiple local minima in the loss function surface leading to optimized parameters that are not physically meaningful.
    This, in combination with the likelihood of improper implementation of free parameters increasing with the number of free parameters, results in an unconstrained $B/D$ fitting scheme being highly susceptible to unrealistic best-fitting values.
\end{itemize}

Our first-pass isolates the outer 75 per cent of the surface brightness profile ($0.25R_{\rm max} < R \leq R_{\rm max})$, a regime that is typically disk dominated, and fits \Eq{BD_disk} for the $I_0$ and $R_{\rm d}$ free parameters.
We took the innermost point of the disk isolated profile as an initial estimate for $I_0$.
We calculated $R_{\rm d,ini}$ using the mean of the $R_{\rm d}$ values derived from \Eq{BD_disk} for each point in the profile.
We limited both free parameters to positive values, and limit $R_{\rm d}$ to $10R_{\rm d,ini}$ at the upper end.
No upper bound was placed on $I_0$.

The remaining inner 25 per cent is then fit as a bulge through Eq.~\ref{eq:BD_bulge} with $I_{\rm e}$, $R_{\rm e}$, and $n$ as free parameters.
We took the innermost point of the bulge isolated profile as initial estimates for $I_{\rm e}$ and $R_{\rm e}$.
For their $B/D$ decompositions of late-type spiral galaxies (similar morphological distribution to ours), 
\cite{MacArthur2003} showed that the peak of the $n$ distribution occurs at $n = 1$. 
We therefore adopted this as an initial estimate for our first-pass $n$.
We limited $I_{\rm e}$ and $R_{\rm e}$ to positive values again, with no upper bound on $I_{\rm e}$.
We imposed the bound that $R_{\rm e} < R_{\rm d}$, based on the assumption that bulges should be more compact than disks.
We allowed $n$ to range between $0 \leq n \leq 4$ (the limits here corresponding to a pure bulge-less disk and a classic de~Vaucouleurs bulge). 

Using the fitted parameters from the first-pass as initial estimates, we performed the five parameter second-pass for a complete profile fit.
Both disk parameters were allowed to float $\pm50$ per cent relative to the first-pass estimates.
In addition to the original $0\leq n \leq4$ range, the $n$ parameter was limited to $\pm0.5$ of its first-pass value so the shape of the bulge does not greatly change in the second-pass.
The remaining parameters were given broader bounds, and all limits are summarized in \Table{BD_bounds}.

\begin{table}
\centering
\singlespace
\begin{tabular}{lllll}
\hline
\multicolumn{1}{c}{}          & \multicolumn{2}{c}{First Pass}                        & \multicolumn{2}{c}{Second Pass}                           \\ \hline
\multicolumn{1}{c}{Parameter} & \multicolumn{1}{c}{Lower} & \multicolumn{1}{c}{Upper} & \multicolumn{1}{c}{Lower}   & \multicolumn{1}{c}{Upper}   \\ \hline
$I_0$                         & 0                         & $\infty$                  & $0.5\,I_{0,\rm FP}$         & $1.5\,I_{0,\rm FP}$         \\
$R_{\rm d}$                   & 0                         & $5\,R_{\rm d,ini}$        & $0.5\,R_{\rm d,FP}$         & $1.5\,R_{\rm d,FP}$         \\
$I_{\rm e}$                   & 0                         & $\infty$                  & $0.1\,I_{\rm e,FP}$         & $3\,I_{\rm e,FP}$           \\
$R_{\rm e}$                   & 0                         & $R_{\rm d,FP}$            & $0.33\,R_{\rm e,FP}$        & $R_{\rm d,FP}$              \\
$n$                           & 0                         & 4                         & $\max{(n_{\rm FP}-0.5, 0)}$ & $\min{(n_{\rm FP}+0.5, 4)}$ \\ \hline
\end{tabular}
\caption{Bounds for each free parameter during the two-pass B/D decomposition fitting procedure. 
The five parameters are disk central intensity ($I_{0}$), disk scale length ($R_{\rm d}$), bulge effective intensity ($I_{\rm e}$), bulge effective radius ($R_{\rm e}$), and bulge S\'ersic index ($n$). 
The subscript FP denotes ``first-pass''.
}
\label{tab:BD_bounds}
\end{table}

Our results were filtered obeying two criteria: the first is a fit quality metric based on the root mean squared (RMS) deviation of the fit relative to observed data, the second is based on the bulge-to-disk light ratio.
We calculated the RMS deviation for the inner 85 per cent of a profile, as the outermost data points may be affected by remaining background noise or the features of a stellar halo.
Galaxies with fits resulting in higher than 0.5 r-\magss RMS error were removed from analysis (N = 160).
The bulge-to-disk light ratio ($B/D$) was used to assess the dominance of the galaxy's bulge (see \Eq{bulge_total} and \Eq{disk_total} below).
The URC parameterization that we tested used $R_{\rm d}$, and since galaxies with a minor disk component would likely have erroneous $R_{\rm d}$ values (that  misrepresent the underlying mass distribution), we discarded galaxies with $B/D > 10$ (N = 24).
Our $B/D$ decomposition procedure failed completely for 55 of our sample galaxies, reducing the total number of galaxies from 4298 to 4049.

\begin{figure}
\singlespace
\centering
\includegraphics[width=\textwidth]{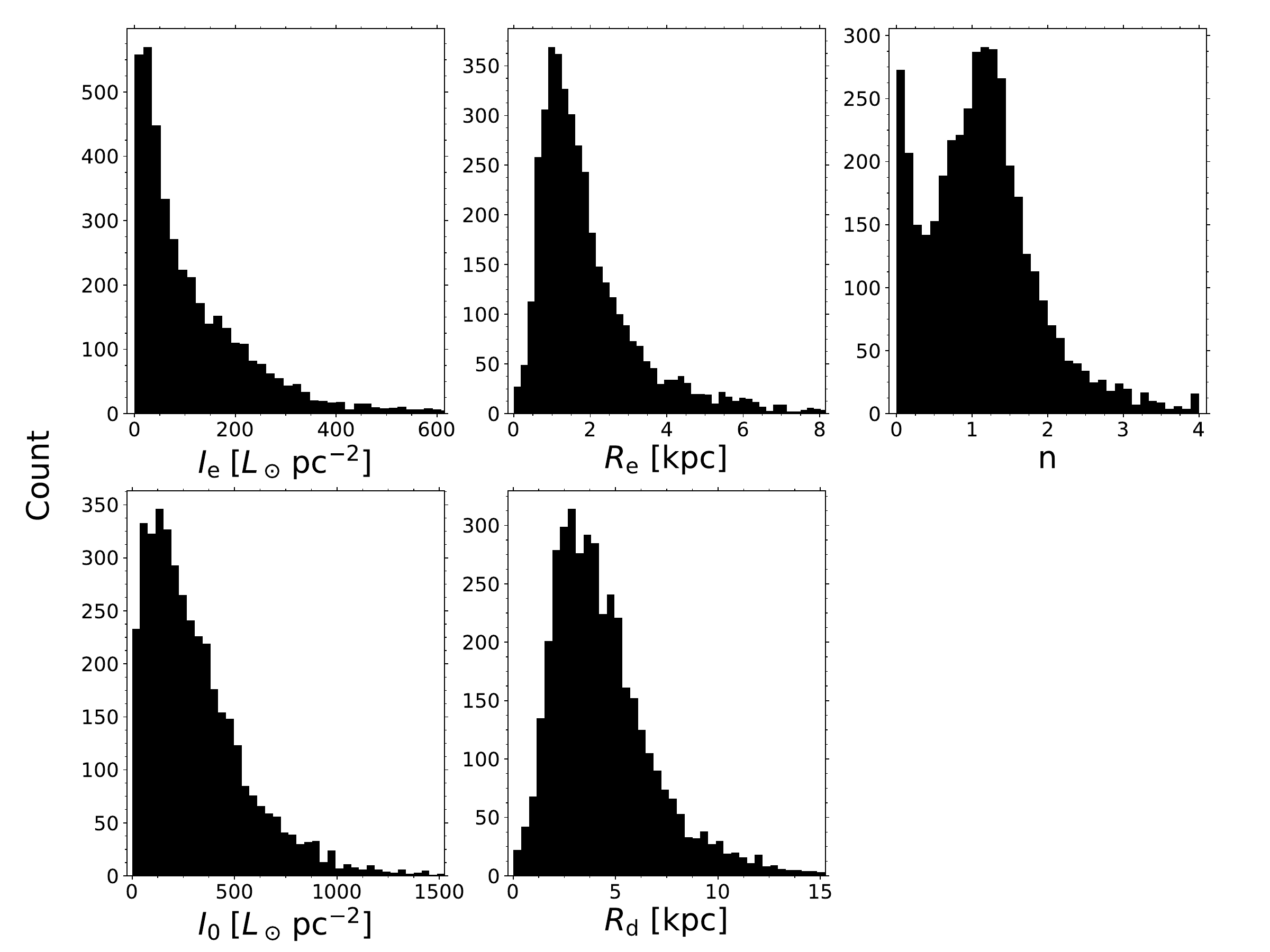}
\caption{Distribution of best-fitting values for each bulge-disk decomposition parameter. 
The extended tails of each distribution (with the exception of $n$, which is bounded between 0 and 4) are not shown to highlight the range where most of the distribution lies. 
The parameters are described in the caption for \Table{BD_bounds}.}
\label{fig:BD_params}
\end{figure}

The resulting distributions of each $B/D$ parameter are plotted in \Fig{BD_params}.
All distributions occupy a reasonable range in parameter space thus validating our  $B/D$ decomposition procedure.
Using these parameters, we calculated some key properties: bulge-to-disk light ratio ($B/D$) and bulge-to-total light ratio ($B/T$).
The total bulge light is calculated as:
\be
\label{eq:bulge_total}
    L_{\rm b} = \frac{2\pi I_{\rm e} R_{\rm e}^2 e^{b_n} n \Gamma(2n)}{b_n^{2n}} ,
\ee
where $\Gamma(2n)$ is the gamma function evaluated at $2n$, the total disk light calculated as:
\be
\label{eq:disk_total}
    L_{\rm d} = 2 \pi I_0 R_{\rm d}^2 ,
\ee
and so:
\be
    \frac{B}{D} = \frac{L_{\rm b}}{L_{\rm d}} ,
\ee
and: 
\be
\label{eq:BT_ratio}
    \frac{B}{T} = \frac{L_{\rm b}}{L_{\rm b} + L_{\rm d}} .
\ee

As well, we can calculate the bulge + disk stellar circular velocity contribution as a function of radius:
\be
\label{eq:stellar_V}
    V_*^2(R) = V^2_{\rm b}(R) + V^2_{\rm d}(R) ,
\ee
where $V_{\rm b}(R)$ is the bulge velocity contribution and $V_{\rm d}(R)$ is the disk velocity contribution (see \Eq{PSS96_disk_final}).
Similar to the disk velocity contribution, the mass enclosed within $R$ for the bulge component can be converted to circular velocity:
\be
    V_{\rm b}^2(R) = \Upsilon_* \frac{2\pi I_{\rm e} R_{\rm e}^2 e^{b_n} n}{b_n^{2n}}  \gamma{\left(2n, b_n \left( \frac{R}{R_{\rm e}} \right)^{1/n} \right)} \frac{G}{R},
\ee
where $\gamma$ is the incomplete gamma function.

The $n$ parameter shows a strong peak at $n\approx1.15$ in \Fig{BD_params}.
Our comparison sources all agree with this result \citep{MacArthur2003, Spitzer_BD_decomp, PyMorph, PyMorph-DR17}.
The bulges of disk galaxies are often well described by an exponential $n\approx1$ profile \citep{CourteauBroeils1996}.
The small peak at $n=4$ corresponds to the bound placed during the fitting process and is not a concern or cause to relax our bounds. 

A relevant consideration for disk scale lengths is their photometric band dependency. 
Disk scale lengths are greatly affected by dust extinction and stellar populations, which are themselves band-dependent \citep{Fathi+2010}. 
Our use of the redder $r$-band is an attempt to lessen the impact of these biases on our light profiles by diminishing dust extinction effects and using an optical band that traces most galactic stellar mass. 

\begin{figure}
\singlespace
\centering
\includegraphics[width=\textwidth]{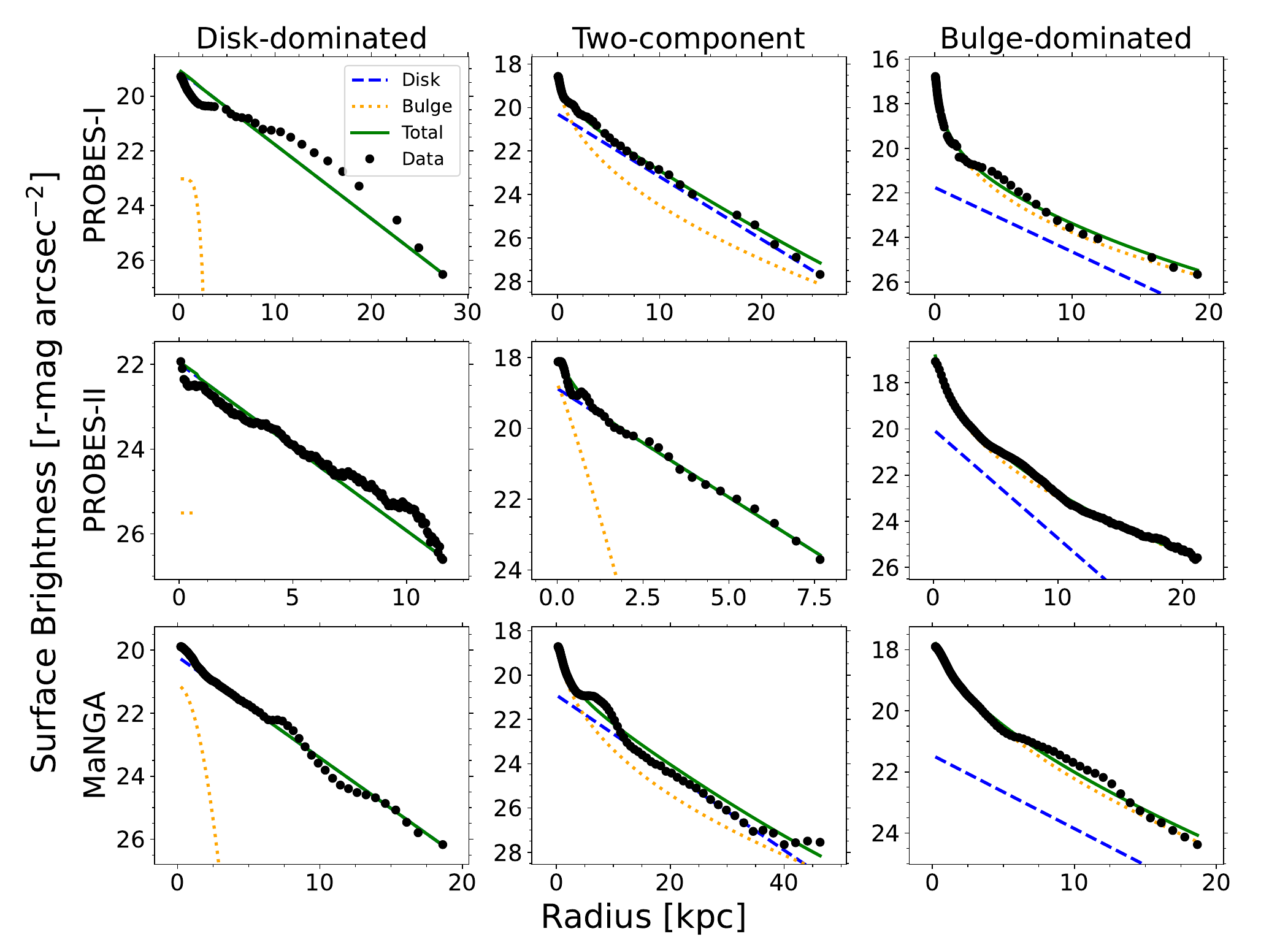}
\caption{Typical bulge-disk decompositions from each data set. 
From top row to bottom: PROBES-I, PROBES-II, MaNGA galaxies. 
Each column shows a different subset of the population. 
The left column shows galaxies largely described by their disk component, the middle column shows profiles where both components play an important role, and the right column shows examples with bulge dominance over the disk.
The light contribution from each component, as well as the combined contribution, are shown as dashed (disk), dotted (bulge), and solid (combined) lines.}
\label{fig:BD_examples}
\end{figure}

Another concern is the degeneracy between bulge-disk parameters.
For most one-dimensional surface brightness profiles, there is not an unique set of $B/D$ parameters that best describes it.
Often, some parameters can be adjusted arbitrarily, and as long as that change is accounted for by the other $B/D$ parameters, the overall fit quality will be similar.
With our bounded fitting process, we have attempted to assert the observed trend of inner bulge and outer disk, but we acknowledge this degeneracy.
Good agreement with the literature was found in \Sec{BD_decomps}, thus providing further support for our $B/D$ parameters.
\Fig{BD_examples} shows sample profiles for each of our morphological subsets.

\end{document}